\documentclass[fleqn,usenatbib]{mnras}

\usepackage{newtxtext,newtxmath}

\usepackage[T1]{fontenc}
\usepackage{hyperref}
\DeclareRobustCommand{\VAN}[3]{#2}
\let\VANthebibliography\thebibliography
\def\thebibliography{\DeclareRobustCommand{\VAN}[3]{##3}\VANthebibliography}

\usepackage{graphicx}	
\usepackage{amsmath}	

\usepackage{ulem} 

\begin{document}
\title[The importance of thermal torques]{The importance of thermal torques on the migration of planets growing by pebble accretion}

\author[O. M. Guilera et al.]{Octavio M. Guilera,$^{1,2,3}$\thanks{E-mail: oguilera@fcaglp.unlp.edu.ar},
Marcelo M. Miller Bertolami$^{1,4}$,
Frederic Masset$^{5}$,
\newauthor Jorge Cuadra$^{6,3}$,
Julia Venturini$^{7}$,
Mar\'{\i}a P. Ronco$^{2,3}$
\\ \\
$^{1}$ Instituto de Astrof\'{\i}sica de La Plata, CONICET-UNLP, La Plata, Argentina. \\
$^{2}$ Instituto de Astrof\'{\i}sica, Pontificia Universidad Cat\'olica de Chile, Santiago,  Chile. \\
$^{3}$ N\'ucleo Milenio de Formaci\'on Planetaria (NPF), Chile. \\
$^{4}$ Facultad de Ciencias Astron\'omicas y Geof\'{\i}sicas, UNLP, La Plata, Argentina. \\
$^{5}$ Instituto de Ciencias F\'{\i}sicas, Universidad Nacional Aut\'onoma de M\'exico, Av. Universidad s/n, 62210 Cuernavaca, Mor., Mexico, \\
$^{6}$ Departamento de Ciencias, Facultad de Artes Liberales, Universidad Adolfo Ib\'a\~nez, Avenida Padre Hurtado 750, Vi\~na del Mar, Chile. \\
$^{7}$ International Space Science Institute, Hallerstrasse 6, CH-3012 , Bern, Switzerland.
}

\date{Accepted 1BC. Received 2BC; in original form 3BC}

\pubyear{1AD}

\label{firstpage}
\pagerange{\pageref{firstpage}--\pageref{lastpage}}
\maketitle

\begin{abstract}
A key process in planet formation is the exchange of angular momentum between a growing planet and the protoplanetary disc, which makes the planet migrate through the disc. Several works show that in general low-mass and intermediate-mass planets migrate towards the central star, unless corotation torques become dominant. Recently, a new kind of torque, called the thermal torque, was proposed as a new source that can generate outward migration of low-mass planets. While the Lindblad and corotation torques depend mostly on the properties of the protoplanetary disc and on the planet mass, the thermal torque depends also on the luminosity of the planet, arising mainly from the accretion of solids. Thus, the accretion of solids plays an important role not only in the formation of the planet but also in its migration process. In a previous work, we evaluated the thermal torque effects on planetary growth and migration mainly in the planetesimal accretion paradigm. In this new work, we study the role of the thermal torque within the pebble accretion paradigm. Computations are carried out consistently in the framework of a global model of planet formation that includes disc evolution, dust growth and evolution, and pebble formation. We also incorporate updated prescriptions of the thermal torque derived from high resolution hydrodynamical simulations. Our simulations show that the thermal torque generates extended regions of outward migration in low viscosity discs. This has a significant impact in the formation of the planets. 
\end{abstract}




\section{Introduction}
\label{intro}

The theory of planet formation, in the framework of the core accretion mechanism, is currently being actively revised  \citep[see][for recent reviews]{Venturini20b, Liu2020}. While pioneer works proposed that at early stages planets grow by the accretion of planetesimals \citep{Safronov1969, P96}, most of the recent works consider that the planet core grows by the accretion of pebbles. This change of paradigm is based mainly on two results. The first one is the fact that pebbles are accreted more efficiently than kilometer-sized planetesimals \citep{OrmelK10, Lambrechts12, Lambrechts14}. The second one is that the streaming instability \citep{Youdin05, Johansen07} has been widely accepted as the most favourable mechanism to form kilometric planetesimals, and possibly embryo seeds, as a consequence of the direct gravitational collapse of the accumulation of pebbles. To trigger the streaming instability, the dust-to-gas ratio has to be locally enhanced in some place in the protoplanetary disc over the typical value of 0.01 \citep[e.g.][]{Carrera2015, Yang2017, Li2021}. By this mechanism, planetesimals and/or embryo seeds can form in preferential locations in the disc, but a significant amount of the solid mass of the disc still remains in dust and pebbles \citep[e.g][]{Drazkowska16, Drazkowska2017, Ormel2017, Drazkowska2018}.

Two important mechanisms in pebble accretion theory are the dust growth and the dust evolution, which determine two key quantities in the pebble accretion rates: the pebble Stokes numbers along the disc, and the pebble mass flux \citep{Lambrechts14}. Currently only a few works deal with the problem of planet formation by pebble accretion taking into account detailed models of dust growth and evolution \citep[e.g.][]{Guilera20, Venturini20c, Venturini20d, Drazkowska21}. These works show that the pebble mass flux and the pebble Stokes numbers along the disc, and hence the pebble accretion rates and planet formation, strongly depend on the properties and evolution of the gas and dust components of the protoplanetary disc.

Another key process in planet formation is the exchange of angular momentum between the planet and the disc that generates a change in the semi-major axis or migration of the planet. Generally, the total torque onto the planet is considered as the sum of two contributions, the Lindblad torque and the corotation torque. In typical protoplanetary discs wherein gas surface density and mid-plane temperature decay as one moves away from the central star, the total torque exerted by the disc over the planet is negative and the planet migrates inwards, towards the central star \citep[e.g.][]{Tanaka02}. When corotation torques become dominant, the total torque over the planet can become positive and regions of planet outward migration appear \citep[e.g.][]{Paardekooper2011, jm2017}. If discs are neither isothermal nor adiabatic, thermal diffusion introduces an additional torque, called the {\it thermal torque} \citep[e.g.][]{masset2017, VelascoRomero2020, Hankla2020, Chametla2021}. If the planet is not luminous, the thermal torque is known as the {\it cold torque} because the surrounding gas is cooler and denser than it would be if it behaved adiabatically, generating an additional torque \citep{Lega2014, masset2017} that is in general negative. If the planet is luminous and hence releases heat to the surrounding gas, an additional torque contribution appears, known as the {\it heating torque}. This contribution is in general positive \citep{Benitez-llambay2015, masset2017, Hankla2020, Chametla2021}. 

In our previous work \citep[][hereafter Paper I]{Guilera2019}, we incorporated in our global code of planet formation, called {\scriptsize PLANETALP} \citep{Ronco2017, Guilera2017b}, the prescriptions for the thermal torque derived from \citet[][hereafter M17]{masset2017}. In that work, we first constructed planet migration maps, assuming for simplicity constant solid accretion rates. We showed that the inclusion of the thermal torque --in particular the heating torque-- in the total torque over the planet, generates a new extended region of outward migration. When the thermal torque is not considered, the total torque is given by the type I migration recipes for non-isothermal discs derived by \citet[][hereafter JM17]{jm2017}. Then, we computed planet formation tracks considering that planets grow by the accretion of pebbles or planetesimals of a single size. We showed that planet formation tracks and final masses and semi-major axis of the planets can be very different if the thermal torque is included. 

More recently, in \citet{Guilera20} and \citet{Venturini20c,Venturini20d} we included in {\scriptsize PLANETALP} a detailed model of dust growth and evolution, and pebble formation. These works, as previous ones \citep[e.g.][]{Drazkowska16, Drazkowska2017}, showed that the time evolution of the Stokes numbers and pebble mass flux strongly depend on the properties and evolution of the gas and dust components of the disc. In addition, \citet{Venturini20c} showed that as time advances, the dust surface density significantly decays due to the finite size of the disc, highly affecting the time evolution of the pebble mass flux and the pebble acretion rates. Furthermore, they showed that for large values of the $\alpha$-viscosity ($\alpha \gtrsim 10^{-3}$) the efficiency in planet formation decays. This happens because the values of the Stokes numbers reached by the pebbles are low, and that the accretion rate of pebbles generally occurs in the 3D regime because of the scale-height of the pebbles is greater than the Hill radius of the planet \citep[e.g.][]{Morby15, Ormel2018}. 

In this work we study the importance of the thermal torque on the migration of low- and intermediate-mass planets growing by pebble accretion. Simulations are performed in the framework of our global model of planet formation, which now includes a detailed model of dust growth and evolution. In addition, we also incorporate the new cut-off functions of the thermal torque developed by \citet[][hereafter VRM20]{VelascoRomero2020}. In Paper I, we adopted for simplicity the conservative approach of dropping to zero the thermal torque when the planet mass becomes greater than the critical thermal mass. However, VRM20 recently showed that the decrease in the thermal torque above the critical thermal mass is not abrupt, and they provide analytical expressions for the cut-off functions of the heating and cold torque. 

The paper is organized in the following way: in Sec.~\ref{sec2}, we briefly describe our global model; in Sec.~\ref{sec3}, we compute planet migration maps for different $\alpha$-viscosity parameters; in Sec.~\ref{sec_4}, we compute planet formation tracks to study the impact of the updated expressions of the thermal torque; in Sec.~\ref{sec_5} we present a summary of our work and discussions about the implications of our results; finally, in Sec.~\ref{sec_6}, we leave our conclusions. 

\section{Brief model description}
\label{sec2} 

In order to construct the planet migration maps and to compute later the planet formation tracks, we use {\scriptsize PLANETALP}, our global code of planet formation and disc evolution described in detail in \citet{Ronco2017}, \citet{Guilera2017b}, Paper I, \citet{Guilera20}, and \citet{Venturini20c,Venturini20d}. 

In our model, the gaseous disc evolves in time by viscous accretion and X-ray photoevaporation due to the central star. We consider the disc is in vertical hydrostatic equilibrium, and we solve the classical disc vertical structure and transport equations. The heating sources are viscous heating and the irradiation from the central star. The heat is vertically transported by radiation and convection. 

The evolution of the dust and pebbles along the disc is computed following the approach given in \citet{Guilera20} and \citet{Venturini20c, Venturini20d}. We consider a discrete solid size distribution between 1~$\mu$m and a maximum size using 200 size bins \citep{Drazkowska16}. The maximum size of the solid particles at each radial bin is limited by dust coagulation, radial drift and fragmentation \citep{Birnstiel12}. We also note that when the viscosity in the disc midplane becomes very low, fragmentation can be driven by differential drift, which is also included in our model \citep[see Eqs. 8 -- 12 in][]{Guilera20}. The dust properties change at the water ice line, which is defined at the place where the midplane temperature equals 170 K (we note that this location changes in time due to disc evolution). We consider that silicate particles have a threshold fragmentation velocity of 1~m/s inside the ice line, while the ice-rich particles beyond the ice line have a threshold fragmentation velocity of 10~m/s \citep{Gundlach2015}. To compute the time evolution of the solid surface density we solve the corresponding advection-diffusion equation using the particle mass weighted mean drift velocities and the mass weighted mean Stokes numbers \citep{Drazkowska2016, Drazkowska2017}. We consider the material accreted by the planets and ice sublimation as sink terms. 

Regarding planet formation, we follow the growth of an initial Moon-mass embryo ($\text{M}_{\text{p}}= 0.01~\text{M}_{\oplus}$) by the concurrent accretion of pebbles and the surrounding gas as in \citet{Venturini20c, Venturini20d}. One important concept in the framework of pebble accretion is the pebble isolation mass. When the planet reaches the pebble isolation mass, it is able to significantly perturb the surrounding disc generating two pressure maxima at both sides of its orbit. The pressure maximum located beyond the planet's orbit acts as a pebble trap halting the pebble radial drift. Thus, pebble accretion stops when the planet reaches the pebble isolation mass. In this work, we adopt two different prescriptions for the pebble isolation mass, one given by \citet{Bitsch18}
\begin{eqnarray}
\text{M}_{\text{iso}}&=& 25 \left(\frac{h}{0.05} \right)^3  \left[  0.34 \left( \frac{\log(0.001)}{\log(\alpha)} \right)^4 + 0.66 \right] \times \nonumber \\
&& \left[ 1- \frac{ \frac{\partial {\log} P}{ \partial {\log} R }+ 2.5}{6} \right]~\text{M}_{\oplus},
\label{eq1_sec2}
\end{eqnarray}
and the other one given by \citet{Ataiee18}
\begin{eqnarray}
\text{M}_{\text{iso}}=   \left(\frac{\text{M}_{\star}}{\text{M}_{\oplus}}\right) \, h^3  \sqrt{82.33 \, \alpha + 0.03}~\text{M}_{\oplus},
\label{eq2_sec2}
\end{eqnarray}
$h$ being the disc aspect ratio, $P$ represents the disc pressure, and $\text{M}_{\star}$ denotes the mass of the central star ($\text{M}_{\star}= 1~\text{M}_{\odot}$ in the present paper)\footnote{We note there is a typo in the definition of Eq.~\ref{eq2_sec2} in \citet{Venturini20c}, where in Eq.~20 of that work the term $25~(h/0.05)^3$ should be $(\text{M}_{\star}/\text{M}_{\oplus})~h^3$.}. 

Finally, the gravitational interactions between the disc and the planet generate a net torque over the planet producing its migration along the disc. As in Paper I, we consider that the total torque over the planet is given by
\begin{eqnarray}
\Gamma_{\text{tot}}= \Gamma_{\text{type I}} + \Gamma_{\text{thermal}}.
\label{eq3_sec2}
\end{eqnarray}
$\Gamma_{\text{type I}}= \Gamma_{\text{Lind}} + \Gamma_{\text{cor}}$ represents the classical resonant torque associated to the type I migration which is the sum of the Lindblad and corotation torques. As in Paper I, we adopt the recipes from JM17 to compute $\Gamma_{\text{type I}}$. The thermal torque is given by the sum of the cold and heating torques $\Gamma_{\text{thermal}}= \Gamma_{\text{cold}} + \Gamma_{\text{heating}}$ following M17. The recipes from M17 were developed in the framework of a linear perturbation theory. \citet{Masset_VelascoRomero17} argued that the estimations of the thermal torque obtained by a linear analysis are valid while the planet mass remains smaller than the critical thermal mass
\begin{eqnarray}
\text{M}_{\text{crit}}= \chi c_s / G,
\label{eq4_sec2}
\end{eqnarray}
where $\chi$ is the disc thermal diffusivity, $c_s$ is the adiabatic sound speed, and $G$ is the gravitational constant. Thus, in Paper I we adopted the conservative approach of dropping to zero the thermal torque after the planet mass becomes greater than the critical thermal mass. However,  VRM20 recently found through high resolution hydrodynamical simulations, that while a planet with a mass smaller than the critical thermal mass is subjected to a thermal force with a magnitude in good agreement with the linear theory developed by M17, the ratio of the heating torque to its linear estimate (the cut off function) decays slowly for masses in excess of the critical thermal mass. In addition, VRM20 give an approximate expression that fits the numerical results:
\begin{eqnarray}
\Gamma_{\text{thermal}}= \Gamma_{\text{heating}} \dfrac{4\text{M}_{\text{crit}}}{\text{M}_{\text{p}} +  4\text{M}_{\text{crit}}} + \Gamma_{\text{cold}} \dfrac{2\text{M}_{\text{crit}}}{\text{M}_{\text{p}} + 2\text{M}_{\text{crit}}}, 
\label{eq5_sec2}
\end{eqnarray}
$\text{M}_{\text{p}}$ being the mass of the planet. We make use of Eq.~(\ref{eq5_sec2}) in the present work instead of the abrupt cut off used in Paper I. Transition to the type II migration regime occurs once the planet opens a gap in the gaseous disc, following the criterion derived by \citet{Crida2006}, and  migrates according to \citet{IdaLin2004a} or \citet{Mordasini2009} depending on whether it is in the "disc-dominated" or "planet-dominated" regime.

\section{Migration maps}
\label{sec3}

\begin{figure}
    \centering
    \includegraphics[width=\columnwidth]{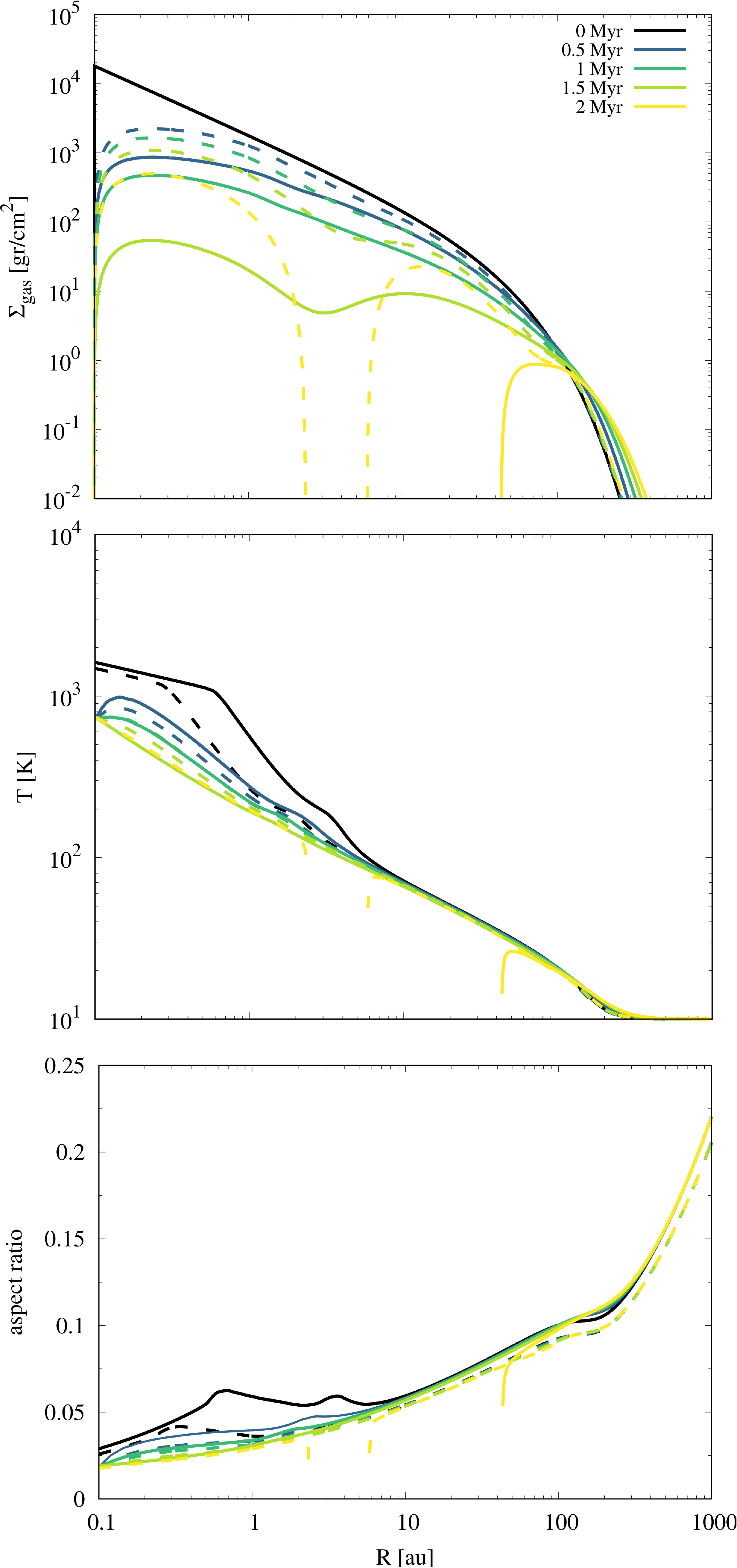}
    \caption{Comparison of the time evolution of the radial profiles of the gas surface density (top panel), mid-plane temperature (center panel), and disc aspect ratio (bottom panel) between simulations adopting the fiducial disc with $\alpha= 10^{-3}$ (solid lines) and $\alpha= 10^{-4}$ (dashed lines).} 
    \label{fig1_sec3}
\end{figure}

\begin{figure}
    \includegraphics[angle= 270,     width=\columnwidth]{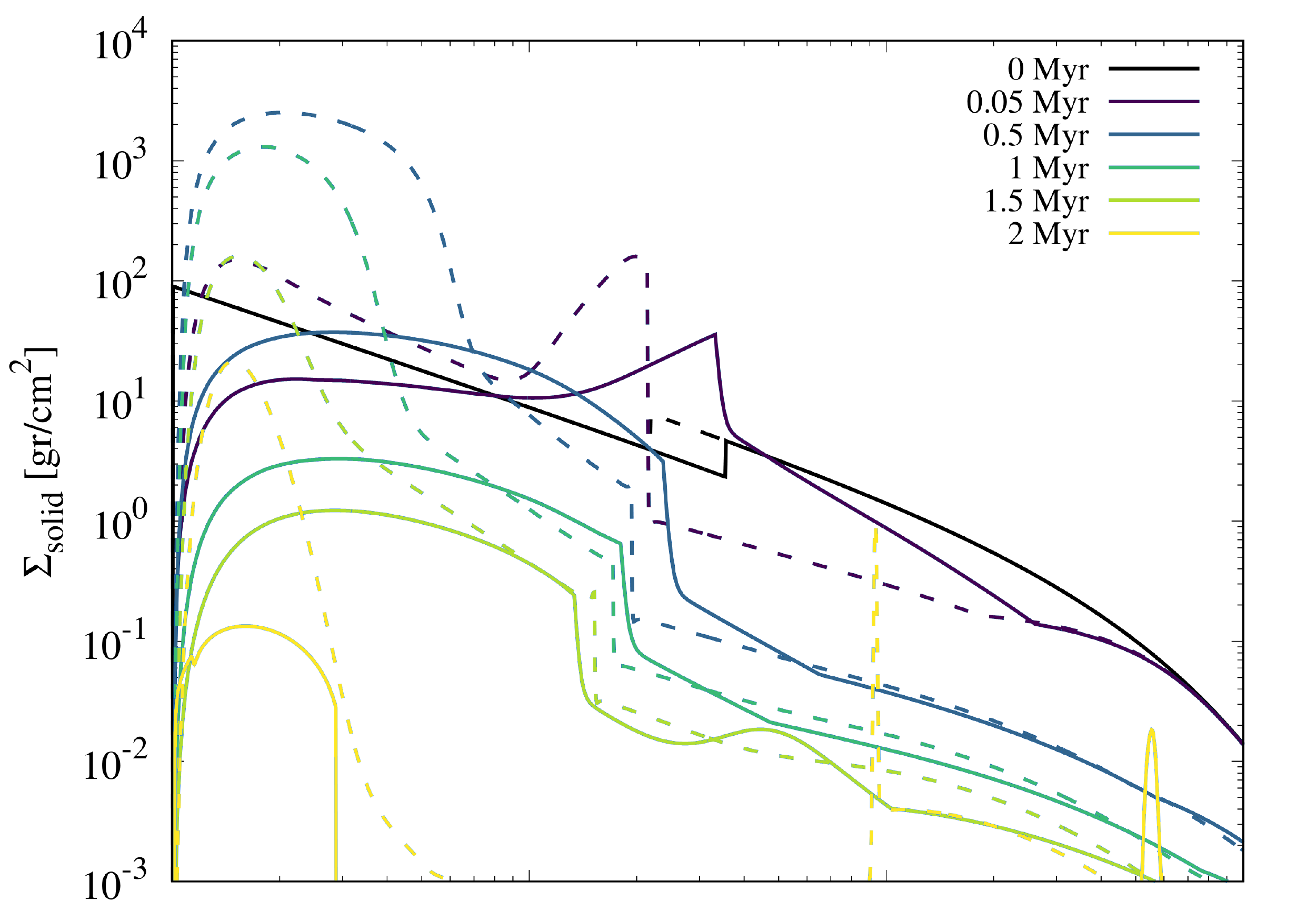} \\
    \includegraphics[angle= 270, width=\columnwidth]{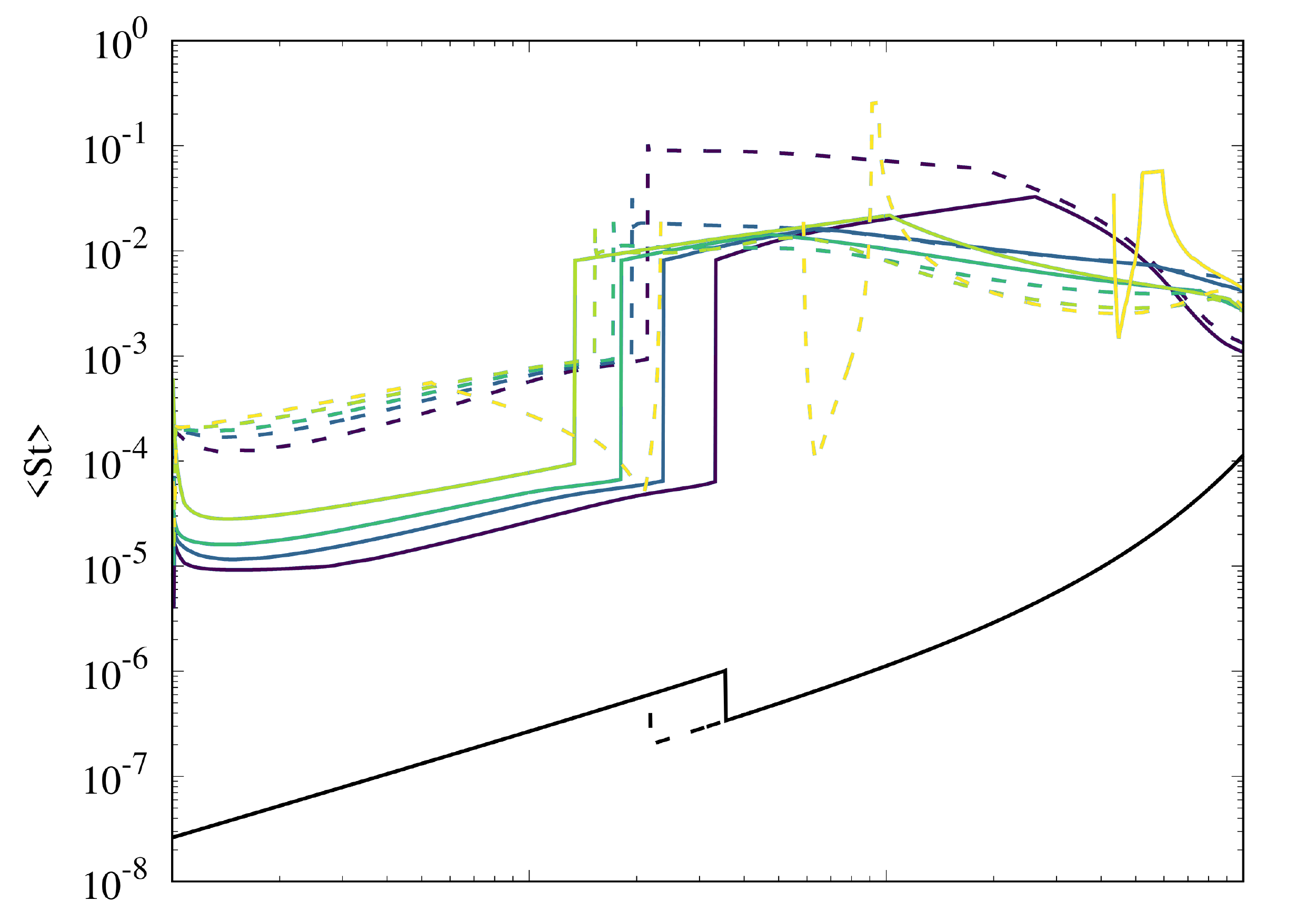} \\
    \includegraphics[angle= 270,     width=\columnwidth]{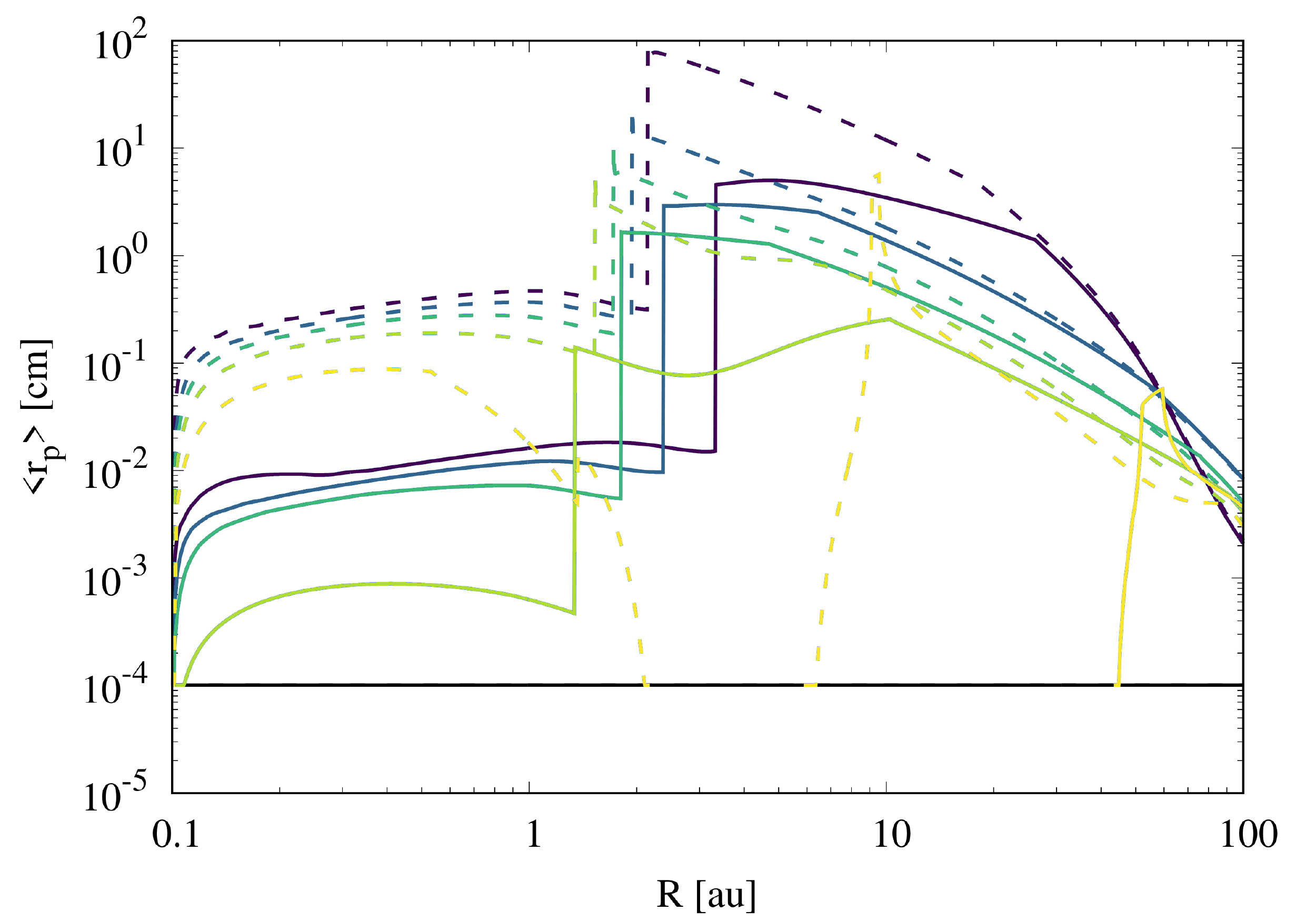}
    \caption{Time evolution of the radial profiles of the solid particles surface density (top panel), the weighted mean Stokes numbers (center panel) and weighted mean solid particles sizes (bottom panel) for the simulations using the fiducial disc with $\alpha= 10^{-3}$ (solid lines) and $\alpha= 10^{-4}$ (dashed lines).}
    \label{fig2_sec3}
\end{figure}

\begin{figure}
    \includegraphics[angle= 270,     width=\columnwidth]{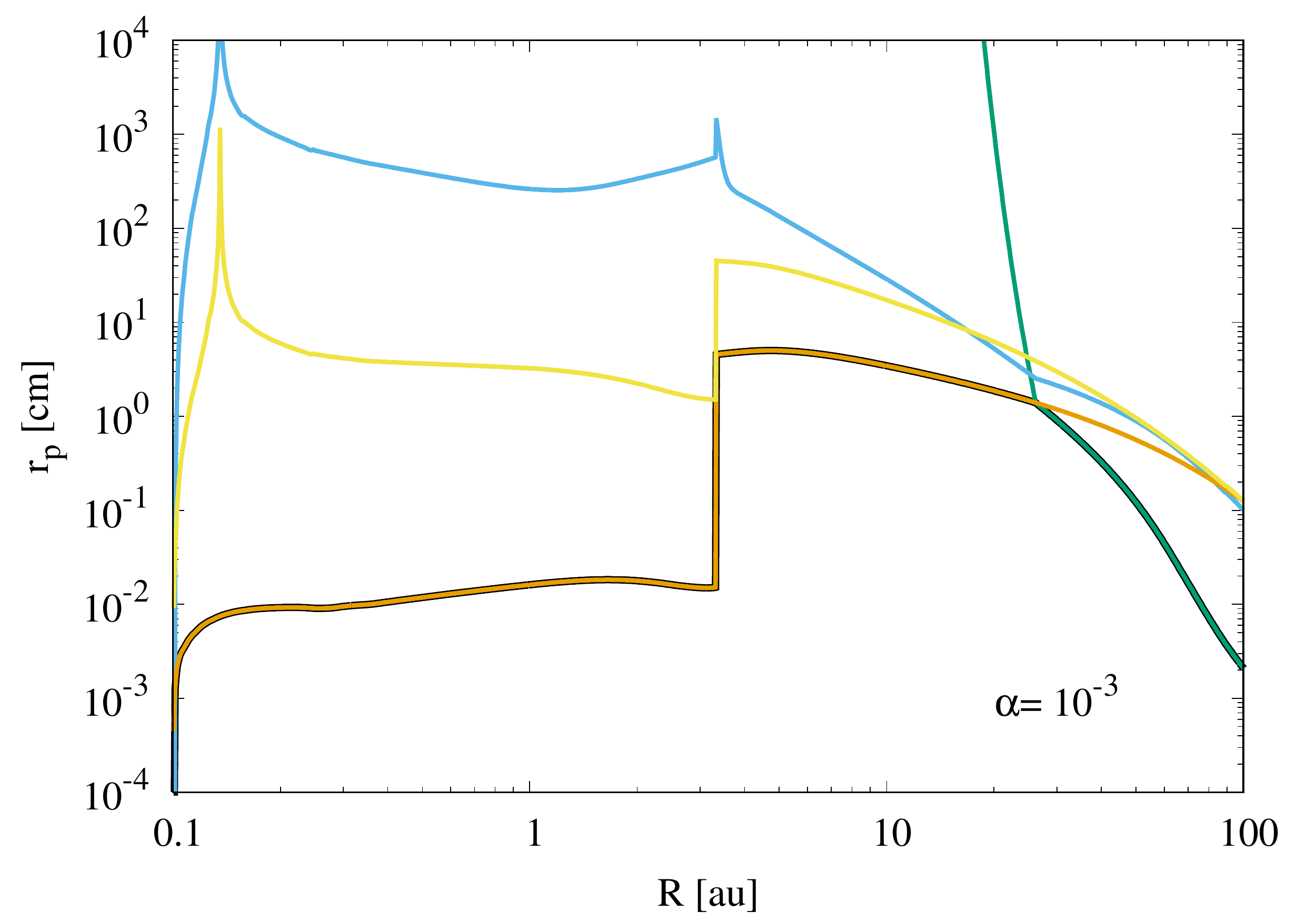} \\
    \includegraphics[angle= 270, width=\columnwidth]{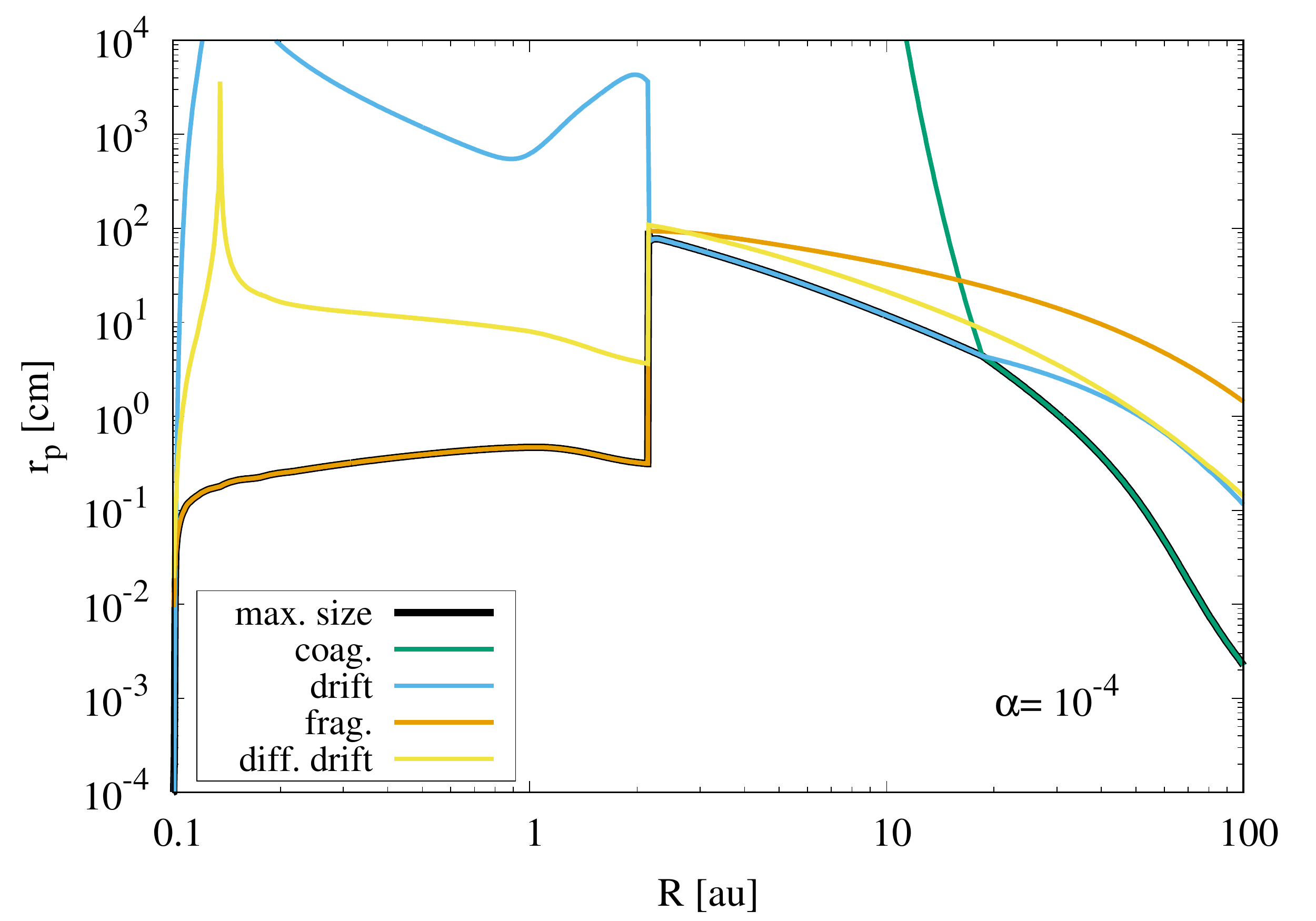} 
    \caption{The different dust growth barriers for the simulations using the fiducial disc with $\alpha= 10^{-3}$ (top panel) and $\alpha= 10^{-4}$ (bottom panel) at 50~Kyr.}
    \label{fig2.2_sec3}
\end{figure}

\begin{figure*}
    \includegraphics[width=1.\columnwidth]{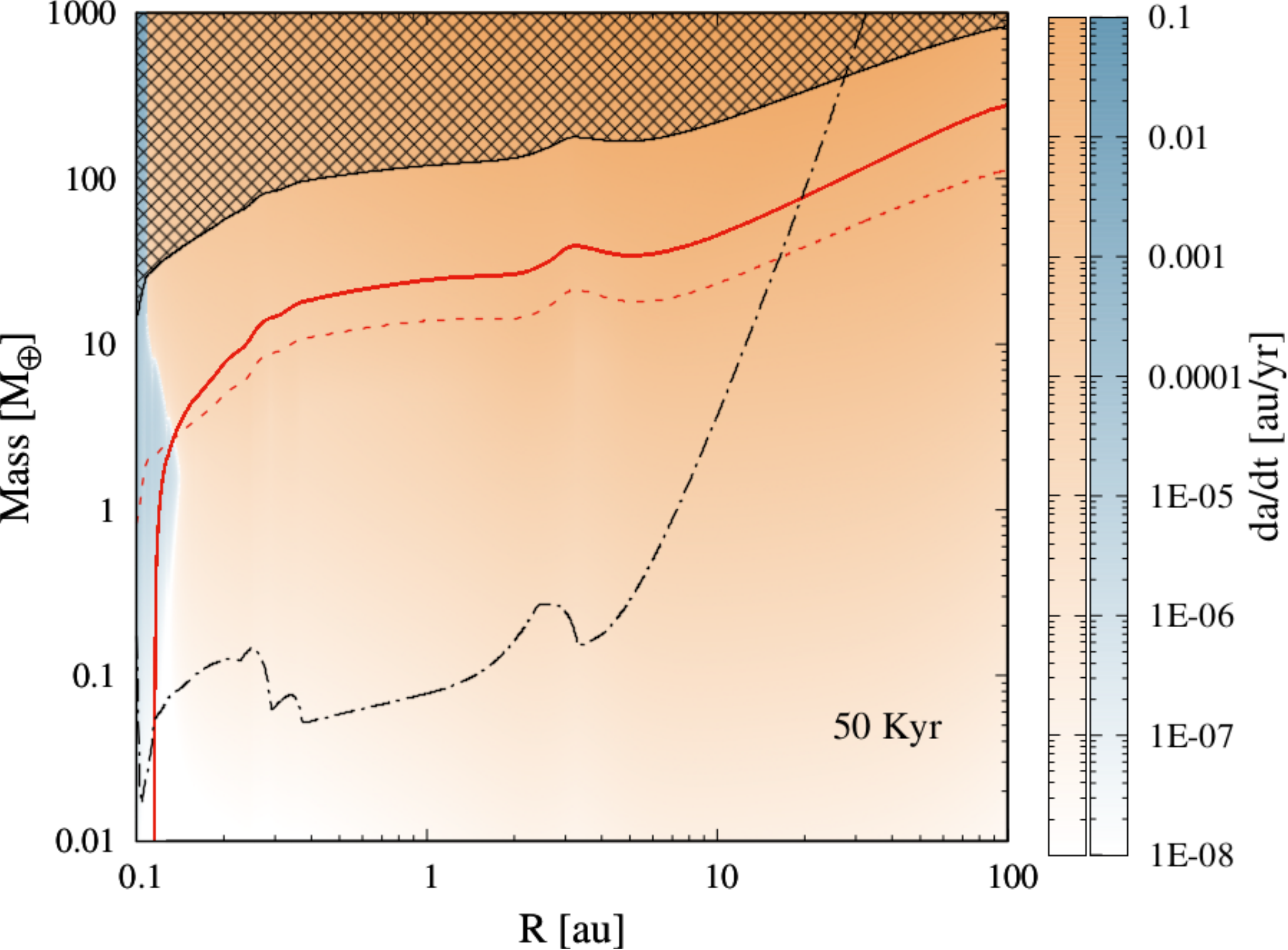} 
    \includegraphics[width=1.\columnwidth]{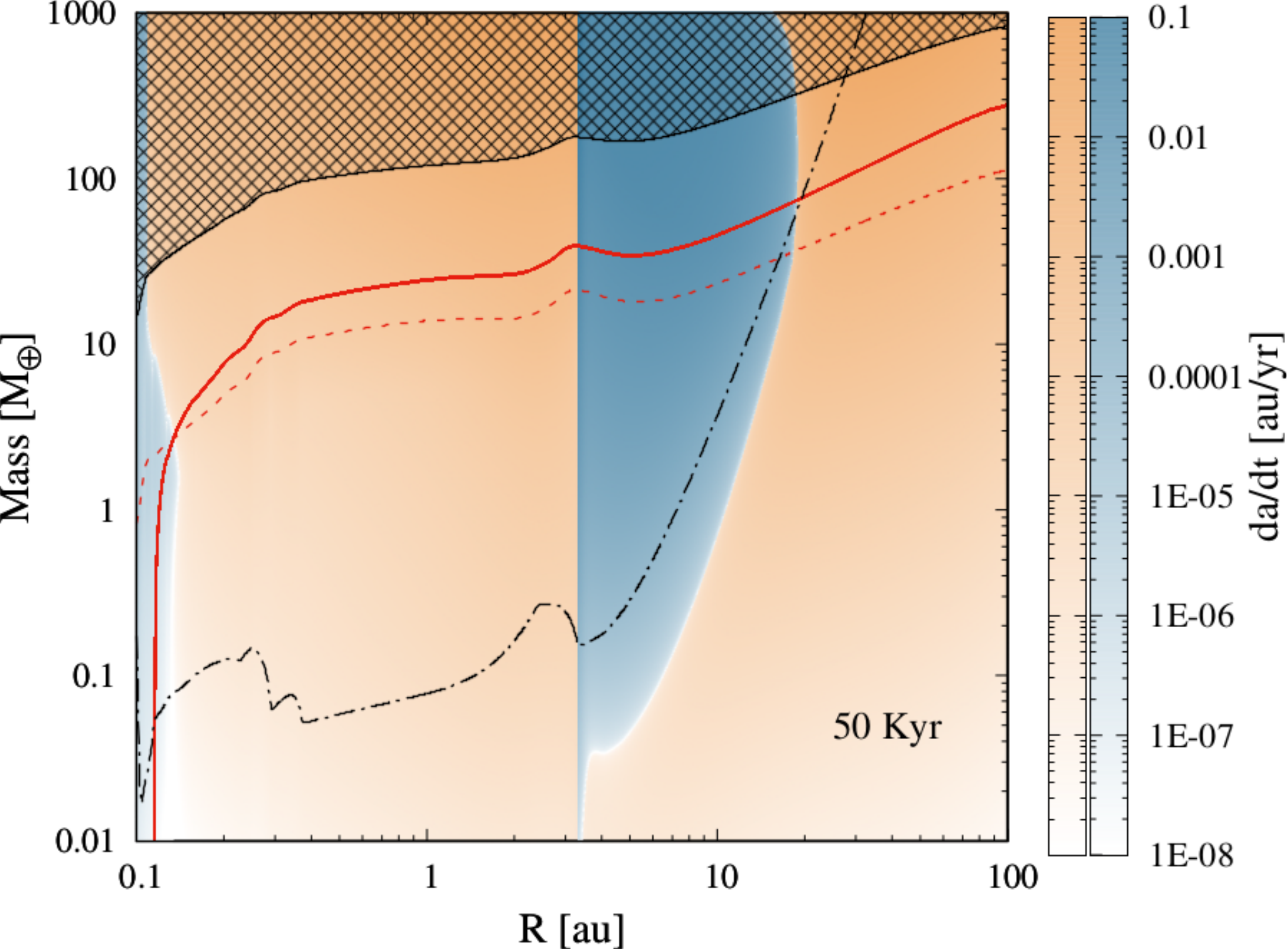} \\
    \includegraphics[width=1.\columnwidth]{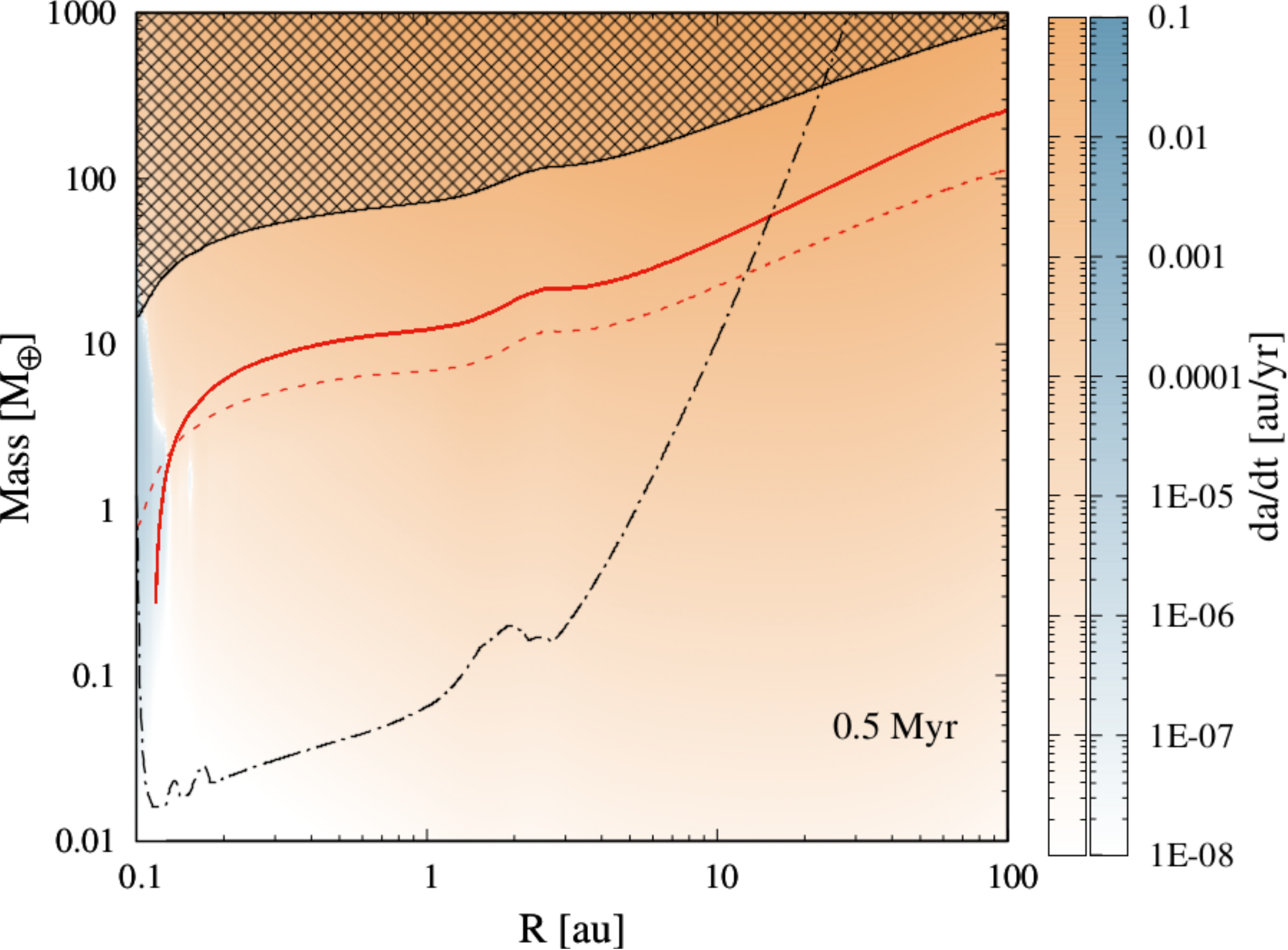} 
    \includegraphics[width=1.\columnwidth]{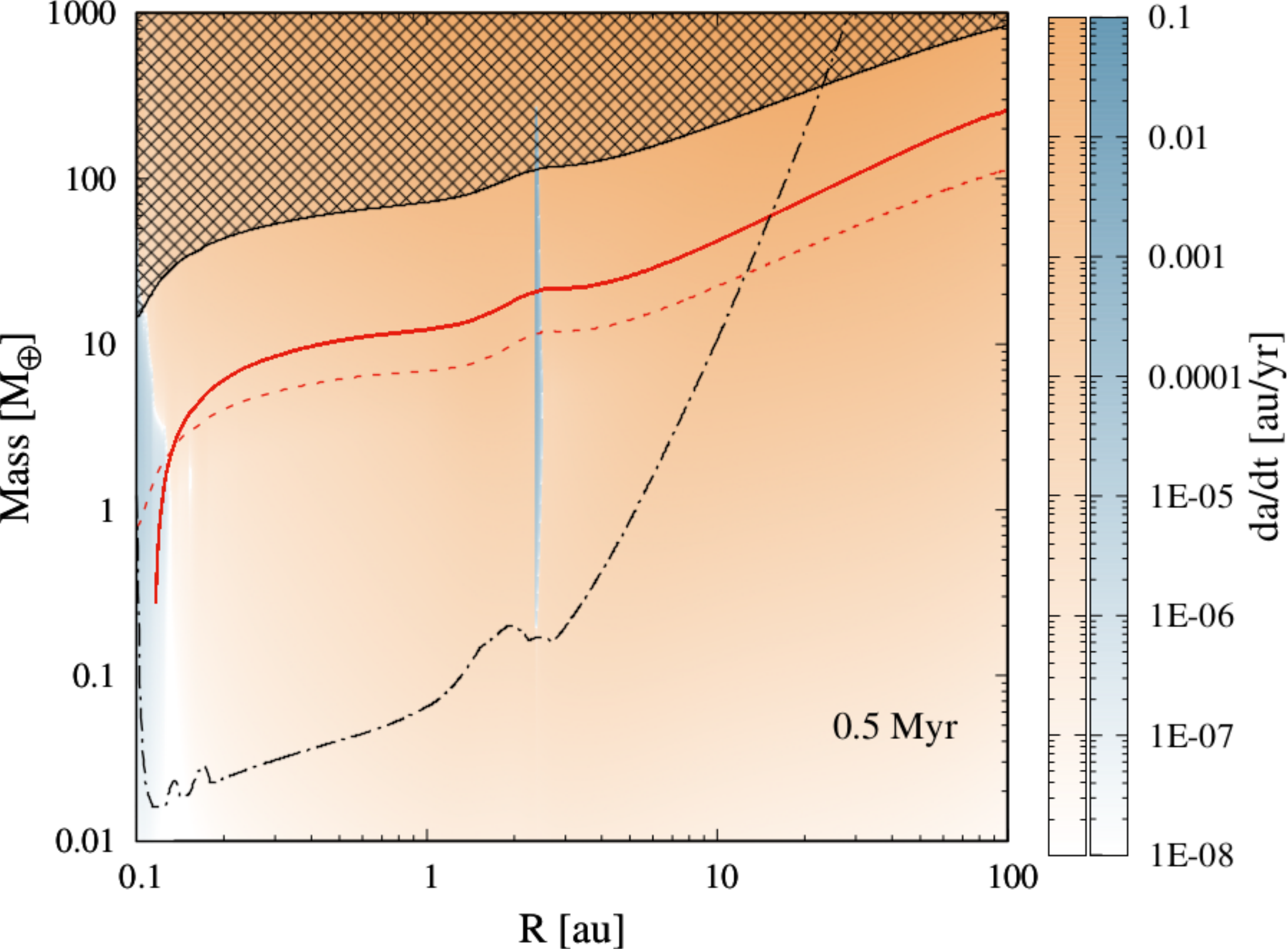} \\
    \includegraphics[width=1.\columnwidth]{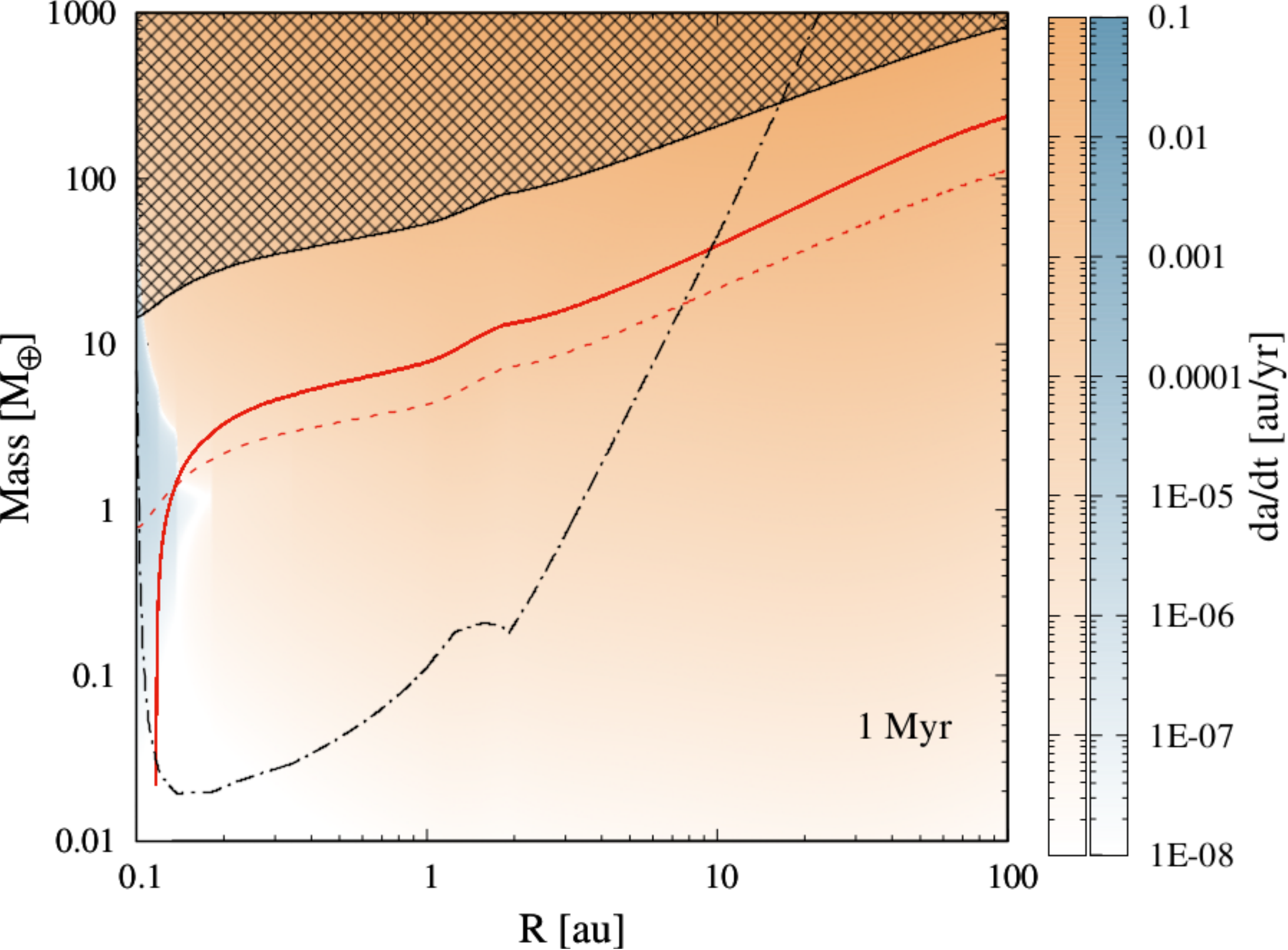} 
    \includegraphics[width=1.\columnwidth]{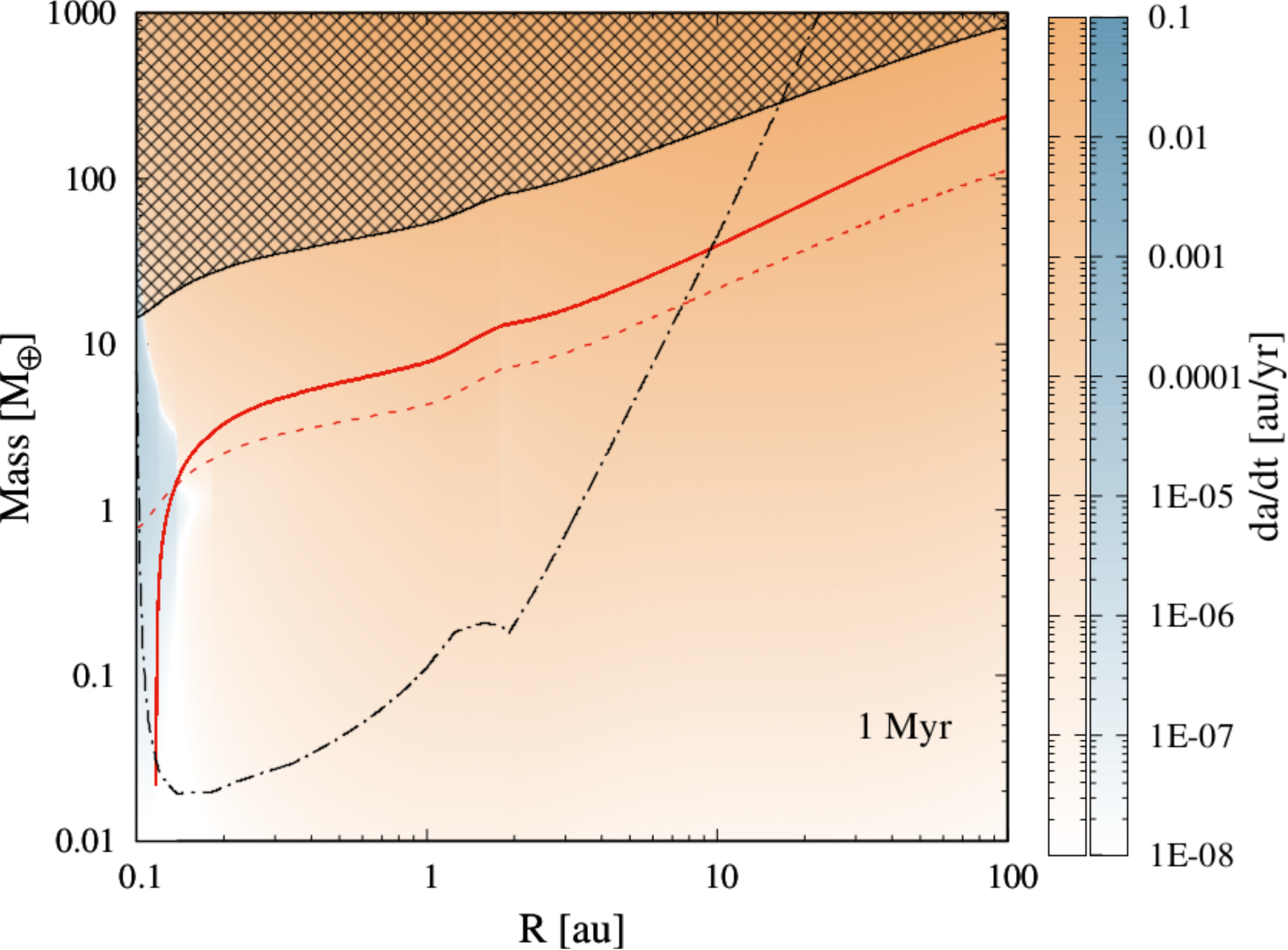}
    \caption{Planet migration maps for the simulation using $\alpha= 10^{-3}$ at three different times of disc evolution: 50~Kyr (top panels), 0.5~Myr (center panels), and 1~Myr (bottom panels), considering the type I migration rates from M17 (left column), and the combined ones from JM17 and VRM20 (right panel), which include the thermal torque. The colour scale represents the intensity of the migration rates where orange represents inward planet migration while blue represents outward migration, while the zero torque boundaries are represented by the white contours. The hatched regions correspond to the transition to the type II migration. The black dash-dotted lines represent the critical thermal mass defined by M17, while the solid and dashed red lines represent the pebble isolation mass given by \citet{Bitsch18} and \citep{Ataiee18}, respectively.}
    \label{fig3_sec3}
\end{figure*}

\begin{figure*}
    \includegraphics[width=1.\columnwidth]{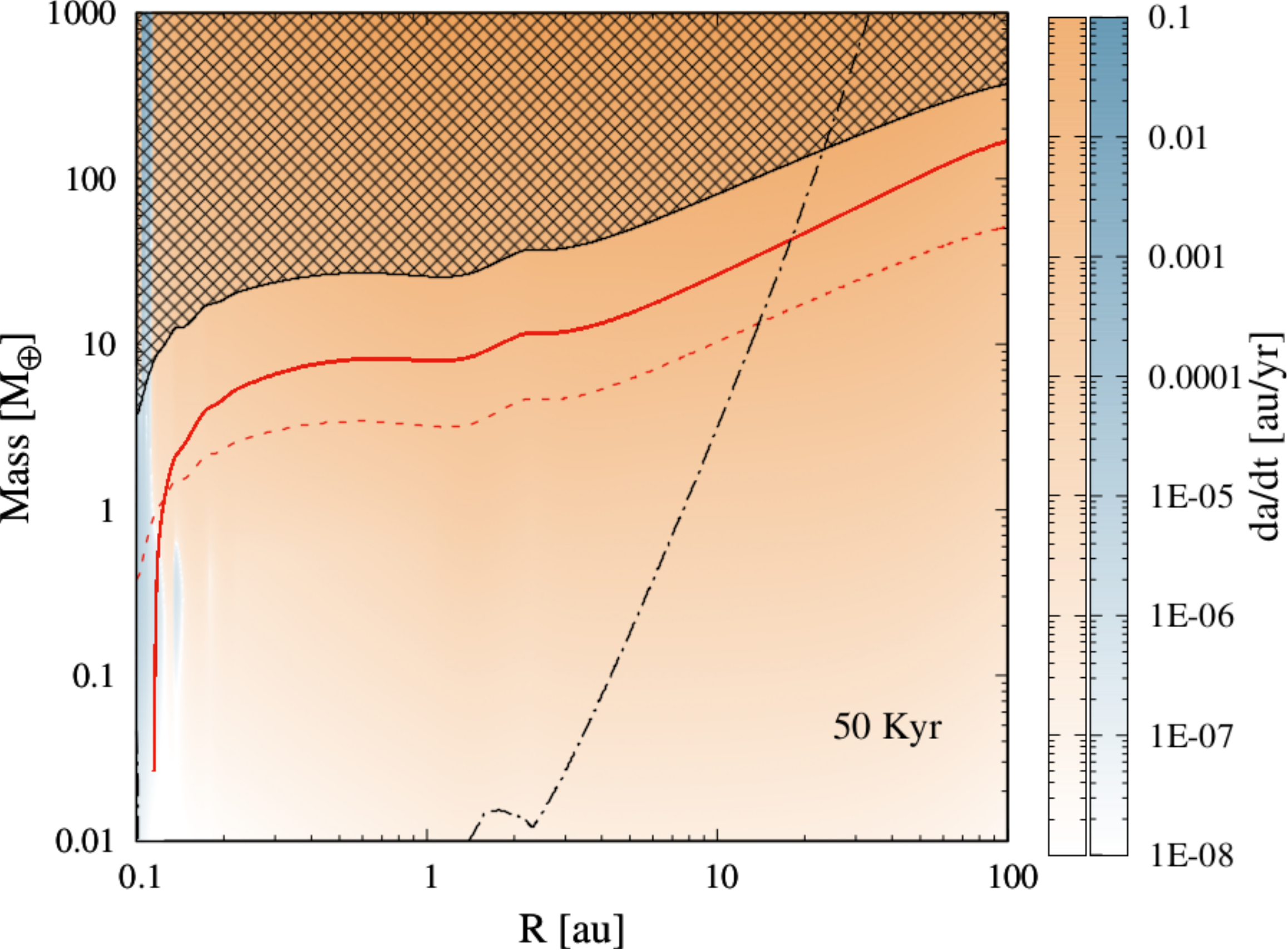} 
    \includegraphics[width=1.\columnwidth]{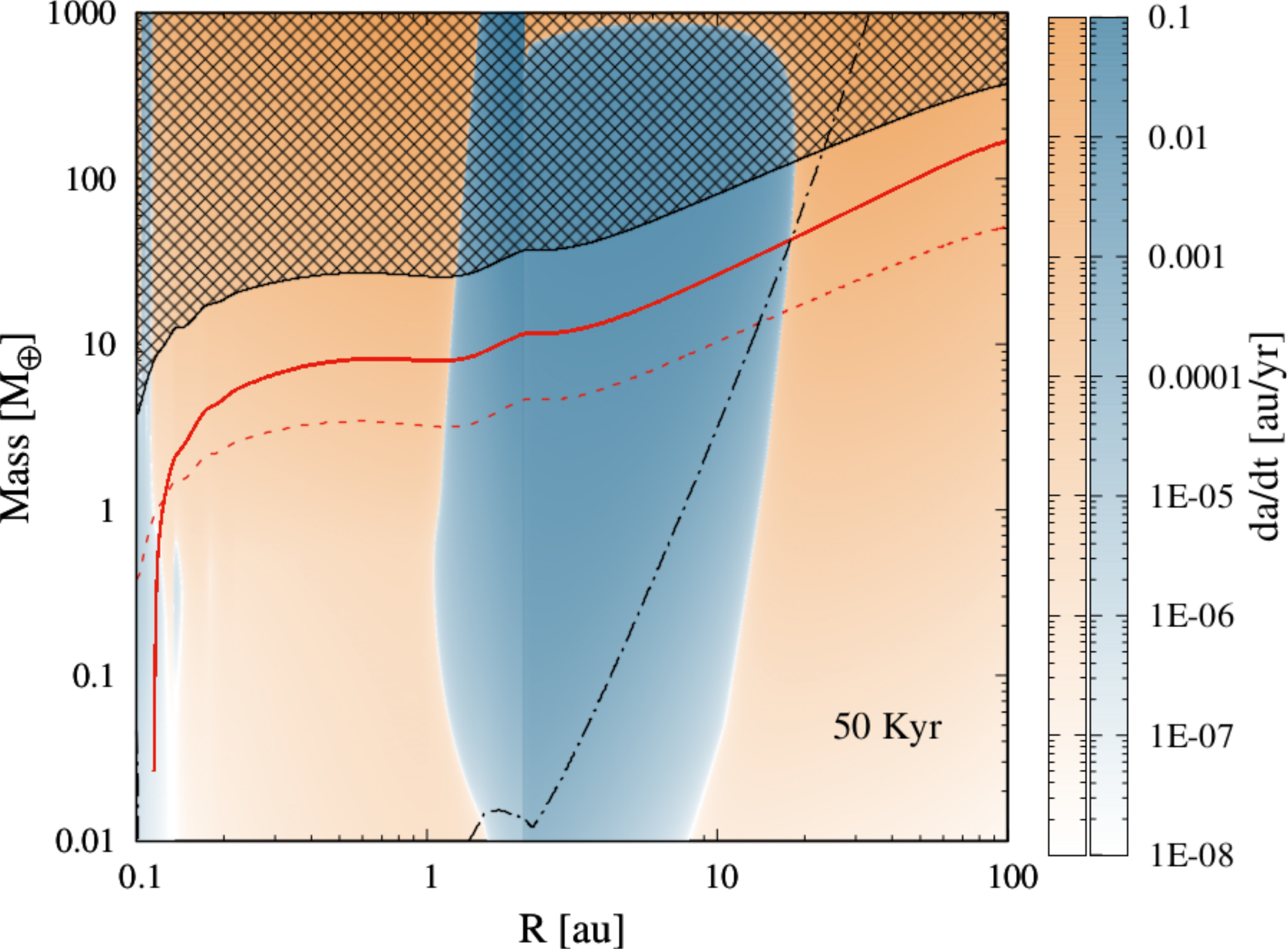} \\
    \includegraphics[width=1.\columnwidth]{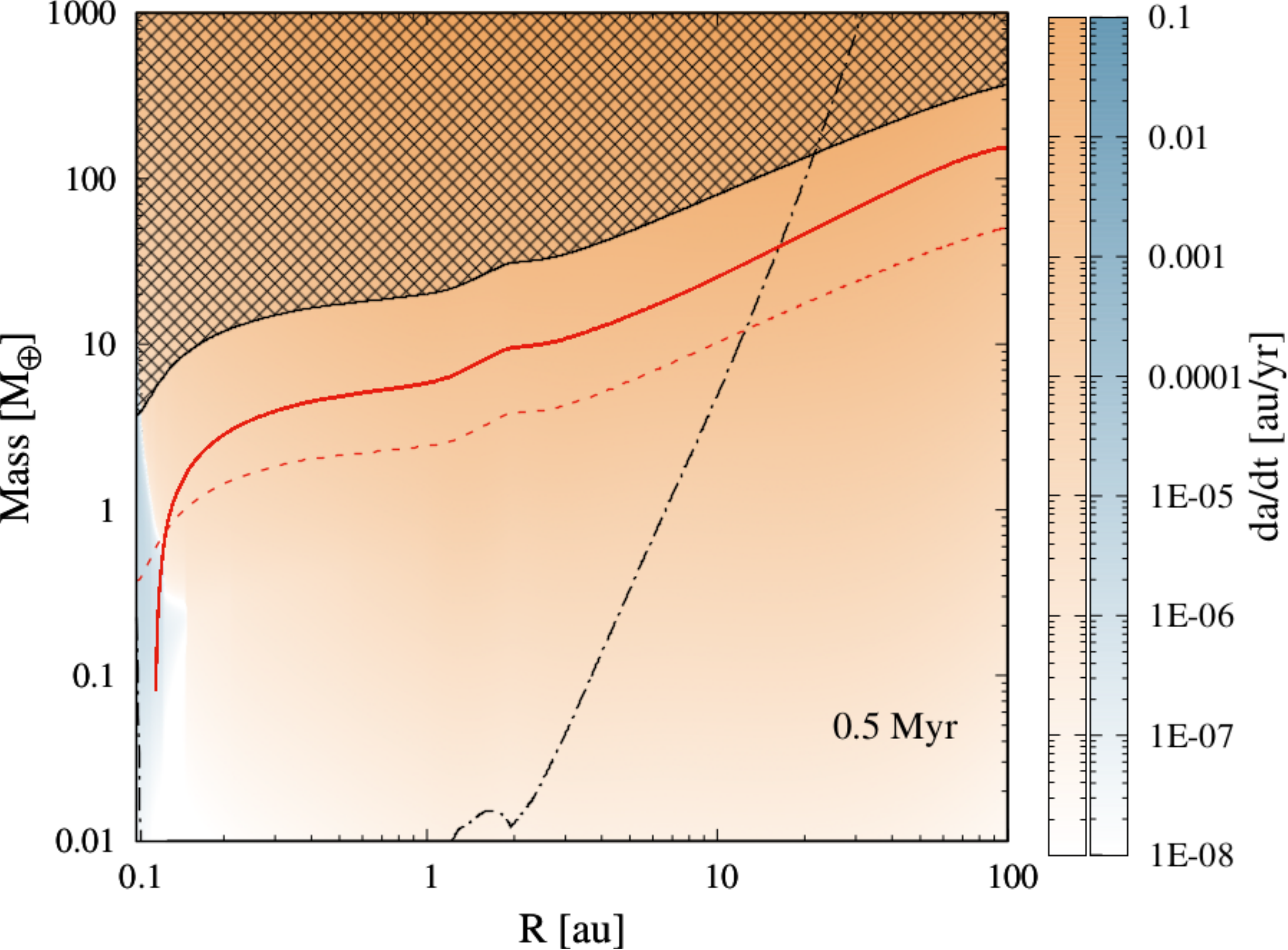} 
    \includegraphics[width=1.\columnwidth]{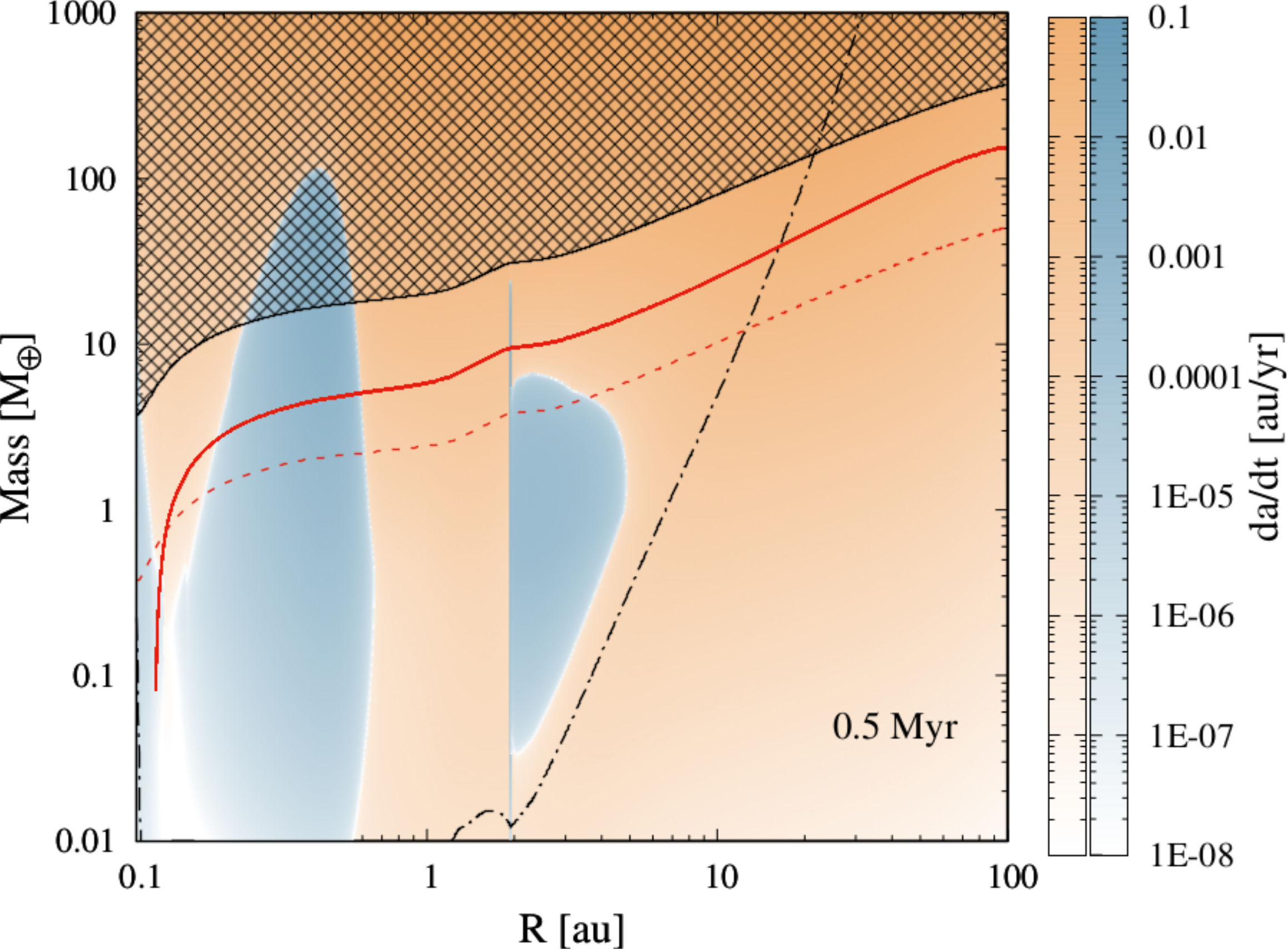} \\
    \includegraphics[width=1.\columnwidth]{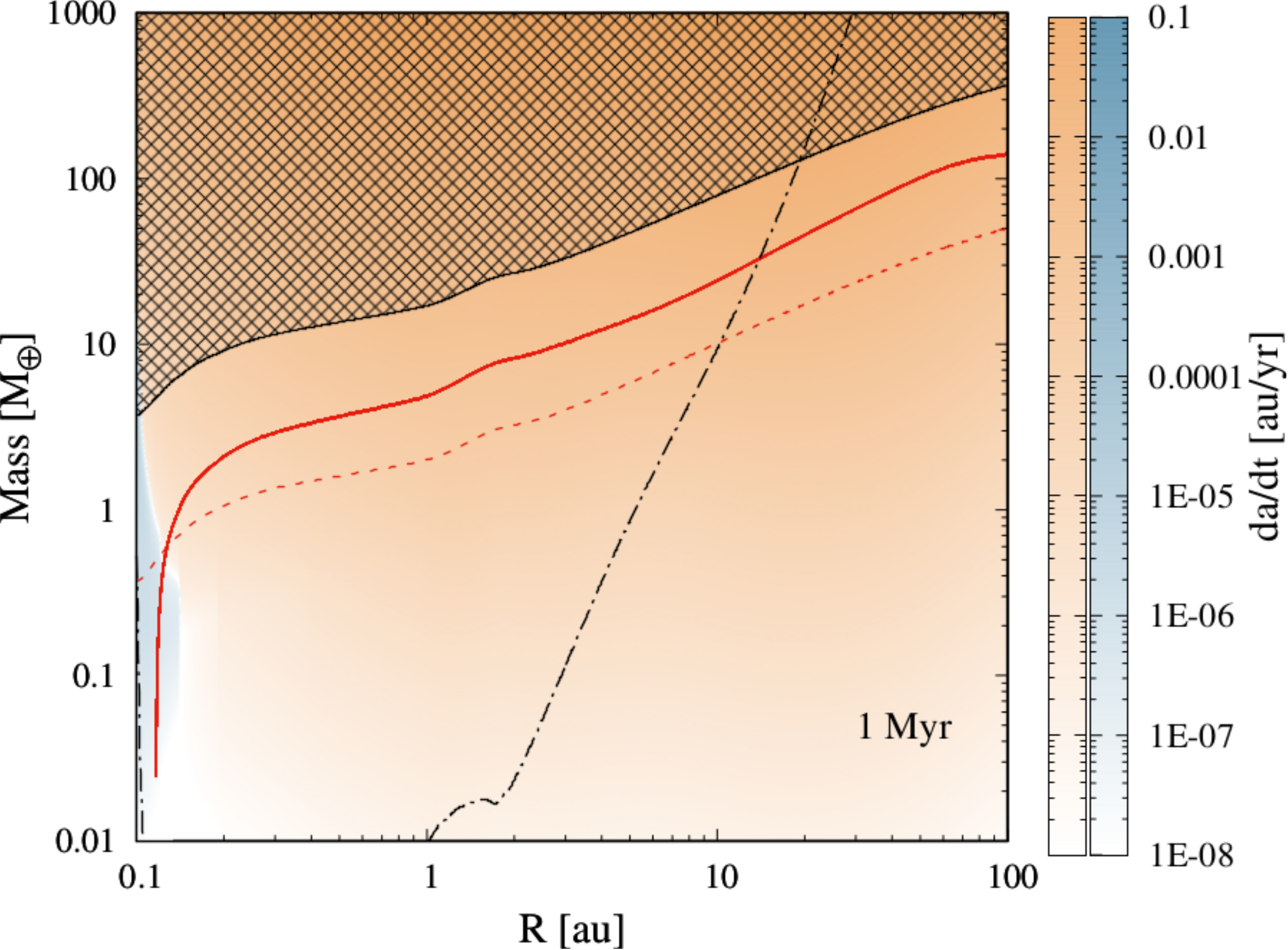} 
    \includegraphics[width=1.\columnwidth]{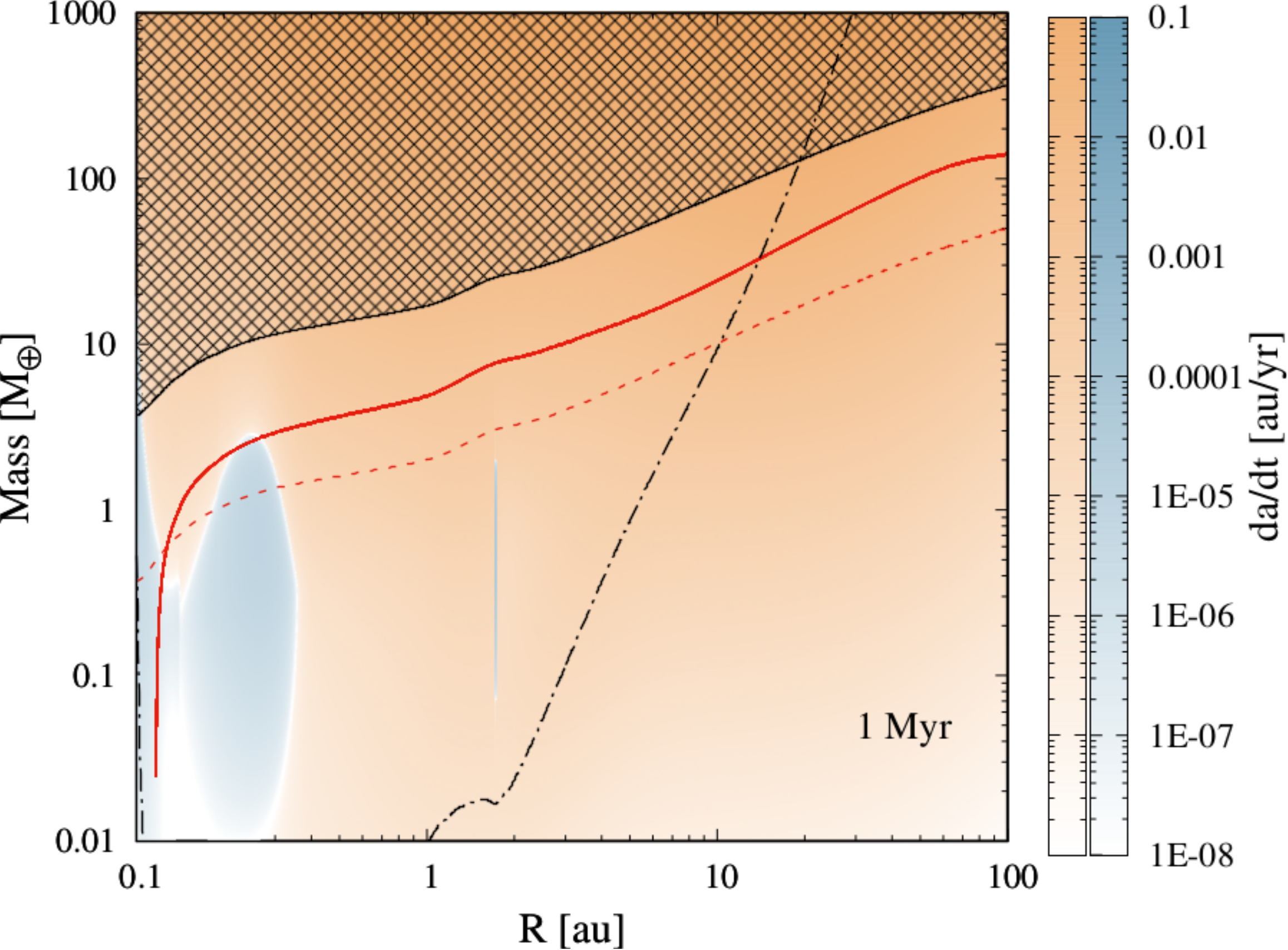}
    \caption{Same as Fig.\ref{fig3_sec3}, but for the simulation using $\alpha= 10^{-4}$.}
    \label{fig4_sec3}
\end{figure*}

In this section we compute the migration maps employing the same methodology as in Paper I. For our fiducial simulations, we use the same disc parameters as in Paper I. We define an initial gas surface density 
\begin{eqnarray}
  \Sigma_{\text{g}}= \Sigma_{\text{g}}^0 \left( \frac{R}{R_c} \right)^{-\gamma} e^{-(R/R_c)^{2-\gamma}}, 
  \label{eq1_sec4}
\end{eqnarray}
adopting $\gamma= 1$ and $R_c= 39$~au \citep{Andrews10}. The normalisation constant $\Sigma_{\text{g}}^0$ is computed for a fiducial disc of mass $\text{M}_{\text{d}}= 0.05 ~\text{M}_{\odot}$ as in Paper I. We compute the time evolution of the disc adopting two different values of the $\alpha$-viscosity parameter, $\alpha= 10^{-3}$ and $\alpha= 10^{-4}$.  

Figure~\ref{fig1_sec3} shows the comparison of the time evolution of the radial profiles of the gas surface density (top panel), mid-plane temperature (center panel), and disc aspect ratio (bottom panel). The solid lines represent the simulation using $\alpha= 10^{-3}$, while the dashed lines correspond to the case of $\alpha= 10^{-4}$. In the first case, due to the higher viscosity the gas surface densities decrease more quickly as time advances. In both cases, photoevaporation due to the central star opens a gap at a few au between 1.5~Myr and 2~Myr. For the case of $\alpha= 10^{-3}$, at 2~Myr all the inner gas disc disappears and the disc becomes a transitional disc with a cavity of 40~au. We can also see that the disc for the simulation using $\alpha= 10^{-3}$ is hotter and has larger aspect ratios with respect to the case of the simulation with $\alpha= 10^{-4}$.  

The figure~\ref{fig2_sec3} plots the time evolution of the radial profiles of the solid surface density (top panel), the weighted mean Stokes numbers (center panel) and weighted mean particle sizes (bottom panel). For the case of  $\alpha= 10^{-3}$, we can see that initially the ice line, which corresponds to the location beyond water condensates and there is a jump in the initial solid surface density, is further away (at $\sim 3.5$~au) with respect to the case of $\alpha= 10^{-4}$ (where ice line is initially located at $\sim 2$~au). In both cases, there is a significant solid accumulation near the ice line at very early times (see profiles at $t= 0.05\,$Myr). As time advances, the drift of pebbles significantly reduces the solid surface densities beyond the ice line, which increases in the inner part of the disc, especially for the case of  $\alpha= 10^{-4}$. When photoevaporation opens a gap in the gaseous disc, the supply of solids to the inner region from the outer part of the disc is halted. The remaining solid material inside the gap is quickly lost, while the one in the outer part of the disc drifts and accumulates in the pressure maximum generated at the edge of the gas cavity. 

Regarding the weighted mean Stokes numbers and weighted mean sizes of the solid population, we see that at $t= 0$~Myr all the dust is 1~$\mu$m size, and the Stokes numbers are very low (the discontinuity at the ice lines is related to the difference in the density for silicate and ice-rich dust). However, in both cases, dust grows very quickly (in less than 50~Kyr) forming pebbles larger than 1~cm between the ice line and $\sim 20$ -- 30~au. For the case of $\alpha=10^{-3}$, the size of the dust is limited by fragmentation in this region, while for $\alpha= 10^{-4}$ the dust size is limited by the radial drift (see Fig.~\ref{fig2.2_sec3}). Thus, particles grow larger and have larger Stokes numbers for the case of $\alpha= 10^{-4}$. As we can see in Fig.~\ref{fig2.2_sec3}, beyond $\sim 20$ -- 30~au, the size of the dust is limited by the coagulation growth timescale at this early time, and the particle sizes and the corresponding Stokes numbers are similar between the cases of $\alpha= 10^{-3}$, and $\alpha= 10^{-4}$. We also note that the size of the dust is limited by fragmentation in both cases in the inner part of the disc. This leads to larger differences in the sizes and Stokes numbers inside the ice line between the cases with $\alpha= 10^{-3}$, and $\alpha= 10^{-4}$. We can see that the smaller the $\alpha$-viscosity parameter is, the larger the sizes and Stokes numbers are. The abrupt change in the sizes and in the Stokes numbers at the ice line is due to the fact that, as we mentioned before, silicate dust has a threshold fragmentation velocity of 1~m/sec, while ice-rich dust has a fragmentation velocity of 10~m/sec. This leads particles to grow larger and have higher Stokes numbers beyond the ice line. Finally, as time advances the sizes and Stokes numbers decrease, especially beyond the ice line, as a consequence of the reduction in the solid surface densities.  

In Fig.~\ref{fig3_sec3} and Fig.~\ref{fig4_sec3} we plot the planet migration maps at 50~Kyr (top row), 0.5~Myr (middle row), and 1~ Myr (bottom row) for the simulations using $\alpha= 10^{-3}$, and $\alpha= 10^{-4}$, respectively. In the left panels we use the migration rates from JM17, while the right panels represent the cases where the combined migration from JM17 and VRM20 (which include the thermal torque) were used. The orange colour represents inward planet migration while blue represents outward migration, and the transitions between both regimes (or the zero torque boundaries) are shown by the white contours. The black dash-dotted line represents the critical thermal mass defined by M17. The hatched region corresponds to the type II migration regime \citep{Crida2006}. The red solid line represents the pebble isolation mass given by \citet{Bitsch18}, while the red dashed line corresponds to the pebble isolation mass given by \citet{Ataiee18}. When the thermal torque is not considered, we can see that outward migration regions are only generated at the inner edge of the disc, independently of the $\alpha$-viscosity parameter. This is related with the decrease in the gas surface density and the disc temperature at such location (see Fig~\ref{fig1_sec3}). This indicates that in most of the extension of the disc, the total torque on the planets is dominated by the Lindblad torque and thus planets migrate inward during the first million year of disc evolution. However, this situation significantly changes when the thermal torque is considered. In the simulation using $\alpha= 10^{-3}$, we can see that at 50~Kyr the outward migration region generated by the thermal torque extends between the ice line (at $\sim 3$~au) and $\sim 15-20$~au for planet masses between $\sim 0.05~\text{M}_{\oplus}$ and the pebble isolation mass. We note that the blue region above the red curves that represent the different pebble isolation masses is not physically meaningful. Pebble accretion is halted for planet masses above the pebble isolation mass, thus the planet does not release more heat into the surrounding disc. However, we also note that we are considering that the luminosity of the planet is only due to the accretion of pebbles, neglecting the luminosity arising from gas accretion and from the envelope contraction, and planetesimal accretion. For massive cores with a non-negligible envelope, or if a hybrid accretion scenario is considered, these contributions could be important, and will be explored in a future work. The abrupt transition between the outward and inward migration regions at $\sim 3$~au is due to the drop in the Stokes number inside the ice line by the changes in the dust properties, as we mentioned before. An important result is that the planet outward migration is present now also for planet masses well above the critical thermal mass. This implies that the conservative approach used in Paper I, where we dropped to zero the thermal torque after the planet mass becomes greater than the critical thermal mass was far too restrictive, limiting considerably the effect of the thermal torque. As time advances, the region of outward migration is significantly reduced. At 0.5~Myr only a very narrow region of outward migration survives just beyond the ice line, while at 1~Myr, the thermal torque does not generate regions of outward migration, and the corresponding migration map is very similar to the case where the thermal torque is not considered. 

In the simulation using $\alpha= 10^{-4}$ (Fig.~\ref{fig4_sec3}), the impact of the thermal torque is more relevant. At 50 Kyr, we can see that the region of outward migration expands now inside the ice line, from $\sim 1$~au at the inner edge, to $\sim 15-20$~au for the outer edge. The region extends in mass between the mass of Moon and the pebble isolation mass. The extension of the region is due to the accumulation of solid material just inside the ice line as a consequence of the particles radial drift, despite the decrease of the Stokes numbers when particles cross the ice line. At 0.5 Myr, the thermal torque develops two regions of outward migration. One between $\sim 0.2$~au and $\sim 0.5$~au, and the other between the ice line (at $\sim 2$~au) and $\sim 5$~au. Both regions extend for planet masses between $\sim 0.01$ -- $0.05~\text{M}_{\oplus}$ and the pebble isolation mass. Finally, at 1~Myr both regions almost disappear, remaining a very narrow outward migration region just beyond the ice line, and an inner region between $\sim 0.2$~au and $\sim 0.3$~au, for planet masses up to only $\sim 3~\text{M}_{\oplus}$. This evolution of the regions of outward migration is related to the evolution of the solid surface densities as a consequence of the accumulation of solid material in the inner part of the disc. The growth of the dust and the particles radial drift deplete the outer region of the disc, increasing the solid surface density in the inner one. 

There are two main factors that account for the differences between the simulations with $\alpha= 10^{-3}$ and $\alpha= 10^{-4}$. The first one is that the dust growth is limited by fragmentation, especially at early times and from the inner disc edge up to tens of au. Thus, for smaller values of the $\alpha$ parameter, the particles having larger Stokes numbers grow larger. This implies that pebble accretion rates becomes larger for smaller values of $\alpha$. The other one is that, as shown by several works \citep[e.g.][]{Morby15, Ormel2018, Venturini20c}, when $\alpha \gtrsim 10^{-3}$, pebbles tend to be accreted in the 3D regime --because the pebbles scale height becomes larger than the Hill radius of the planet-- decreasing even more the pebble accretion rates. In addition, we find that the thermal torques do not play a relevant role for $\alpha > 10^{-3}$. 
 
\subsection{Comparison with \citet{Benitez-llambay2015}}

\begin{figure*}
    \includegraphics[width=1.\columnwidth]{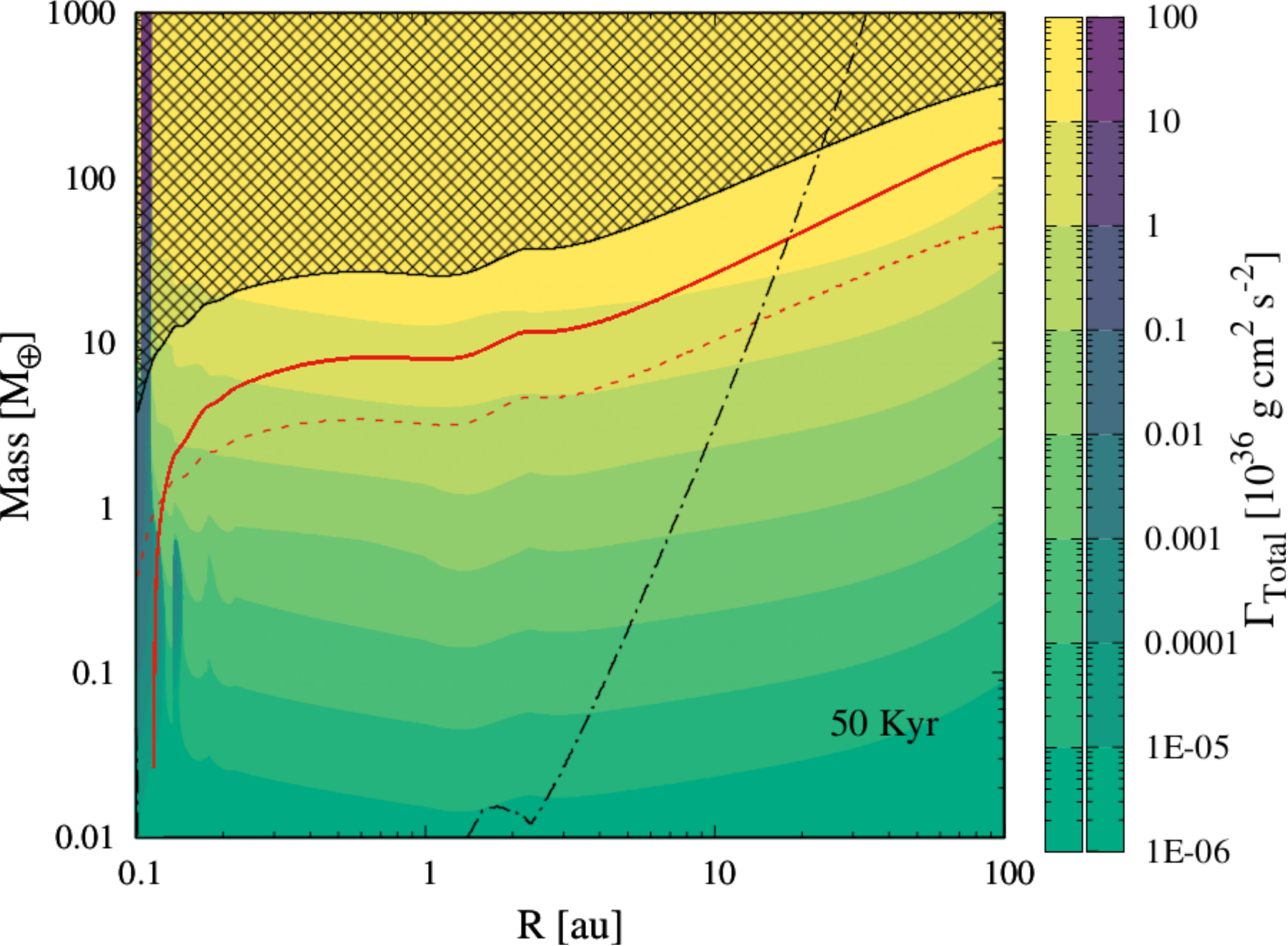} 
    \includegraphics[width=1.\columnwidth]{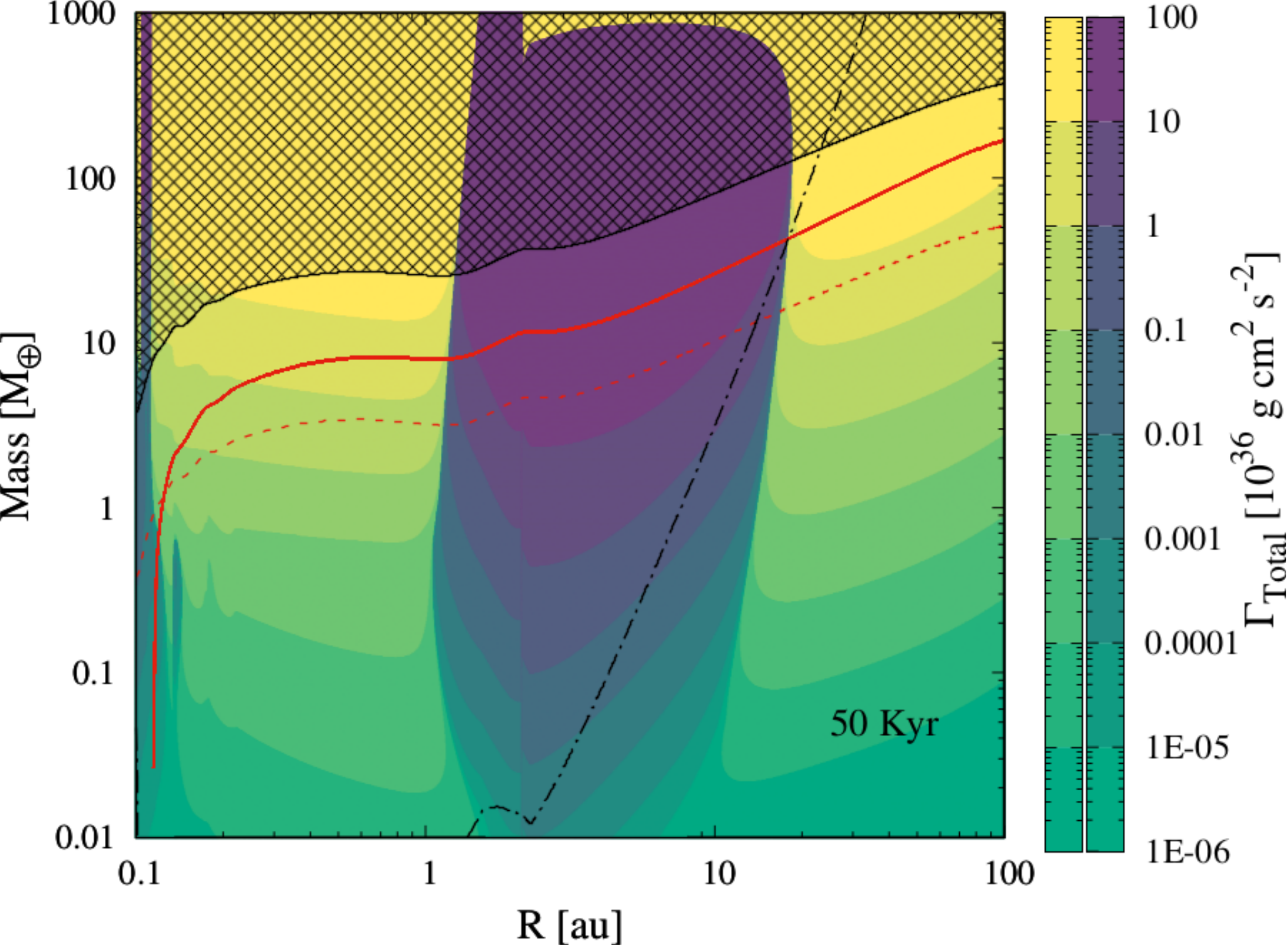} 
    \caption{Total torque maps for the simulations using $\alpha= 10^{-4}$ at 50~Kyr and 0.5~Myr. As before, the left column represents the case without the inclusion of the thermal torque, while the right column includes the thermal torque. The colour scales represent the magnitude of the total torque, where the green -- yellow represents a negative total torque and the green -- violet indicates a postive one. The rest of the curves are the same as in Fig.~\ref{fig4_sec3}.}
    \label{fig5_sec3}
\end{figure*}

 The analytical description of thermal torques derived in the linear analysis of M17 has been confirmed by several high resolution local hydrodynamical simulations \citep[][and VRM20]{Hankla2020, Chametla2020}. These simulations not only confirm the accuracy of the linear theory but also the simulations performed by VRM20 indicate how the linear estimate has to be modified for planetary masses higher than the critical thermal mass (see Eq.~\ref{eq5_sec2}). Besides this agreement it would be interesting to compare the total predicted torques by our simulations with those of predicted by detailed hydrodynamical simulations. Unfortunately, due to their local nature, these simulations are unable to capture all the components of the total torque. The only global hydrodynamical simulations available in the literature for comparison of the total torque are those of \citet{Benitez-llambay2015}. Even in this case, direct comparisons are not straightforward due to the use of different microphysics and differences in the arising thermal and density structures of the discs. Moreover, M17 showed that the heating torque measured by \citet{Benitez-llambay2015} in their numerical experiments is about one order of magnitude lower than that predicted by the linear theory. This was confirmed by the local simulations of \citet{Hankla2020}, who attributed this differences to the unavoidable low resolution in the global simulations performed by \citet{Benitez-llambay2015}. With these caveats in mind, in what follows we compare our simulation with those of \citet{Benitez-llambay2015}.

\citet{Benitez-llambay2015} performed simulations considering a $3~\text{M}_{\oplus}$ planet at the location of Jupiter ($\sim 5$~au), and assuming different solid accretion rates. One of their simulations, corresponding to a solid accretion rate of $10^{-4}~\text{M}_{\oplus}/\text{yr}$, describes a situation which is very similar to that found in our $\alpha= 10^{-4}$ simulation at an early age of 50 Kyr. In our $\alpha= 10^{-4}$ simulation at 50 Kyr, we find that a $3~\text{M}_{\oplus}$ planet located at 5~au has a pebble accretion rate of $1.05 \times 10^{-4}~\text{M}_{\oplus}/\text{yr}$. This fortunate coincidence allow us to quantitatively compare the torques computed in our simulations to those of \citet{Benitez-llambay2015}.

Fig.~\ref{fig5_sec3} shows the total torque maps for the simulations using $\alpha= 10^{-4}$ at 50~Kyr with and without the inclusion of the thermal torques. As in the case of Fig.~\ref{fig3_sec3} and Fig.~\ref{fig4_sec3}, the maps are only valid while the mass of the planet is lower than the pebble isolation mass. As expected from previous sections, in the absence of thermal torques the total torques are always negative (green-yellow colour scale). When the thermal torque is included the heating torque dominates and the total torque becomes positive (green-violet colour scale). For a $3~\text{M}_{\oplus}$ planet located at 5~au, we find a total torque of $-2.835 \times 10^{35}~\text{g}~\text{cm}^2~\text{s}^{-2}$ when the thermal torque is not included (left panel in Fig.~\ref{fig5_sec3}), and a total torque of $8.728 \times 10^{36}~\text{g}~\text{cm}^2~\text{s}^{-2}$ when the thermal torque is considered (right panel in Fig.\ref{fig5_sec3}). Thus, for the $3~\text{M}_{\oplus}$ planet located at 5~au, under a pebble accretion rate of $1.05 \times 10^{-4}~\text{M}_{\oplus}/\text{yr}$, the thermal torque is about $9 \times 10^{36}~\text{g}~\text{cm}^2~\text{s}^{-2}$ \footnote{In this case the contribution of the cold torque is negligible and the thermal torque is given basically by the heating torque.}. For their simulation with a solid accretion rate of $10^{-4}~\text{M}_{\oplus}/\text{yr}$ \citet{Benitez-llambay2015} find a total torque on the planet of $\sim 2.4 \times 10^{35}~\text{g}~\text{cm}^2~\text{s}^{-2}$. Given that, in the absence of the heating torque they find a total torque of $-2.8 \times 10^{35}~\text{g}~\text{cm}^2~\text{s}^{-2}$ (see Fig. 2 of that work), this implies that the heating torque takes a value of about $5.2 \times 10^{35}~\text{g}~\text{cm}^2~\text{s}^{-2}$ for a $10^{-4}~\text{M}_{\oplus}/\text{yr}$ solid accretion rate. This value is about 17 times smaller than that found in our simulations. Consequently, and in line with the estimation of M17, we are also finding a difference of about one order of magnitude between the heating torques reported by \citet{Benitez-llambay2015} and those derived from the linear theory and local numerical simulations.

\section{Planet formation tracks}
\label{sec_4}

\begin{figure}
    \includegraphics[angle=270, width=\columnwidth]{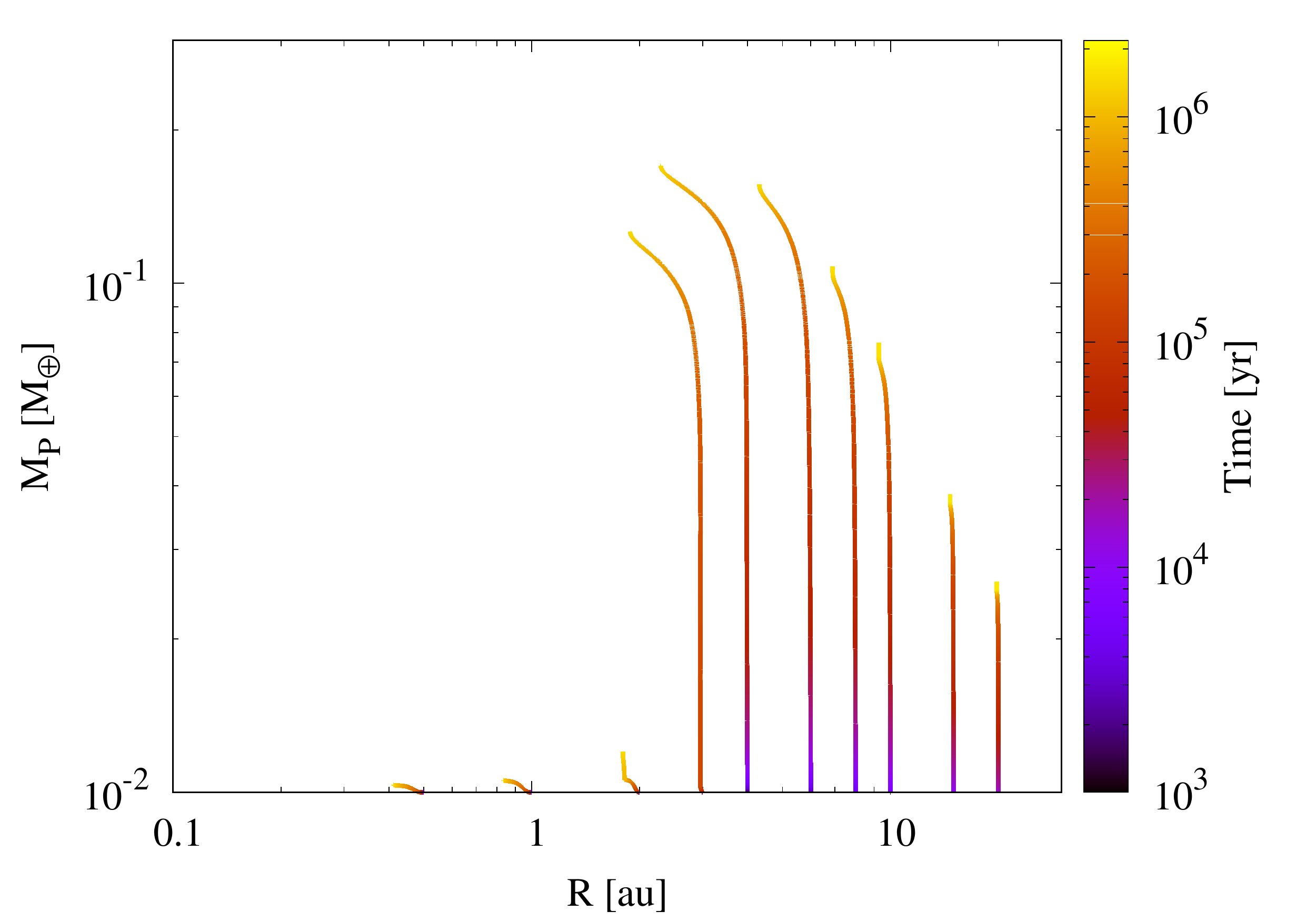}
    \caption{Planet formation tracks --planet mass vs. semi-major axis-- for the simulation using the fiducial disc and $\alpha= 10^{-3}$.}
    \label{fig1_sec4}
\end{figure}

In this section we compute planet formation tracks to analyse the impact of the different migration recipes in the formation and migration history of a growing planet by pebble accretion. We compute the formation of a single planet per disc, considering that the embryo has initially a core of $0.01~\text{M}_{\oplus}$ with a negligible gaseous envelope ($\sim 10^{-10}~\text{M}_{\oplus}$), and semi-major axes of 0.5, 1, 2, 3, 4, 6, 8, 10, 15, and 20~au.

In Fig.~\ref{fig1_sec4}, we plot the planet formation tracks for the case of $\alpha= 10^{-3}$. In this case, as planets practically do not grow because of pebble accretion is very inefficient, there are no differences in the planet formation tracks using the different migration recipes. Despite that, for $\alpha= 10^{-3}$ the heating torque is able to generate an outward migration region at 50 Kyr for planets with masses larger than $\sim 0.05~\text{M}_{\oplus}$ (see Fig.~\ref{fig3_sec3}), we can see in Fig.~\ref{fig1_sec4} that the timescale to growth to $\sim 0.05~\text{M}_{\oplus}$ from the mass of the Moon is greater than 50~Kyr. Thus, as in \citet{Venturini20c}, we find that in order for planet formation by pebble accretion to be efficient  --when a model of dust growth and evolution is considered-- the $\alpha$-viscosity parameter has to be low, $\alpha \lesssim 10^{-4}$.

\begin{figure}
    \includegraphics[angle=270, width=\columnwidth]{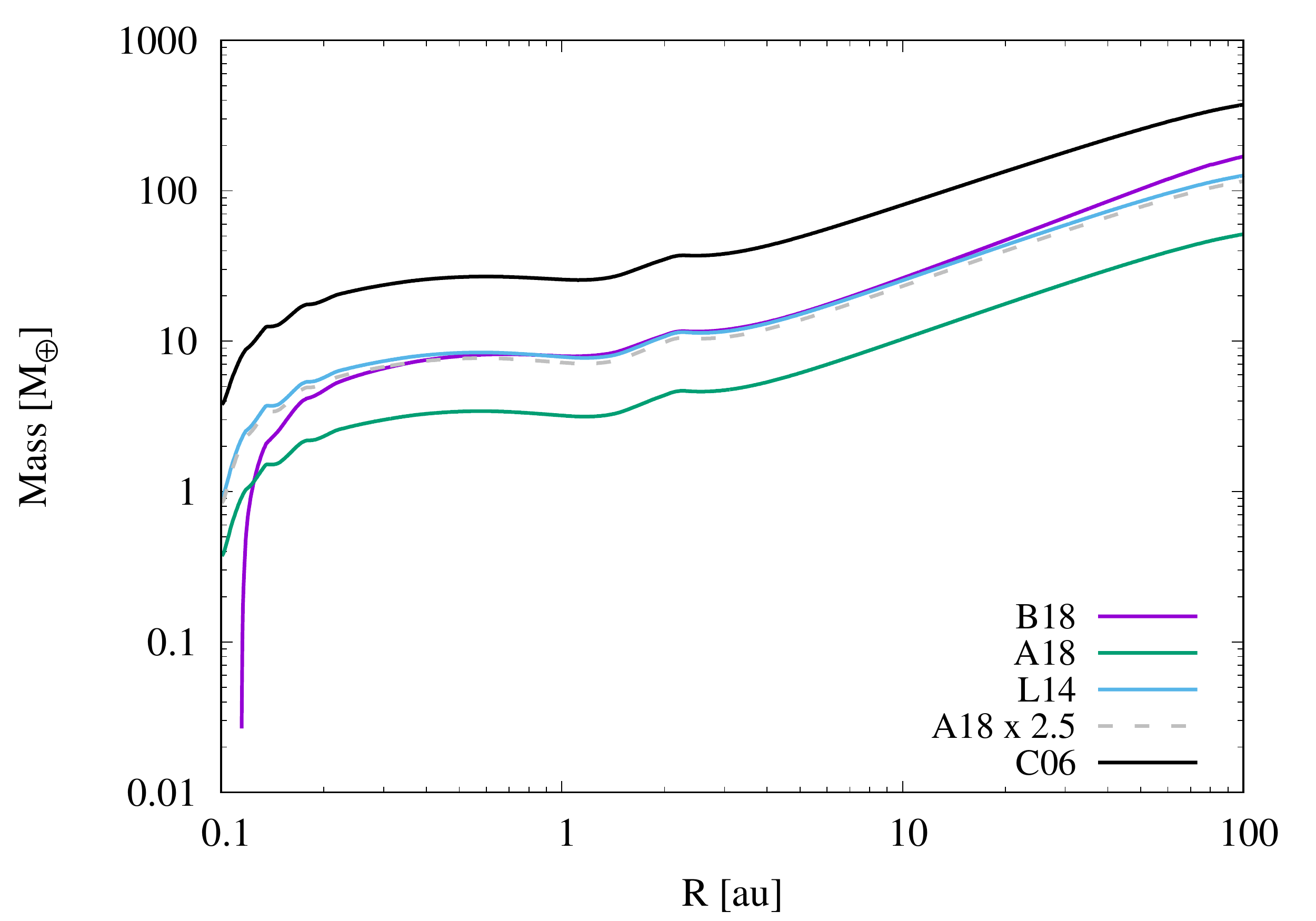} 
    \caption{Pebble isolation masses (PIM) vs. distance to the central star for the simulation using the fiducial disc and $\alpha= 10^{-4}$ at 50~Kyr. The violet curve represents the PIM using the recipes from \citet{Bitsch18}, the green curve corresponds to the PIM using the recipes from \citet{Ataiee18}, while the blue light curve shows the PIM adopting the prescriptions from \citet{Lambrechts14}. The grey dashed line corresponds to a PIM 2.5 times larger than the corresponding ones from   \citet{Ataiee18}. Finally, the black line represents the mass needed from a planet to open a full gap in the disc and to switch to the type II migration regime, from the recipes by \citet{Crida2006}.} 
    \label{fig2_sec4}
\end{figure}

Before showing the planet formation tracks for the case of $\alpha= 10^{-4}$, we briefly discuss  the differences in the pebble isolation mass (PIM) prescriptions. As we can see in Fig.~\ref{fig3_sec3} and Fig.~\ref{fig4_sec3} the PIM given by \citet{Ataiee18} are systematically lower than the corresponding ones obtained using the recipes from \citet{Bitsch18}. In Fig.~\ref{fig2_sec4}, we plot the PIM computing the fiducial disc with $\alpha= 10^{-4}$ at 50~Kyr. We can see that the PIM obtained adopting the recipes from \citet{Bitsch18} (the violet curve) are about 2.5 times larger than the ones obtained using the prescriptions from \citet{Ataiee18} in almost the whole extension of the disc. We note we find practically the same factor between both recipes at any time for the simulation using $\alpha= 10^{-4}$ (this factor is lower for the case of $\alpha= 10^{-3}$). We also note that the PIM computed from \citet{Bitsch18} drops to zero before reaching the disc inner edge. Because planets stop accreting pebbles after they reach the PIM, significant differences between the different recipes can lead to very different outcomes. In Fig.~\ref{fig2_sec4} we also plot in light blue the PIM obtained from \citet{Lambrechts14}, which is given by
\begin{eqnarray}
\text{M}_{\text{iso}}&=& 20 \left(\frac{h}{0.05} \right)^3~\text{M}_{\oplus}.  
\end{eqnarray}
Despite the fact that this last recipe does not depend on $\alpha$, it matches very well the PIM from \citet{Bitsch18} for the simulation using $\alpha= 10^{-4}$. In addition, the PIM computed from \citet{Lambrechts14} do not drop to zero before reaching the disc inner edge, and the curve has the same functional form than the one derived from \citet{Crida2006} (the black curve) which indicates the masses needed for the planets to open a full gap and switch to the type II migration regime. Thus, to compute the planet formation tracks using $\alpha= 10^{-4}$, we use the PIM given by \citet{Lambrechts14} as the fiducial ones\footnote{\citet{Drazkowska21} used also this PIM for different values of the $\alpha$-viscosity parameter.}, and in App. 1, we repeat some computations employing the PIM given by \citet{Ataiee18}. 

\begin{figure}
    \includegraphics[angle= 270, width=\columnwidth]{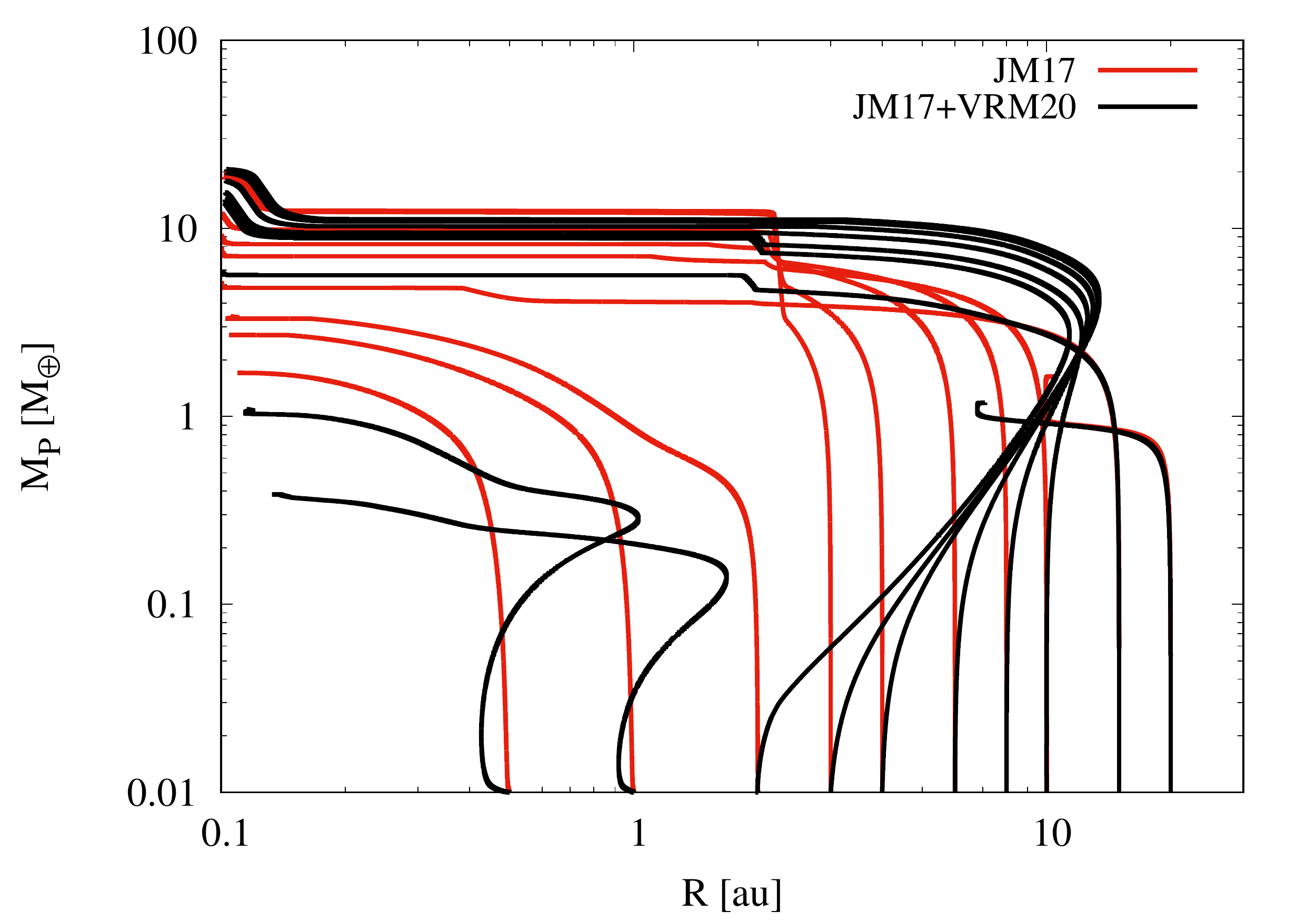} 
    \caption{Planet formation tracks for our fiducial disc using $\alpha=10^{-4}$. While red lines correspond to the simulations using the migration recipes from JM17, the black ones represent the simulation adopting the combined migration prescriptions from JM17 and VRM20 which take into account the computation of the thermal torque.}
    \label{fig3_sec4}
\end{figure}

Figure~\ref{fig3_sec4} shows the planet formation tracks adopting the migration recipes from JM17 (red lines), and the new combined rates from JM17 and VRM20 (black lines). When the migration recipes from JM17 are used, the planets always migrate inwards. For the planets initially located inside the ice line ($a_{\text{P}} \le 2$~au), they grow and migrate gradually until they end close to the disc inner edge with masses between $\sim 2$ -- $4~\text{M}_{\oplus}$. For planets initially located beyond the ice line --except for the outermost--, they grow faster to larger core masses, and when they reach the ice line they accrete very efficiently the solid accumulated at such location due to the solid radial drift  (see top panel of Fig.~\ref{fig2_sec3}). This allows these planets to significantly increase their core masses in a very short timescale, until they reach their corresponding PIM (the PIM are about $10~\text{M}_{\oplus}$ for the more massive planets). Then, they quickly migrate inwards, practically at constant mass, and the most massive planets are able to accrete significant amounts of gas ending with $\sim 20~\text{M}_{\oplus}$ close to the disc inner edge. 

For the simulations where the thermal torque is considered (the black lines), we can see that planet formation tracks are quite different. First, we note that for the planets initially located inside the ice line not only the planet formation tracks are very different, but also their final masses. In addition, for the planets initially located at 0.5 and 1~au, planets migrate first inwards but quickly they start to migrate outward due to the heating torque. This behaviour was not found in Paper I, because in the inner part of the disc the planet mass is generally greater than the critical thermal mass (see Fig.\ref{fig3_sec3} and Fig.\ref{fig4_sec3}). This situation, as we show below, completely changes the formation timescale of the planets resulting in less massive ones. For the planets initially located beyond the ice line (except for the two outermost ones), and the one located just inside it, at 2~au, they significantly migrate outward due to the heating torque. At $\sim 10$ -- 15~au, the outward migration is reversed and the planets migrate inwards reaching their PIM of about $10~\text{M}_{\oplus}$ near the ice line or just crossing it, and all of them accrete between 5 and $10~\text{M}_{\oplus}$ of gas near the disc inner edge. We remark that when the planet reaches the PIM we set to zero the thermal torque. This is due to the fact that we are considering that the heat that the planet releases to the surrounding gas is only due to the accretion of solids, which is halted from this moment. We can see that despite planet formation tracks being quite different, final masses and semi-major axes are similar, considering or not the thermal torques. This is due to the fact that the PIM obtained in both sets of simulations result to be very similar. In the case of the two outermost planets the thermal torque does not play an important role. The one located at 15~au is able to reach the inner part of the disc before photoevaporation opens a gap in the gas disc, while the last one does not.  

\begin{figure}
    \includegraphics[angle= 270, width=\columnwidth]{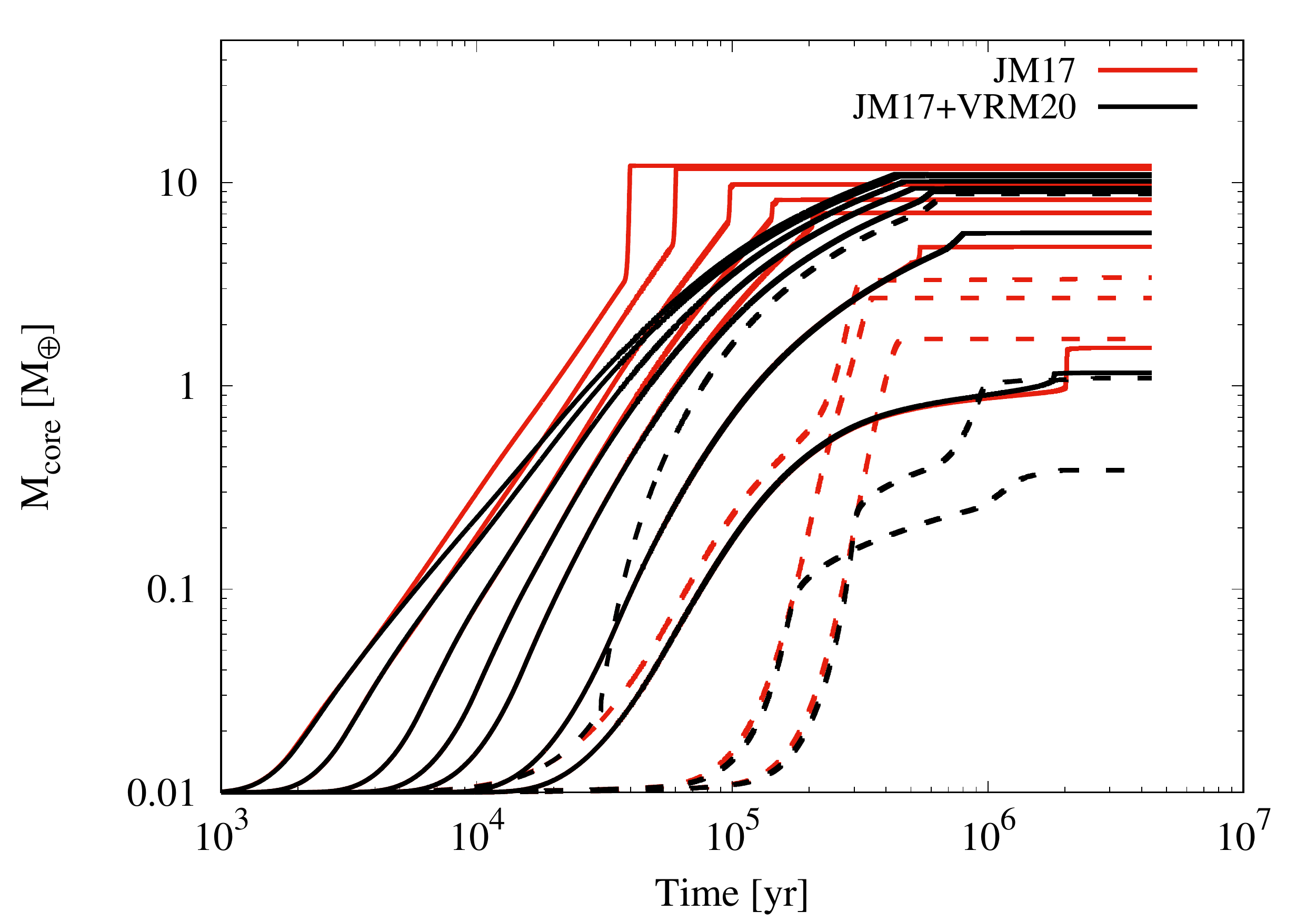} 
    \caption{Core masses as a function of time for all the simulations with the fiducial disc using $\alpha=10^{-4}$. The red lines correspond to the simulations using the migration recipes from JM17, while the black lines represent the simulation adopting the combined migration prescriptions from JM17 and VRM20. The dashed and solid lines correspond to the planets initially located inside and beyond the corresponding ice line, respectively. }
    \label{fig4_sec4}
\end{figure}

In Fig.~\ref{fig4_sec4}, we plot the mass of the cores as a function of time. We note that the timescale to reach the PIM is shorter when the thermal torque is not considered (the red lines). For the planets initially located beyond the ice line (the solid lines), except the ones located at 15 and 20~au, formation timescales are very short, between $\sim 4 \times 10^4$ and $2 \times 10^5$~yr. These formation timescales are in very good agreement with the ones found by \citet{Drazkowska21} for their simulations using $\alpha=10^{-4}$ and a dust fragmentation threshold velocity of 10~m/s. We note that \citet{Drazkowska21} did not neither consider planet migration, nor a change in the dust properties at the ice line. In our  simulations, the planets migrating inwards significantly increase their core masses in a very short timescale near the ice line due to the solid accumulation at this location by the radial drift. For the planets initially located inside the ice line (dashed lines), formation timescales are longer, between $\sim 2 \times 10^5$ and $4 \times 10^5$~yr. For the simulations where the thermal torque is included (black lines), the outward migration of the planets increase the formation timescales. For the planets initially located outside the ice line (except the two outermost) and the one at 2~au, which are the planets that significantly migrate outward, formation timescales are between $\sim 4 \times 10^5$ and $7 \times 10^5$~yr. For the planets initially located at 0.5 and 1~au the delays in the formation are longer, with formation timescales of about 1~Myr. Finally, the two outermost planets have similar  formation timescales in both set of simulations.  

\subsection{Dependence with the time at which embryos are inserted in the disc}

\begin{figure}
    \includegraphics[angle= 270, width=\columnwidth]{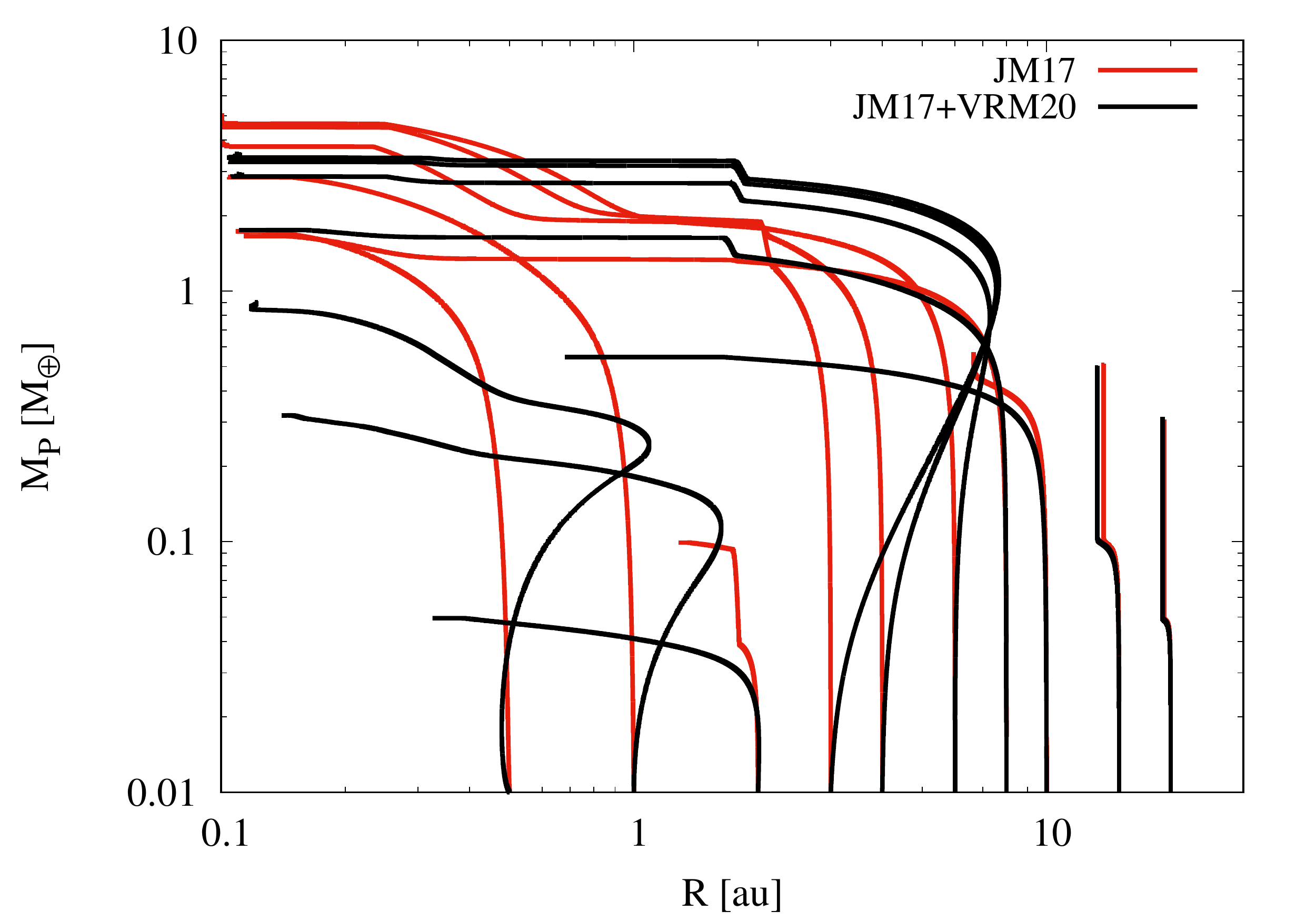} 
    \caption{Planet formation tracks for our fiducial disc using $\alpha=10^{-4}$ and introducing the embryos in the disc after $10^5$~yr of disc evolution.  While red lines correspond to the simulations using the migration recipes from JM17, the black ones represent the simulation adopting the combined migration prescriptions from JM17 and VRM20 which take into account the computation of the thermal torque.}
    \label{fig1_sec4.1}
\end{figure}

In this section, we analyse the planet formation tracks, but now inserting the planets in the disc after $10^5$~yr of disc evolution, shown in Fig.~\ref{fig1_sec4.1}. As before, the black and red lines represent the simulations where the thermal torque is and is not considered, respectively. In this case, for the planets initially located beyond the ice line, the heating torque only generates an outward migration for those planets located at 4, 6 and 8~au. For the planet initially located at 10~au the cold torque generates an extra inward migration, while for the outermost planets they only grow to about $0.5\text{M}_{\oplus}$ and thermal torque does not play a relevant role. While the most massive planets reach the PIM, these are significantly lower than the corresponding ones when embryos are inserted at the beginning of the simulation. This is a different result than the one found by \citet{Drazkowska21}, where they obtained that planets inserted after $10^5$~yr of disc evolution reach practically the same PIM and over the same timescale than planets inserted at $t=0$. However, we note that \citet{Drazkowska21} used a very massive disc of $0.2~\text{M}_{\odot}$, which implies a total solid mass of $650~\text{M}_{\oplus}$, four times larger than in our fiducial disc. We also note that, as  before, due to the fact that PIM are similar with and without the thermal torque, planets end near the disc inner edge with similar final masses. For the planets initially located inside the ice line, we find again the largest differences. For those initially  located at 0.5 and 1~au, the heating torque generates an outward migration and they finish with smaller masses than the corresponding ones without thermal torque, which always migrate inwards. In the case of the planet initially located at 2~au, when the thermal torque is considered, the cold torque generates an extra inward migration leaving the planet closer to the central star and with a lower mass. 

Finally, we note that if embryos are inserted in the disc after 0.5~Myr of disc evolution, they practically do not grow. This is due to the fact that the solid surface density significantly decreases due to the solid radial drift, especially beyond the ice line, and also to the fact that inside the ice line, Stokes numbers are very low. We note that \citet{Drazkowska21} found a similar result but introducing the embryos at 1~Myr of disc evolution.  

\subsection{Dependence with the disc mass and metallicity}

As the heating torque depends directly on the planet's luminosity, and hence on the solid accretion rate, we analyse in this section two quantities that increase the pebble accretion in comparison with that of our fiducial disc: the mass of the disc and its metallicity. We consider two cases: the fiducial disc but with $M_{\text{d}} = 0.1M_\odot$, and the fiducial disc but with an initial metallicity of $Z_0= 0.03$.

\begin{figure}
    \includegraphics[angle= 270, width=\columnwidth]{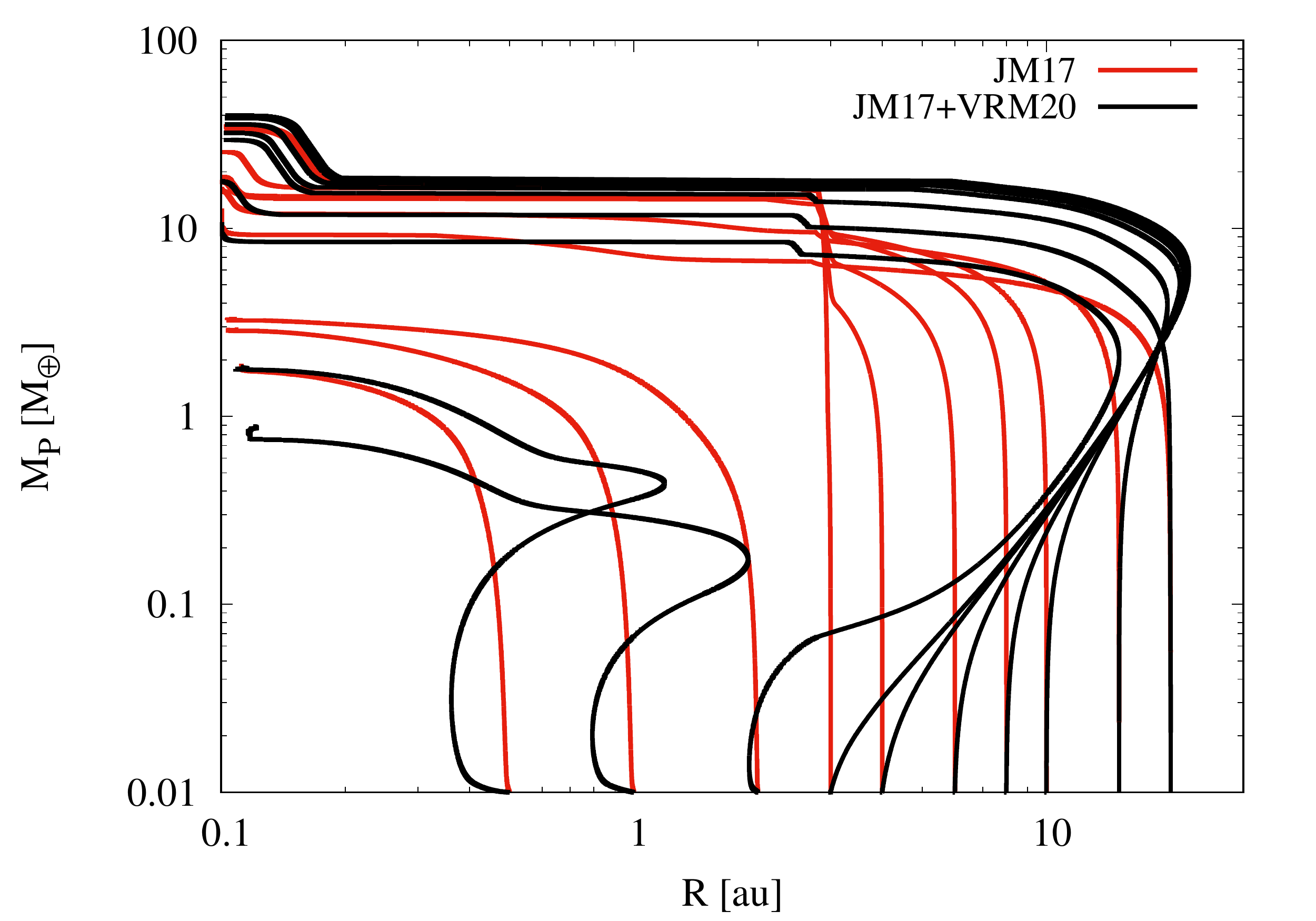} 
    \caption{Planet formation tracks for a massive disc of $0.1~\text{M}_{\odot}$ using the same disc parameters as in the  fiducial case, and adopting $\alpha=10^{-4}$. The red lines correspond to the simulations using the migration recipes from JM17, while the black ones represent the simulations adopting the combined migration prescriptions from JM17 and VRM20.}
    \label{fig1_sec4.2}
\end{figure}

In Fig.~\ref{fig1_sec4.2}, we show the formation tracks for the set of simulations using a massive disc of $0.1~\text{M}_{\odot}$ and $\alpha=10^{-4}$. For the set of simulations with thermal torque, all the planets migrate inward until they reach the disc inner edge. Those planets initially located beyond the ice line, which in this case is initially located at $\sim 2.9$~au, reach the PIM (between $\sim 15~\text{M}_{\oplus}$ and $\sim 19~\text{M}_{\oplus}$) near the ice line (with the exception of the two outermost planets that reach the PIM inside the ice line). Then, these planets migrate inward accreting considerable amounts of gas near the inner disc edge ending up with total masses between $\sim 20~\text{M}_{\oplus}$ and  $\sim 35~\text{M}_{\oplus}$. For the planet formation tracks adopting the combined migration recipes from JM17 and VRM20, the planets migrate outward until $\sim 20$ -- 22~au, except for the outermost one that always migrates inward. At this point, the heating torque does not dominate the total torque anymore and the planet migration is reversed. Except for the planet initially located at 16~au, which reaches a PIM of $\sim 17~\text{M}_{\oplus}$ near the ice line, the rest of the planets reach a PIM of about $\sim 20~\text{M}_{\oplus}$ at locations between $\sim 5$ -- 6~au. These larger cores allow most of the planets to accrete more gas, ending up generally with larger total masses between $\sim 30~\text{M}_{\oplus}$ and $\sim 40~\text{M}_{\oplus}$. Moreover, the more massive planets change to the type II migration regime at $\sim 0.2$~au and accrete significant amounts of gas. The subsequent asymptotic growth of the planets is due to the limitation in gas accretion after the planets open a gap in the disc \citep[see][for details]{Ronco2017}. We can see here again that the planet formation tracks for the planets initially located inside the ice line are those that that present the largest differences. When the thermal torque is considered, planets initially migrate inward more efficiently by the contribution of the cold torque, but inward migration is quickly reversed by the heating torque. Particularly interesting is the case of the planet initially located at 2~au, which migrates outward until about 15~au crossing the ice line, and ending with a final mass of $\sim 10~\text{M}_{\oplus}$. In this case, not only the final mass of the planets is very different with respect to the runs without the thermal torque, but also the final composition differs significantly. When the thermal torque is not considered, the planet ends with a dry composition after accreting $\sim 3~\text{M}_{\oplus}$ of dry pebbles always inside the ice line. When the thermal torque is taken into account, most of the mass of the planet is accreted beyond the ice line (the planet crosses the ice line when it has about the mass of Mars), ending as a water rich planet. 

\begin{figure}
    \includegraphics[angle= 270, width=\columnwidth]{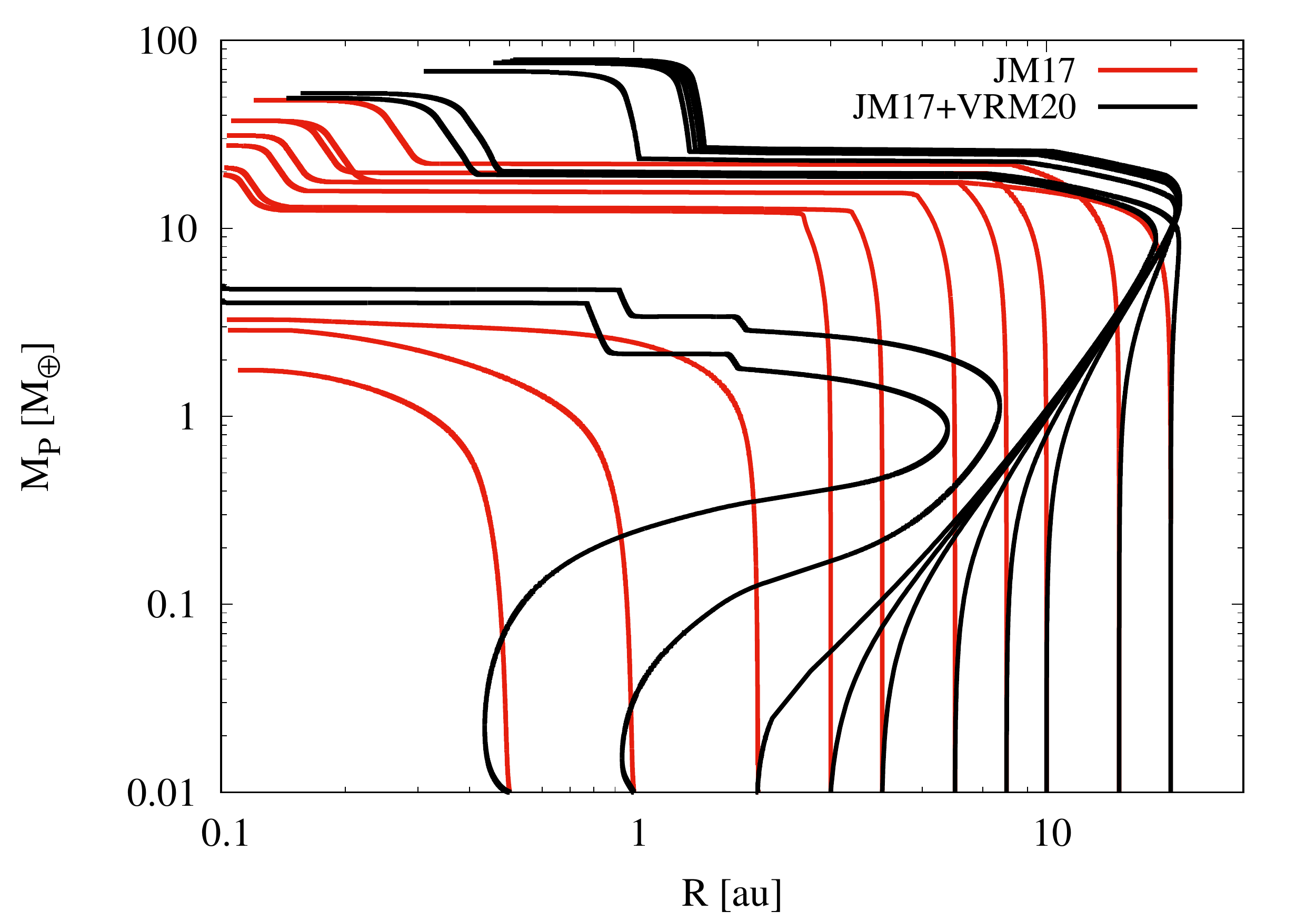} 
    \caption{Planet formation tracks for our fiducial disc using $\alpha=10^{-4}$ but now considering an initial dust-to-gas ratio of 0.03. Again, the red lines correspond to the simulations using the migration recipes from JM17, and the black ones represent the simulation adopting the combined migration prescriptions from JM17 and VRM20.}
    \label{fig2_sec4.2}
\end{figure}

\begin{figure}
    \includegraphics[angle= 270, width=\columnwidth]{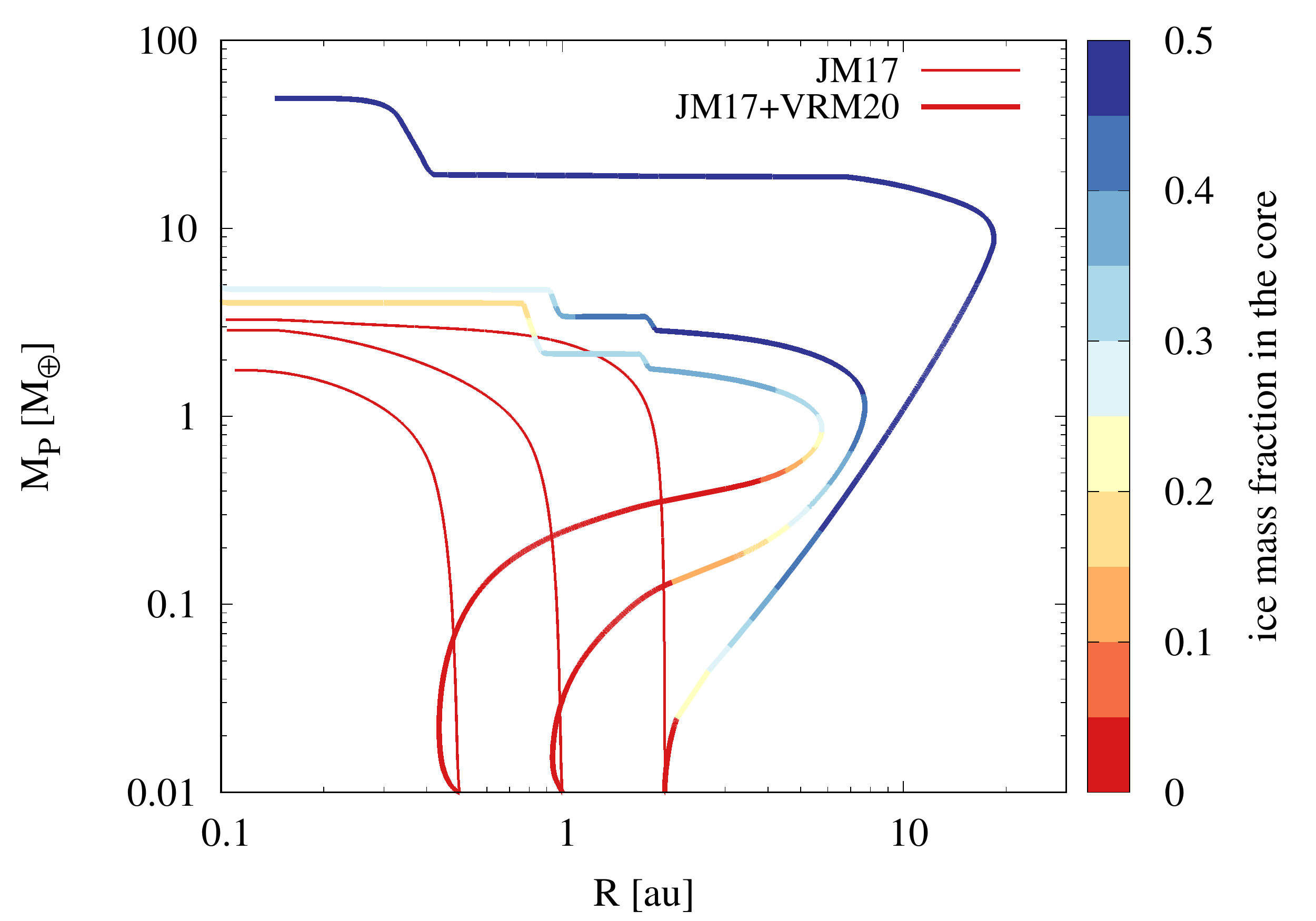} 
    \caption{Planet formation tracks for the planets initially located inside the ice line, for the case of an initial dust-to-gas ratio of 0.03. The thin lines correspond to the simulations using the migration recipes from JM17, and the thick ones represent the simulation adopting the combined migration prescriptions from JM17 and VRM20.}
    \label{fig3_sec4.2}
\end{figure}

Figure~\ref{fig2_sec4.2} shows the planet formation tracks for the case of the fiducial disc except for an initial metallicity, or dust-to-gas ratio, of 0.03. We can see that in this case the differences between both sets of simulations are larger. For the planets initially located beyond the ice line where the thermal torque is not considered, the timescale to reach the PIM is very short. This is due to the fact that there is three times more solids than in the fiducial case. In addition, the planets up to 10~au grow practically in situ. The planets reach a PIM of about 10 -- $20~\text{M}_{\oplus}$. After that, pebble accretion is halted and they quickly migrate inward accreting significant amounts of gas in the inner part of the disc. For the case where the thermal torque is considered, planets efficiently migrate outward until about 20~au, at which point their migration is reversed. They reach a PIM between $\sim 15$ -- $25~\text{M}_{\oplus}$ around 7 -- 10 au. For the more massive planets, this allows them to switch to type II regime at $\sim 1.5$~au and end up with final masses of about $90~\text{M}_{\oplus}$ at 0.5~au. As before, for the planets initially located inside the ice line, the thermal torque generates an outward migration, and planets cross the ice line changing  not only the final masses but also the compositions with respect to the cases where the thermal torque is not considered (see Fig~\ref{fig3_sec4.2}). 

 \section{Discussion}
\label{sec_5}

When the heating torque phenomenon was presented for the first time by \citet{Benitez-llambay2015}, it appeared as a possible solution to the well known problem of the fast inward migration of low- and intermediate-mass planets. Subsequently, M17 gave analitycal expressions of the thermal torque using linear perturbation theory, which were confirmed by numerical simulations of \citet{Hankla2020} and \citet{Chametla2021}. These expressions allow to incorporate thermal torque in global models of planet formation. In Paper I we included, for the first time, the expressions derived by M17 in a planet formation code and computed planet migration maps and planet formation tracks to analyse the role of the thermal torque on the migration of growing planets. Considering different planetesimal sizes and pebbles of 1 cm size, we showed that the thermal torque, and especially the heating torque, generates significant planet outward migration, drastically changing the planet formation tracks. On the contrary, \citet{Baumann2020} found a negligible impact of the thermal torque on planets growing purely by pebble accretion. These authors also included the expressions of the thermal torque derived by M17 in the planet formation and disc evolution model developed by \citet{Bitsch15b}. They studied the role of thermal torque on planets growing only by the accretion of pebbles, concluding that the inclusion of the thermal torque only generates small changes in the final masses and semi-major axis of the planets. However, we note that \citet{Baumann2020} also adopted the conservative approach of dropping to zero the thermal torque when the planet mass becomes larger than the critical thermal mass, underestimating the effect of the thermal torque. In addition, two important simplifications were considered in that work in order to compute the accretion rate of pebbles: a fixed and constant in time Stokes number of 0.1 for all pebbles along the disc, and an arbitrary initial pebble flux that simply decays exponentially with time.

Taking advantage of the recent incorporation of a model of dust growth and evolution in our global model of planet formation \citep{Guilera20, Venturini20c, Venturini20d}, here we studied the impact of the thermal torque in planets growing purely by pebble accretion, computing now, in a more consistent way, the dust and pebble evolution, i.e. the time evolution of Stokes numbers along the disc and the pebble mass flux. In addition, we also incorporated the cut-off functions of the thermal torque obtained by VRM20 from their high resolution simulations. First, we adopted a disc similar to that of Paper I, to construct planet migration maps for different values of the $\alpha$-viscosity parameter. As showed by \citet{Venturini20c,Venturini20d} and \citet{Drazkowska21}, planet formation by pebble accretion is very sensitive to the disc properties, especially to the $\alpha$-viscosity parameter, when dust growth and evolution is considered. We found that for $\alpha= 10^{-3}$ the thermal torques generate outward migration regions only at early times. This is due to the fact that Stokes numbers are low, and the solid surface density decreases quickly due to radial drift. In addition, pebble accretion tends to occur in the 3D regime for this value of $\alpha$, leading to practically no planet growth, as previously found by \citet{Venturini20c}. On the contrary, we found that the heating torque becomes important generating a significant region of outward migration when low values of the $\alpha$-viscosity parameter are adopted ($\alpha \lesssim 10^{-4}$). We note that such small values of $\alpha$ are needed to reproduce some of the observed disc ring structures \citep{Dullemond2018}, and that there is a growing body of evidence that turbulence levels, in protoplanetary discs, should in most cases be lower than that correpsonding to $\alpha \sim 10^{-4}$ \citep[][and references there in]{Flaherty2020}. Using the new prescriptions estimated by VRM20, the thermal torque generates extended regions of outward migration that survive for $\sim$ 1 Myr. These regions evolve in time moving towards the central star and they cover a mass range between about the mass of the Moon and the pebble isolation mass. 

Then, we computed planet formation tracks in order to study the impact of the new recipes from VRM20. We note that we do not perform a population synthesis analysis as \citet{Baumann2020} did. We adopted the same disc as in Paper I, and we focused in a detailed comparison between the simulations considering or not the thermal torque, computing  the formation of the planets at different initial locations. For $\alpha = 10^{-4} $, we found that the thermal torque significantly modifies the planet formation tracks. For $\alpha= 10^{-3}$, planets practically do not grow over most of the disc. As we mentioned before, pebble accretion tends to be very inefficient for moderate and high values of $\alpha$. We note here that \citet{Drazkowska21} found that planets can reach the pebble isolation masses in the inner part of the disc using $\alpha= 10^{-3}$ and an initial massive disc of $0.2~\text{M}_{\odot}$. Despite that initial disc masses are still an open question, disc mass estimations from ALMA observations tend to indicate lower values \citep[see][for a recent review]{Andrews2020}, although some controversy persists \citep[e.g.][]{Manara2018, Binkert2021}.  

For planets with initial locations inside the ice line, we found that the heating torque generates an outward migration, changing mainly the final mass of the planets. In addition, for massive discs and high metallicities, in which the thermal torque is large, these planets can cross the ice line, accrete ice-rich pebbles and change their composition. This result could have important implications on the composition of super-Earths and mini-Neptunes
 \citep[e.g][]{Izidoro19, Venturini20c}. We would expect more massive and metallic stars (i.e, younger systems) to harbour more ice-rich mini-Neptunes.
 
 In the case of planets initially located beyond the ice line, when the thermal torque is considered, planets initially migrate outward upto $\sim 15$ -- 20~au. In the case of the fiducial disc, the planets reach similar PIM in both sets of simulations, i.e. with and without the thermal torque. Thus, the final masses of the planets tend to be similar. However, the timescales and locations where planets reach the PIM are different. In the case of massive discs and high metallicities, the thermal torque allows planets to generally reach larger PIM and end up also with larger masses. In addition, in the case of a disc with an initial dust-to-gas ratio of 0.03, and when the thermal torque is included, the planets open a gap in the disc and switch to type II migration regime at larger distances. Planets end up in this case with significant larger final masses and at larger final semi-major axes. 

Another important result is that, if planets do not begin their formation at early times, they practically do not grow. This result is directly linked to the fact that dust evolves very quickly by radial drift, so in less than about 1 Myr the solid surface density significantly decreases and most of the solid mass is lost \citep[see for example ][]{Drazkowska2016, Drazkowska2017, Drazkowska21}. This made us realize that in order to obtain reliable results for planet formation by pebble accretion with models that include dust growth and evolution, realistic initial conditions are needed for the discs. We note that our initial disc profiles are based on observations of class I -- class II young stellar objects. In addition, instead of inserting the embryos in the disc at a given time and location, it would be desirable to track where in the disc the conditions for planetesimal and embryo formation are fulfilled in order to have a better characterisation of the impact of the thermal torque on planet formation by pebble accretion. Along this line, we also note that planet formation tracks become different when we use the pebble isolation masses estimated by \citet{Ataiee18} instead of the fiducial ones derived by \citet{Lambrechts14} (which, as mentioned before, are very similar to the ones derived by \citet{Bitsch18} for the case of $\alpha= 10^{-4}$). While for high values of the $\alpha$-viscosity parameter the estimations of the pebble isolation mass derived by \citet{Ataiee18} and \citet{Bitsch18} are broadly similar, for lower $\alpha$ the pebble isolation masses computed from \citet{Ataiee18} are significantly smaller. This has a strong impact on the planet formation tracks, leading to lower final masses for the planets that are initially located beyond the ice line and quickly reach the pebble isolation mass.  

Another interesting point that deserves future investigation is the role played by the thermal torque on a hybrid scenario of pebble and planetesimal accretion \citep{A18, Venturini20}. \citet{Guilera20} showed that after the planet reaches the pebble isolation mass, the accretion of planetesimals is important. In the hybrid scenario, the energy released by the planetesimal accretion significantly delays the onset of runaway gas accretion. The accretion of planetesimals after the planet reaches the pebble isolation mass could also yield a non-negligible heating torque. Such hybrid accretion scenario could be favoured, for example, by  planetesimal formation models based on the flux of pebbles like the one presented by \citet{Lenz19}. In addition, \citet{Lichtenberg2021} recently proposed that the formation of planetesimals at the ice line during the class I and class II phases of the proto Solar nebula \citep[employing the model developed in][]{Drazkowska2018} could explain some constraints from accretion chronology, thermo-chemistry, and the mass divergence of inner and outer Solar System. In this model, about a Jupiter mass in planetesimals is formed during the class II phase, in which the ice line moves inward. Thus, hybrid accretion could also apply in such scenario. 

We note that our approach does not consider the growth of the planet eccentricity due to the thermal torque. \citet{Eklund2017} and \citet{Fromenteau2019} showed that the heating torque can increase the planet eccentricity. When this happens, the migration path can differ significantly from that predicted by the thermal torque evaluated assuming a circular orbit. Also, \citet{Liu2018} showed that pebble accretion efficiency increases for higher eccentricities, leading to more luminous planets, and, in consequence, a larger thermal torque. Thus, the simultaneous computation of both phenomena is needed in a more detailed model in order asses their impact on the results presented here.

We also remark that the thermal torque depends on the disc thermodynamics, especially on the thermal diffusivity. Thus, a detailed description of the disc thermodynamics is desirable to obtain reliable values of the thermal torque. Both, the cooling of the disc and the value of the thermal diffusivity, depend on the disc opacity. We note that in our model, we use the grain and molecular opacities, depending on the density and temperature, from \citet{BL94}. In general, grain opacities dominate in protoplanetary discs except in the hottest region where grains can be fully vaporised. The opacities from \citet{BL94} are suitable for typical grain sizes in the interstellar medium of about micro-meter sizes. We note that in the case of a dust growth and evolution model, grain opacities should be computed taking into account the dust size distribution. In addition, \citet{Savvidou2020} showed using a dust size distribution, for a stationary model, that the disc opacities can be lower from those of \citet{BL94}, especially for low values of the $\alpha$-viscosity parameter where dust grows to larger sizes. However, the incorporation of the computation of the opacities from a dust size distribution at each time step during disc evolution in our 1D+1D model is a very complex and computationally expensive task, well beyond the scope of the present work.  

Finally, we point out that most of the planets end up near, or at the disc inner edge. This is related with the formation timescales of the planets, especially those planets initially located beyond the ice line. We show that planets reach the pebble isolation masses in a timescale shorter than $5\times10^{5}$~yr \citep[in line with the results found by][]{Drazkowska21}, when there is still plenty of gas in the disc. Due to the fact that the pebble isolation masses are generally $\gtrsim 10~\text{M}_{\oplus}$, type I migration timescales result shorter than gas accretion timescales, and generally the planets open a gap in the gas disc an switch to type II migration close to the central star. We note here that, as in \citet{Venturini20c, Venturini20d}, when planets reach the pebble isolation masses, we use the gas accretion rates from \citet{Ikoma00} without any dust opacity reduction. Thus, if a dust opacity reduction is invoked, gas accretion timescales become shorter and could allow the planets to open a gap at larger distances, avoiding reach the inner disc edge\footnote{Other effect that accelerates gas accretion is the envelope pollution by heavy elements \citep[e.g][]{Venturini16, Ormel2021}.}. We also note that in \citet{Guilera20}, we found larger gas accretion rates (a factor $\sim 3$) respect to the rates given by \cite{Ikoma00}, after the planet reaches the pebble isolation mass. However, the model to compute gas accretion in \citet{Guilera20} crashes some times after the mass of the envelope becomes greater than the core mass, and it is not able to compute all the full planet formation tracks. Only in the case of high metallicities, planets are able to open a gap beyond 1~au and end up with semi-major axis of about 0.5~au. In App. 2 we test this result for the fiducial disc adopting the type I migration recipes from \citet{Paardekooper2011}, instead of the ones from JM17, finding similar results. In this line, another important mechanism that can change the planet formation tracks, preventing the planets to end up very close to the central star, is the migration of intermediate-mass planets in low-viscosity discs. Through high resolution 2D hydrodynamical simulations, \citet{McNally2019} showed that while a $\sim 4~\text{M}_{\oplus}$ planet in a low viscous disc with $\alpha \sim 10^{-4}$ has a migration rate very similar to that given by the classical Type I estimations \citep{Paardekooper2011}, the migration of a $10~\text{M}_{\oplus}$ planet is slower than the expected one from the analytical Type I estimations. This is due to the fact that at this mass and for this $\alpha$ the planet is massive enough to modify the local gas surface density of the disc, generating a partial gap. \citet{McNally2019} showed that for $\alpha \sim 10^{-4}$ the gas surface density is smoothed and vortices can be damped efficiently slowing the migration of the planet. In addition, for $\alpha \lesssim 10^{-5}$, the generation of vortices at the pressure maxima at the edge of the partial gap can induce an outward planet migration, while for $\alpha \gtrsim 10^{-3}$ vortices are suppressed and the analytical Type I migration rates are recovered. Moreover, \citet{McNally2020} showed through 3D hydrodynamical simulations that the planet migration can be more complex due to the presence of buoyancy resonances. Thus, the migration of the planets in our low viscosity discs after they reach the PIM could be substantially smaller that the migration computed in our work, allowing them to end at larger distances.  

In order to obtain more extended outcomes, a planetary population synthesis study is needed (varying the free parameters of the model like the disc parameters, times at which embryos are inserted in the disc, dust opacity reduction, modify type I migration rates, etc.). This will be the subject of future works. However, we remark that our model (computing the thermal torque as in Paper I) is able to reproduce the observed mass -- radius relationship of the exoplanets with orbital periods less than 100~days \citep{Venturini20d}.

\section{Conclusions}
\label{sec_6}

In this work we study the role of the thermal torque in the migration of pebble accreting planets. Our models include dust growth and evolution as well as new prescriptions for the thermal torque given by VRM20. Our main results are:

\begin{itemize}
\item[--] The thermal torque plays an important role on the planet formation tracks for low $\alpha$-viscosity parameters, $\alpha \lesssim 10^{-4}$, especially for massive discs ($\text{M}_{\text{d}} \gtrsim 0.1~\text{M}_{\odot}$) and high metallicities ($Z \gtrsim 0.03$). For values of $\alpha \gtrsim 10^{-3}$, the dust does not grow efficiently, leading to low Stokes numbers, and hence to low core accretion rates and negligible thermal torques.

\item[--] The thermal torque is stronger for planets forming beyond the ice line, where pebble sizes are larger and core accretion rates higher. A planet growing in this region can experience an outward migration of several au.

\item[--] For planets whose formation starts inside the ice line, the final mass is not dramatically modified when comparing to the case where the thermal torque is neglected. However, with the new thermal torque prescriptions from VRM20, we find that the planets can undergo outward planet migration, which modifies the final planet composition. This effect is particularly relevant for the case of massive discs ($\text{M}_{\text{d}} \gtrsim 0.1~\text{M}_{\odot}$) and/or high disc metallicites ($Z \gtrsim 0.03$), where planets can cross the ice line and accrete large amounts of icy pebbles. This result was not found in previous computations, which abruptly halted the thermal torque when the planet mass becomes larger than the critical thermal mass as a conservative approach.

\item[--] Embryos growing by pebble accretion must start accreting at very early times (t$\lesssim 0.5$ Myr) for planets to form. This is due to the rapid pebble drift, which depletes the solids in the discs for later times.

\item[--] The thermal torque can be an important contribution to the total torque over a planet when it grows by pebble accretion, changing its final mass, semi-major axis and composition. We conclude that it should not be neglected in planet formation models.
\end{itemize}

\section*{Acknowledgements}

We thank the anonymous referee for her/his review, which helped us to improve and clarify this work. OMG is partially supported by the PICT 2018-0934 from ANPCyT, Argentina. OMG and M3B are partially supported by the PICT 2016-0053 from ANPCyT, Argentina. OMG, JC and MPR acknowledge support by ANID, -- Millennium Science Initiative Program -- NCN19\_171. OMG acknowledges the hosting by IA-PUC as an invited researcher. MPR acknowledges financial support provided by FONDECYT grant 3190336.

\section*{Data availability}

The data presented in this paper is available from the authors upon reasonable
request.




\bibliographystyle{mnras}
\bibliography{biblio} 

\begin{thebibliography}{}
\makeatletter
\relax
\def\mn@urlcharsother{\let\do\@makeother \do\$\do\&\do\#\do\^\do\_\do\%\do\~}
\def\mn@doi{\begingroup\mn@urlcharsother \@ifnextchar [ {\mn@doi@}
  {\mn@doi@[]}}
\def\mn@doi@[#1]#2{\def\@tempa{#1}\ifx\@tempa\@empty \href
  {http://dx.doi.org/#2} {doi:#2}\else \href {http://dx.doi.org/#2} {#1}\fi
  \endgroup}
\def\mn@eprint#1#2{\mn@eprint@#1:#2::\@nil}
\def\mn@eprint@arXiv#1{\href {http://arxiv.org/abs/#1} {{\tt arXiv:#1}}}
\def\mn@eprint@dblp#1{\href {http://dblp.uni-trier.de/rec/bibtex/#1.xml}
  {dblp:#1}}
\def\mn@eprint@#1:#2:#3:#4\@nil{\def\@tempa {#1}\def\@tempb {#2}\def\@tempc
  {#3}\ifx \@tempc \@empty \let \@tempc \@tempb \let \@tempb \@tempa \fi \ifx
  \@tempb \@empty \def\@tempb {arXiv}\fi \@ifundefined
  {mn@eprint@\@tempb}{\@tempb:\@tempc}{\expandafter \expandafter \csname
  mn@eprint@\@tempb\endcsname \expandafter{\@tempc}}}

\bibitem[\protect\citeauthoryear{Alibert et~al.,}{Alibert et~al.}{2018}]{A18}
Alibert Y.,  et~al., 2018, \mn@doi [Nature Astronomy]
  {10.1038/s41550-018-0557-2}

\bibitem[\protect\citeauthoryear{{Andrews}}{{Andrews}}{2020}]{Andrews2020}
{Andrews} S.~M.,  2020, \mn@doi [\araa] {10.1146/annurev-astro-031220-010302},
  \href {https://ui.adsabs.harvard.edu/abs/2020ARA&A..58..483A} {58, 483}

\bibitem[\protect\citeauthoryear{{Andrews}, {Wilner}, {Hughes}, {Qi}  \&
  {Dullemond}}{{Andrews} et~al.}{2010}]{Andrews10}
{Andrews} S.~M.,  {Wilner} D.~J.,  {Hughes} A.~M.,  {Qi} C.,   {Dullemond}
  C.~P.,  2010, \mn@doi [\apj] {10.1088/0004-637X/723/2/1241}, \href
  {http://adsabs.harvard.edu/abs/2010ApJ...723.1241A} {723, 1241}

\bibitem[\protect\citeauthoryear{{Ataiee}, {Baruteau}, {Alibert}  \&
  {Benz}}{{Ataiee} et~al.}{2018}]{Ataiee18}
{Ataiee} S.,  {Baruteau} C.,  {Alibert} Y.,   {Benz} W.,  2018, \mn@doi [\aap]
  {10.1051/0004-6361/201732026}, \href
  {http://adsabs.harvard.edu/abs/2018A%26A...615A.110A} {615, A110}

\bibitem[\protect\citeauthoryear{{Baumann} \& {Bitsch}}{{Baumann} \&
  {Bitsch}}{2020}]{Baumann2020}
{Baumann} T.,  {Bitsch} B.,  2020, \mn@doi [\aap]
  {10.1051/0004-6361/202037579}, \href
  {https://ui.adsabs.harvard.edu/abs/2020A&A...637A..11B} {637, A11}

\bibitem[\protect\citeauthoryear{{Bell} \& {Lin}}{{Bell} \& {Lin}}{1994}]{BL94}
{Bell} K.~R.,  {Lin} D.~N.~C.,  1994, \mn@doi [\apj] {10.1086/174206}, \href
  {http://adsabs.harvard.edu/abs/1994ApJ...427..987B} {427, 987}

\bibitem[\protect\citeauthoryear{{Ben{\'{\i}}tez-Llambay}, {Masset},
  {Koenigsberger}  \& {Szul{\'a}gyi}}{{Ben{\'{\i}}tez-Llambay}
  et~al.}{2015}]{Benitez-llambay2015}
{Ben{\'{\i}}tez-Llambay} P.,  {Masset} F.,  {Koenigsberger} G.,
  {Szul{\'a}gyi} J.,  2015, \mn@doi [\nat] {10.1038/nature14277}, \href
  {http://adsabs.harvard.edu/abs/2015Natur.520...63B} {520, 63}

\bibitem[\protect\citeauthoryear{{Binkert}, {Szul{\'a}gyi}  \&
  {Birnstiel}}{{Binkert} et~al.}{2021}]{Binkert2021}
{Binkert} F.,  {Szul{\'a}gyi} J.,   {Birnstiel} T.,  2021, arXiv e-prints,
  \href {https://ui.adsabs.harvard.edu/abs/2021arXiv210310177B} {p.
  arXiv:2103.10177}

\bibitem[\protect\citeauthoryear{{Birnstiel}, {Klahr}  \&
  {Ercolano}}{{Birnstiel} et~al.}{2012}]{Birnstiel12}
{Birnstiel} T.,  {Klahr} H.,   {Ercolano} B.,  2012, \mn@doi [\aap]
  {10.1051/0004-6361/201118136}, \href
  {http://adsabs.harvard.edu/abs/2012A%26A...539A.148B} {539, A148}

\bibitem[\protect\citeauthoryear{{Bitsch}, {Lambrechts}  \&
  {Johansen}}{{Bitsch} et~al.}{2015}]{Bitsch15b}
{Bitsch} B.,  {Lambrechts} M.,   {Johansen} A.,  2015, \mn@doi [\aap]
  {10.1051/0004-6361/201526463}, \href
  {http://adsabs.harvard.edu/abs/2015A%26A...582A.112B} {582, A112}

\bibitem[\protect\citeauthoryear{{Bitsch}, {Morbidelli}, {Johansen}, {Lega},
  {Lambrechts}  \& {Crida}}{{Bitsch} et~al.}{2018}]{Bitsch18}
{Bitsch} B.,  {Morbidelli} A.,  {Johansen} A.,  {Lega} E.,  {Lambrechts} M.,
  {Crida} A.,  2018, \mn@doi [\aap] {10.1051/0004-6361/201731931}, \href
  {http://adsabs.harvard.edu/abs/2018A%26A...612A..30B} {612, A30}

\bibitem[\protect\citeauthoryear{{Carrera}, {Johansen}  \& {Davies}}{{Carrera}
  et~al.}{2015}]{Carrera2015}
{Carrera} D.,  {Johansen} A.,   {Davies} M.~B.,  2015, \mn@doi [\aap]
  {10.1051/0004-6361/201425120}, \href
  {https://ui.adsabs.harvard.edu/abs/2015A&A...579A..43C} {579, A43}

\bibitem[\protect\citeauthoryear{{Chametla} \& {Masset}}{{Chametla} \&
  {Masset}}{2021}]{Chametla2021}
{Chametla} R.~O.,  {Masset} F.~S.,  2021, \mn@doi [\mnras]
  {10.1093/mnras/staa3681}, \href
  {https://ui.adsabs.harvard.edu/abs/2021MNRAS.501...24C} {501, 24}

\bibitem[\protect\citeauthoryear{{Chametla}, {D'Angelo}, {Reyes-Ruiz}  \&
  {S{\'a}nchez-Salcedo}}{{Chametla} et~al.}{2020}]{Chametla2020}
{Chametla} R.~O.,  {D'Angelo} G.,  {Reyes-Ruiz} M.,   {S{\'a}nchez-Salcedo}
  F.~J.,  2020, \mn@doi [\mnras] {10.1093/mnras/staa260}, \href
  {https://ui.adsabs.harvard.edu/abs/2020MNRAS.tmp..243C} {p.~243}

\bibitem[\protect\citeauthoryear{{Crida}, {Morbidelli}  \& {Masset}}{{Crida}
  et~al.}{2006}]{Crida2006}
{Crida} A.,  {Morbidelli} A.,   {Masset} F.,  2006, \mn@doi [\icarus]
  {10.1016/j.icarus.2005.10.007}, \href
  {http://adsabs.harvard.edu/abs/2006Icar..181..587C} {181, 587}

\bibitem[\protect\citeauthoryear{{Drazkowska}, {Stammler}  \&
  {Birnstiel}}{{Drazkowska} et~al.}{2021}]{Drazkowska21}
{Drazkowska} J.,  {Stammler} S.~M.,   {Birnstiel} T.,  2021, arXiv e-prints,
  \href {https://ui.adsabs.harvard.edu/abs/2021arXiv210101728D} {p.
  arXiv:2101.01728}

\bibitem[\protect\citeauthoryear{{Dr{\c a}zkowska}, {Alibert}  \&
  {Moore}}{{Dr{\c a}zkowska} et~al.}{2016}]{Drazkowska16}
{Dr{\c a}zkowska} J.,  {Alibert} Y.,   {Moore} B.,  2016, \mn@doi [\aap]
  {10.1051/0004-6361/201628983}, \href
  {http://adsabs.harvard.edu/abs/2016A%26A...594A.105D} {594, A105}

\bibitem[\protect\citeauthoryear{{Dr{\k{a}}{\.z}kowska} \&
  {Alibert}}{{Dr{\k{a}}{\.z}kowska} \& {Alibert}}{2017}]{Drazkowska2017}
{Dr{\k{a}}{\.z}kowska} J.,  {Alibert} Y.,  2017, \mn@doi [\aap]
  {10.1051/0004-6361/201731491}, \href
  {https://ui.adsabs.harvard.edu/abs/2017A&A...608A..92D} {608, A92}

\bibitem[\protect\citeauthoryear{{Dr{\k{a}}{\.z}kowska} \&
  {Dullemond}}{{Dr{\k{a}}{\.z}kowska} \& {Dullemond}}{2018}]{Drazkowska2018}
{Dr{\k{a}}{\.z}kowska} J.,  {Dullemond} C.~P.,  2018, \mn@doi [\aap]
  {10.1051/0004-6361/201732221}, \href
  {https://ui.adsabs.harvard.edu/abs/2018A&A...614A..62D} {614, A62}

\bibitem[\protect\citeauthoryear{{Dr{\k{a}}{\.z}kowska}, {Alibert}  \&
  {Moore}}{{Dr{\k{a}}{\.z}kowska} et~al.}{2016}]{Drazkowska2016}
{Dr{\k{a}}{\.z}kowska} J.,  {Alibert} Y.,   {Moore} B.,  2016, \mn@doi [\aap]
  {10.1051/0004-6361/201628983}, \href
  {https://ui.adsabs.harvard.edu/abs/2016A&A...594A.105D} {594, A105}

\bibitem[\protect\citeauthoryear{{Dullemond} et~al.,}{{Dullemond}
  et~al.}{2018}]{Dullemond2018}
{Dullemond} C.~P.,  et~al., 2018, \mn@doi [\apjl] {10.3847/2041-8213/aaf742},
  \href {https://ui.adsabs.harvard.edu/abs/2018ApJ...869L..46D} {869, L46}

\bibitem[\protect\citeauthoryear{{Eklund} \& {Masset}}{{Eklund} \&
  {Masset}}{2017}]{Eklund2017}
{Eklund} H.,  {Masset} F.~S.,  2017, \mn@doi [\mnras] {10.1093/mnras/stx856},
  \href {https://ui.adsabs.harvard.edu/abs/2017MNRAS.469..206E} {469, 206}

\bibitem[\protect\citeauthoryear{{Flaherty} et~al.,}{{Flaherty}
  et~al.}{2020}]{Flaherty2020}
{Flaherty} K.,  et~al., 2020, \mn@doi [\apj] {10.3847/1538-4357/ab8cc5}, \href
  {https://ui.adsabs.harvard.edu/abs/2020ApJ...895..109F} {895, 109}

\bibitem[\protect\citeauthoryear{{Fromenteau} \& {Masset}}{{Fromenteau} \&
  {Masset}}{2019}]{Fromenteau2019}
{Fromenteau} S.,  {Masset} F.~S.,  2019, \mn@doi [\mnras]
  {10.1093/mnras/stz718}, \href
  {https://ui.adsabs.harvard.edu/abs/2019MNRAS.485.5035F} {485, 5035}

\bibitem[\protect\citeauthoryear{{Guilera}, {Miller Bertolami}  \&
  {Ronco}}{{Guilera} et~al.}{2017}]{Guilera2017b}
{Guilera} O.~M.,  {Miller Bertolami} M.~M.,   {Ronco} M.~P.,  2017, \mn@doi
  [\mnras] {10.1093/mnrasl/slx095}, \href
  {https://ui.adsabs.harvard.edu/abs/2017MNRAS.471L..16G} {471, L16}

\bibitem[\protect\citeauthoryear{{Guilera}, {Cuello}, {Montesinos}, {Miller
  Bertolami}, {Ronco}, {Cuadra}  \& {Masset}}{{Guilera}
  et~al.}{2019}]{Guilera2019}
{Guilera} O.~M.,  {Cuello} N.,  {Montesinos} M.,  {Miller Bertolami} M.~M.,
  {Ronco} M.~P.,  {Cuadra} J.,   {Masset} F.~S.,  2019, \mn@doi [\mnras]
  {10.1093/mnras/stz1158}, \href
  {https://ui.adsabs.harvard.edu/abs/2019MNRAS.486.5690G} {486, 5690}

\bibitem[\protect\citeauthoryear{{Guilera}, {S{\'a}ndor}, {Ronco}, {Venturini}
  \& {Miller Bertolami}}{{Guilera} et~al.}{2020}]{Guilera20}
{Guilera} O.~M.,  {S{\'a}ndor} Z.,  {Ronco} M.~P.,  {Venturini} J.,   {Miller
  Bertolami} M.~M.,  2020, \mn@doi [\aap] {10.1051/0004-6361/202038458}, \href
  {https://ui.adsabs.harvard.edu/abs/2020A&A...642A.140G} {642, A140}

\bibitem[\protect\citeauthoryear{{Gundlach} \& {Blum}}{{Gundlach} \&
  {Blum}}{2015}]{Gundlach2015}
{Gundlach} B.,  {Blum} J.,  2015, \mn@doi [\apj] {10.1088/0004-637X/798/1/34},
  \href {https://ui.adsabs.harvard.edu/abs/2015ApJ...798...34G} {798, 34}

\bibitem[\protect\citeauthoryear{{Hankla}, {Jiang}  \& {Armitage}}{{Hankla}
  et~al.}{2020}]{Hankla2020}
{Hankla} A.~M.,  {Jiang} Y.-F.,   {Armitage} P.~J.,  2020, \mn@doi [\apj]
  {10.3847/1538-4357/abb4df}, \href
  {https://ui.adsabs.harvard.edu/abs/2020ApJ...902...50H} {902, 50}

\bibitem[\protect\citeauthoryear{{Ida} \& {Lin}}{{Ida} \&
  {Lin}}{2004}]{IdaLin2004a}
{Ida} S.,  {Lin} D.~N.~C.,  2004, \mn@doi [\apj] {10.1086/381724}, \href
  {http://adsabs.harvard.edu/abs/2004ApJ...604..388I} {604, 388}

\bibitem[\protect\citeauthoryear{{Ikoma}, {Nakazawa}  \& {Emori}}{{Ikoma}
  et~al.}{2000}]{Ikoma00}
{Ikoma} M.,  {Nakazawa} K.,   {Emori} H.,  2000, \mn@doi [\apj]
  {10.1086/309050}, \href {http://adsabs.harvard.edu/abs/2000ApJ...537.1013I}
  {537, 1013}

\bibitem[\protect\citeauthoryear{{Izidoro}, {Bitsch}, {Raymond}, {Johansen},
  {Morbidelli}, {Lambrechts}  \& {Jacobson}}{{Izidoro}
  et~al.}{2019}]{Izidoro19}
{Izidoro} A.,  {Bitsch} B.,  {Raymond} S.~N.,  {Johansen} A.,  {Morbidelli} A.,
   {Lambrechts} M.,   {Jacobson} S.~A.,  2019, arXiv e-prints, \href
  {https://ui.adsabs.harvard.edu/abs/2019arXiv190208772I} {p. arXiv:1902.08772}

\bibitem[\protect\citeauthoryear{{Jim{\'e}nez} \& {Masset}}{{Jim{\'e}nez} \&
  {Masset}}{2017}]{jm2017}
{Jim{\'e}nez} M.~A.,  {Masset} F.~S.,  2017, \mn@doi [\mnras]
  {10.1093/mnras/stx1946}, \href
  {http://adsabs.harvard.edu/abs/2017MNRAS.471.4917J} {471, 4917}

\bibitem[\protect\citeauthoryear{{Johansen}, {Oishi}, {Mac Low}, {Klahr},
  {Henning}  \& {Youdin}}{{Johansen} et~al.}{2007}]{Johansen07}
{Johansen} A.,  {Oishi} J.~S.,  {Mac Low} M.-M.,  {Klahr} H.,  {Henning} T.,
  {Youdin} A.,  2007, \mn@doi [\nat] {10.1038/nature06086}, \href
  {http://adsabs.harvard.edu/abs/2007Natur.448.1022J} {448, 1022}

\bibitem[\protect\citeauthoryear{{Lambrechts} \& {Johansen}}{{Lambrechts} \&
  {Johansen}}{2012}]{Lambrechts12}
{Lambrechts} M.,  {Johansen} A.,  2012, \mn@doi [\aap]
  {10.1051/0004-6361/201219127}, \href
  {http://adsabs.harvard.edu/abs/2012A%26A...544A..32L} {544, A32}

\bibitem[\protect\citeauthoryear{{Lambrechts}, {Johansen}  \&
  {Morbidelli}}{{Lambrechts} et~al.}{2014}]{Lambrechts14}
{Lambrechts} M.,  {Johansen} A.,   {Morbidelli} A.,  2014, \mn@doi [\aap]
  {10.1051/0004-6361/201423814}, \href
  {http://adsabs.harvard.edu/abs/2014A%26A...572A..35L} {572, A35}

\bibitem[\protect\citeauthoryear{{Lega}, {Crida}, {Bitsch}  \&
  {Morbidelli}}{{Lega} et~al.}{2014}]{Lega2014}
{Lega} E.,  {Crida} A.,  {Bitsch} B.,   {Morbidelli} A.,  2014, \mn@doi
  [\mnras] {10.1093/mnras/stu304}, \href
  {http://adsabs.harvard.edu/abs/2014MNRAS.440..683L} {440, 683}

\bibitem[\protect\citeauthoryear{{Lenz}, {Klahr}  \& {Birnstiel}}{{Lenz}
  et~al.}{2019}]{Lenz19}
{Lenz} C.~T.,  {Klahr} H.,   {Birnstiel} T.,  2019, \mn@doi [\apj]
  {10.3847/1538-4357/ab05d9}, \href
  {https://ui.adsabs.harvard.edu/abs/2019ApJ...874...36L} {874, 36}

\bibitem[\protect\citeauthoryear{{Li} \& {Youdin}}{{Li} \&
  {Youdin}}{2021}]{Li2021}
{Li} R.,  {Youdin} A.,  2021, arXiv e-prints, \href
  {https://ui.adsabs.harvard.edu/abs/2021arXiv210506042L} {p. arXiv:2105.06042}

\bibitem[\protect\citeauthoryear{{Lichtenberg}, {Dr{\k{a}}{\.z}kowska},
  {Sch{\"o}nb{\"a}chler}, {Golabek}  \& {Hands}}{{Lichtenberg}
  et~al.}{2021}]{Lichtenberg2021}
{Lichtenberg} T.,  {Dr{\k{a}}{\.z}kowska} J.,  {Sch{\"o}nb{\"a}chler} M.,
  {Golabek} G.~J.,   {Hands} T.~O.,  2021, \mn@doi [Science]
  {10.1126/science.abb3091}, \href
  {https://ui.adsabs.harvard.edu/abs/2021Sci...371..365L} {371, 365}

\bibitem[\protect\citeauthoryear{{Liu} \& {Ji}}{{Liu} \& {Ji}}{2020}]{Liu2020}
{Liu} B.,  {Ji} J.,  2020, \mn@doi [Research in Astronomy and Astrophysics]
  {10.1088/1674-4527/20/10/164}, \href
  {https://ui.adsabs.harvard.edu/abs/2020RAA....20..164L} {20, 164}

\bibitem[\protect\citeauthoryear{{Liu} \& {Ormel}}{{Liu} \&
  {Ormel}}{2018}]{Liu2018}
{Liu} B.,  {Ormel} C.~W.,  2018, \mn@doi [\aap] {10.1051/0004-6361/201732307},
  \href {https://ui.adsabs.harvard.edu/abs/2018A&A...615A.138L} {615, A138}

\bibitem[\protect\citeauthoryear{{Manara}, {Morbidelli}  \& {Guillot}}{{Manara}
  et~al.}{2018}]{Manara2018}
{Manara} C.~F.,  {Morbidelli} A.,   {Guillot} T.,  2018, \mn@doi [\aap]
  {10.1051/0004-6361/201834076}, \href
  {https://ui.adsabs.harvard.edu/abs/2018A&A...618L...3M} {618, L3}

\bibitem[\protect\citeauthoryear{{Masset}}{{Masset}}{2017}]{masset2017}
{Masset} F.~S.,  2017, \mn@doi [\mnras] {10.1093/mnras/stx2271}, \href
  {http://adsabs.harvard.edu/abs/2017MNRAS.472.4204M} {472, 4204}

\bibitem[\protect\citeauthoryear{{Masset} \& {Velasco Romero}}{{Masset} \&
  {Velasco Romero}}{2017}]{Masset_VelascoRomero17}
{Masset} F.~S.,  {Velasco Romero} D.~A.,  2017, \mn@doi [\mnras]
  {10.1093/mnras/stw3008}, \href
  {https://ui.adsabs.harvard.edu/abs/2017MNRAS.465.3175M} {465, 3175}

\bibitem[\protect\citeauthoryear{{McNally}, {Nelson}, {Paardekooper}  \&
  {Ben{\'\i}tez-Llambay}}{{McNally} et~al.}{2019}]{McNally2019}
{McNally} C.~P.,  {Nelson} R.~P.,  {Paardekooper} S.-J.,
  {Ben{\'\i}tez-Llambay} P.,  2019, \mn@doi [\mnras] {10.1093/mnras/stz023},
  \href {https://ui.adsabs.harvard.edu/abs/2019MNRAS.484..728M} {484, 728}

\bibitem[\protect\citeauthoryear{{McNally}, {Nelson}, {Paardekooper},
  {Ben{\'\i}tez-Llambay}  \& {Gressel}}{{McNally} et~al.}{2020}]{McNally2020}
{McNally} C.~P.,  {Nelson} R.~P.,  {Paardekooper} S.-J.,
  {Ben{\'\i}tez-Llambay} P.,   {Gressel} O.,  2020, \mn@doi [\mnras]
  {10.1093/mnras/staa576}, \href
  {https://ui.adsabs.harvard.edu/abs/2020MNRAS.493.4382M} {493, 4382}

\bibitem[\protect\citeauthoryear{{Morbidelli}, {Lambrechts}, {Jacobson}  \&
  {Bitsch}}{{Morbidelli} et~al.}{2015}]{Morby15}
{Morbidelli} A.,  {Lambrechts} M.,  {Jacobson} S.,   {Bitsch} B.,  2015,
  \mn@doi [\icarus] {10.1016/j.icarus.2015.06.003}, \href
  {http://adsabs.harvard.edu/abs/2015Icar..258..418M} {258, 418}

\bibitem[\protect\citeauthoryear{{Mordasini}, {Alibert}, {Benz}  \&
  {Naef}}{{Mordasini} et~al.}{2009}]{Mordasini2009}
{Mordasini} C.,  {Alibert} Y.,  {Benz} W.,   {Naef} D.,  2009, \mn@doi [\aap]
  {10.1051/0004-6361/200810697}, \href
  {http://adsabs.harvard.edu/abs/2009A$%$26A...501.1161M} {501, 1161}

\bibitem[\protect\citeauthoryear{{Ormel} \& {Klahr}}{{Ormel} \&
  {Klahr}}{2010}]{OrmelK10}
{Ormel} C.~W.,  {Klahr} H.~H.,  2010, \mn@doi [\aap]
  {10.1051/0004-6361/201014903}, \href
  {http://adsabs.harvard.edu/abs/2010A%26A...520A..43O} {520, A43}

\bibitem[\protect\citeauthoryear{{Ormel} \& {Liu}}{{Ormel} \&
  {Liu}}{2018}]{Ormel2018}
{Ormel} C.~W.,  {Liu} B.,  2018, \mn@doi [\aap] {10.1051/0004-6361/201732562},
  \href {https://ui.adsabs.harvard.edu/abs/2018A&A...615A.178O} {615, A178}

\bibitem[\protect\citeauthoryear{{Ormel}, {Liu}  \& {Schoonenberg}}{{Ormel}
  et~al.}{2017}]{Ormel2017}
{Ormel} C.~W.,  {Liu} B.,   {Schoonenberg} D.,  2017, \mn@doi [\aap]
  {10.1051/0004-6361/201730826}, \href
  {https://ui.adsabs.harvard.edu/abs/2017A&A...604A...1O} {604, A1}

\bibitem[\protect\citeauthoryear{{Ormel}, {Vazan}  \& {Brouwers}}{{Ormel}
  et~al.}{2021}]{Ormel2021}
{Ormel} C.~W.,  {Vazan} A.,   {Brouwers} M.~G.,  2021, \mn@doi [\aap]
  {10.1051/0004-6361/202039706}, \href
  {https://ui.adsabs.harvard.edu/abs/2021A&A...647A.175O} {647, A175}

\bibitem[\protect\citeauthoryear{{Paardekooper}, {Baruteau}  \&
  {Kley}}{{Paardekooper} et~al.}{2011}]{Paardekooper2011}
{Paardekooper} S.-J.,  {Baruteau} C.,   {Kley} W.,  2011, \mn@doi [\mnras]
  {10.1111/j.1365-2966.2010.17442.x}, \href
  {http://adsabs.harvard.edu/abs/2011MNRAS.410..293P} {410, 293}

\bibitem[\protect\citeauthoryear{{Pollack}, {Hubickyj}, {Bodenheimer},
  {Lissauer}, {Podolak}  \& {Greenzweig}}{{Pollack} et~al.}{1996}]{P96}
{Pollack} J.~B.,  {Hubickyj} O.,  {Bodenheimer} P.,  {Lissauer} J.~J.,
  {Podolak} M.,   {Greenzweig} Y.,  1996, \mn@doi [\icarus]
  {10.1006/icar.1996.0190}, \href
  {http://adsabs.harvard.edu/abs/1996Icar..124...62P} {124, 62}

\bibitem[\protect\citeauthoryear{{Ronco}, {Guilera}  \& {de El{\'\i}a}}{{Ronco}
  et~al.}{2017}]{Ronco2017}
{Ronco} M.~P.,  {Guilera} O.~M.,   {de El{\'\i}a} G.~C.,  2017, \mn@doi
  [\mnras] {10.1093/mnras/stx1746}, \href
  {https://ui.adsabs.harvard.edu/abs/2017MNRAS.471.2753R} {471, 2753}

\bibitem[\protect\citeauthoryear{{Safronov}}{{Safronov}}{1969}]{Safronov1969}
{Safronov} V.~S.,  1969, {Evoliutsiia doplanetnogo oblaka.}

\bibitem[\protect\citeauthoryear{{Savvidou}, {Bitsch}  \&
  {Lambrechts}}{{Savvidou} et~al.}{2020}]{Savvidou2020}
{Savvidou} S.,  {Bitsch} B.,   {Lambrechts} M.,  2020, \mn@doi [\aap]
  {10.1051/0004-6361/201936576}, \href
  {https://ui.adsabs.harvard.edu/abs/2020A&A...640A..63S} {640, A63}

\bibitem[\protect\citeauthoryear{{Tanaka}, {Takeuchi}  \& {Ward}}{{Tanaka}
  et~al.}{2002}]{Tanaka02}
{Tanaka} H.,  {Takeuchi} T.,   {Ward} W.~R.,  2002, \mn@doi [\apj]
  {10.1086/324713}, \href {http://adsabs.harvard.edu/abs/2002ApJ...565.1257T}
  {565, 1257}

\bibitem[\protect\citeauthoryear{{Velasco Romero} \& {Masset}}{{Velasco Romero}
  \& {Masset}}{2020}]{VelascoRomero2020}
{Velasco Romero} D.~A.,  {Masset} F.~S.,  2020, \mn@doi [\mnras]
  {10.1093/mnras/staa1215}, \href
  {https://ui.adsabs.harvard.edu/abs/2020MNRAS.495.2063V} {495, 2063}

\bibitem[\protect\citeauthoryear{{Venturini} \& {Helled}}{{Venturini} \&
  {Helled}}{2020}]{Venturini20}
{Venturini} J.,  {Helled} R.,  2020, \mn@doi [\aap]
  {10.1051/0004-6361/201936591}, \href
  {https://ui.adsabs.harvard.edu/abs/2020A&A...634A..31V} {634, A31}

\bibitem[\protect\citeauthoryear{{Venturini}, {Alibert}  \& {Benz}}{{Venturini}
  et~al.}{2016}]{Venturini16}
{Venturini} J.,  {Alibert} Y.,   {Benz} W.,  2016, \mn@doi [\aap]
  {10.1051/0004-6361/201628828}, \href
  {http://adsabs.harvard.edu/abs/2016A%26A...596A..90V} {596, A90}

\bibitem[\protect\citeauthoryear{{Venturini}, {Ronco}  \&
  {Guilera}}{{Venturini} et~al.}{2020a}]{Venturini20b}
{Venturini} J.,  {Ronco} M.~P.,   {Guilera} O.~M.,  2020a, \mn@doi [\ssr]
  {10.1007/s11214-020-00700-y}, \href
  {https://ui.adsabs.harvard.edu/abs/2020SSRv..216...86V} {216, 86}

\bibitem[\protect\citeauthoryear{{Venturini}, {Guilera}, {Haldemann}, {Ronco}
  \& {Mordasini}}{{Venturini} et~al.}{2020b}]{Venturini20d}
{Venturini} J.,  {Guilera} O.~M.,  {Haldemann} J.,  {Ronco} M.~P.,
  {Mordasini} C.,  2020b, \mn@doi [\aap] {10.1051/0004-6361/202039141}, \href
  {https://ui.adsabs.harvard.edu/abs/2020A&A...643L...1V} {643, L1}

\bibitem[\protect\citeauthoryear{{Venturini}, {Guilera}, {Ronco}  \&
  {Mordasini}}{{Venturini} et~al.}{2020c}]{Venturini20c}
{Venturini} J.,  {Guilera} O.~M.,  {Ronco} M.~P.,   {Mordasini} C.,  2020c,
  \mn@doi [\aap] {10.1051/0004-6361/202039140}, \href
  {https://ui.adsabs.harvard.edu/abs/2020A&A...644A.174V} {644, A174}

\bibitem[\protect\citeauthoryear{{Yang}, {Johansen}  \& {Carrera}}{{Yang}
  et~al.}{2017}]{Yang2017}
{Yang} C.-C.,  {Johansen} A.,   {Carrera} D.,  2017, \mn@doi [\aap]
  {10.1051/0004-6361/201630106}, \href
  {https://ui.adsabs.harvard.edu/abs/2017A&A...606A..80Y} {606, A80}

\bibitem[\protect\citeauthoryear{{Youdin} \& {Goodman}}{{Youdin} \&
  {Goodman}}{2005}]{Youdin05}
{Youdin} A.~N.,  {Goodman} J.,  2005, \mn@doi [\apj] {10.1086/426895}, \href
  {http://adsabs.harvard.edu/abs/2005ApJ...620..459Y} {620, 459}

\makeatother
\end{thebibliography}




\appendix

\section{Comparison between different pebble isolation masses}
\label{appex1}

Here we compute the planet formation tracks for our fiducial disc adopting now the PIM from \citet{Ataiee18}. The aim is to study how different recipes for the PIM impact in our results. 

\begin{figure}
    \includegraphics[angle= 270, width=\columnwidth]{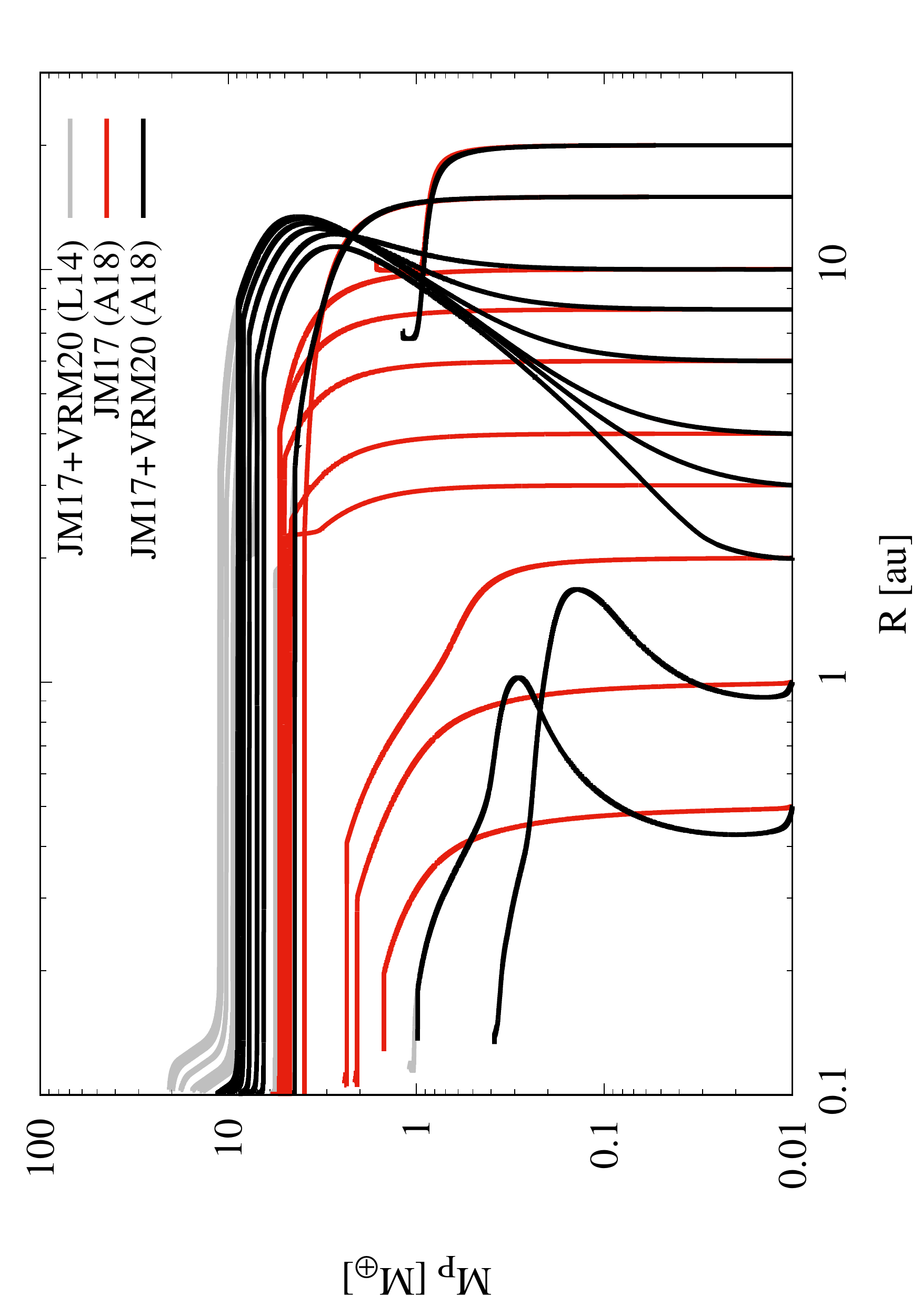}
    \caption{Comparison of the planet formation tracks using different PIM recipes. Simulations correspond to our fiducial disc using $\alpha=10^{-4}$. The red lines correspond to the simulations using the migration recipes from JM17 and PIM from \citet[][A18]{Ataiee18}, the black ones represent the simulation adopting the combined migration prescriptions from JM17 and VRM20 and PIM from \citet[][A18]{Ataiee18}, while the grey lines represent the simulation adopting the combined migration prescriptions from JM17 and VRM20 and PIM from \citet[][L14]{Lambrechts14}.}
    \label{fig1_appex1}
\end{figure}

In Fig.~\ref{fig1_appex1}, we plot the planet formation tracks using different PIM recipes. Simulations correspond to our fiducial disc using $\alpha=10^{-4}$. The general trend of the simulations are the same as discussed above in Sec.~\ref{sec_4}. When the thermal torque is considered, it produces a significant planet outward migration, especially for those planets initially located beyond the ice line. However, we note that in the case where the PIM from \citet{Ataiee18} are used, the differences between simulations with (the red lines) and with out (the black lines) the thermal torque are more evident. In this case, the final masses of the planets when the thermal torque is included are considerably larger. This is related with the fact that when thermal torque is adopted, planets migrate outward and reach in this case larger PIM. For those planets initially located inside the ice line, results are the same as for the case when the PIM from \citet{Lambrechts14} are used, because planets do not reach the PIM, or they reached it at lower masses. When we compare the simulations using the combined migration prescriptions from JM17 and VRM20, we note that in the case of adopting the PIM from \citet{Ataiee18}, planets reach their corresponding PIM at larger distances. However, as this PIM is lower than the corresponding ones from \citet[][the grey lines]{Lambrechts14}, planets end with lower final masses.  

\section{Comparison between different type I migration recipes}
\label{appex2}

In this appendix, we compute the planet formation tracks for our fiducial disc adopting now the type I migration recipes from \citet{Paardekooper2011}. The aim is to study if different type I migration recipes modify our previous results, due to the fact that in Paper I we showed that while the migration recipes from \citet{Paardekooper2011} and JM17 give practically the same values for the Lindblad torques, they differ in the magnitude of the total corotation torque. 

\begin{figure}
    \includegraphics[angle= 270, width=\columnwidth]{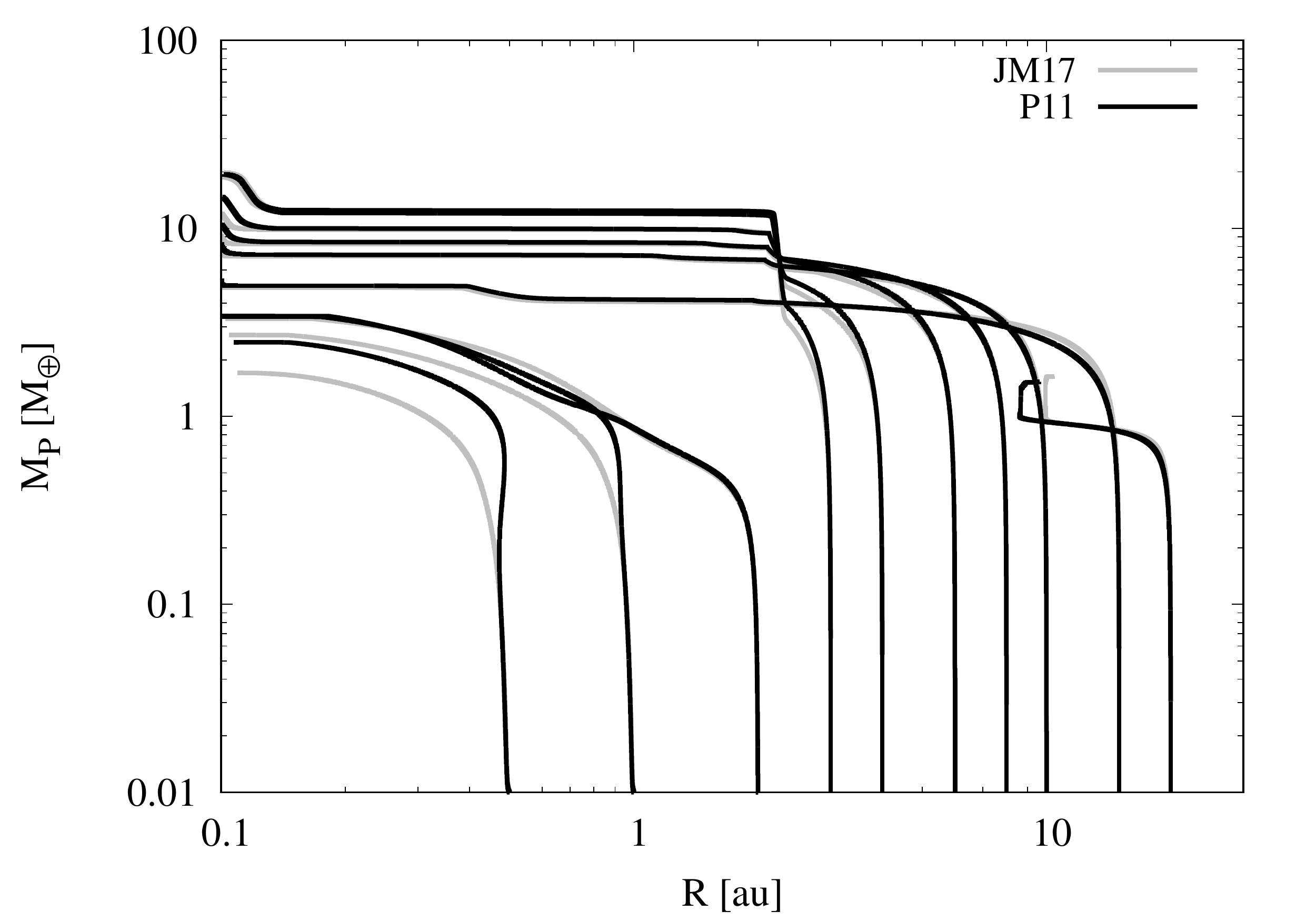} \\
    \includegraphics[angle= 270, width=\columnwidth]{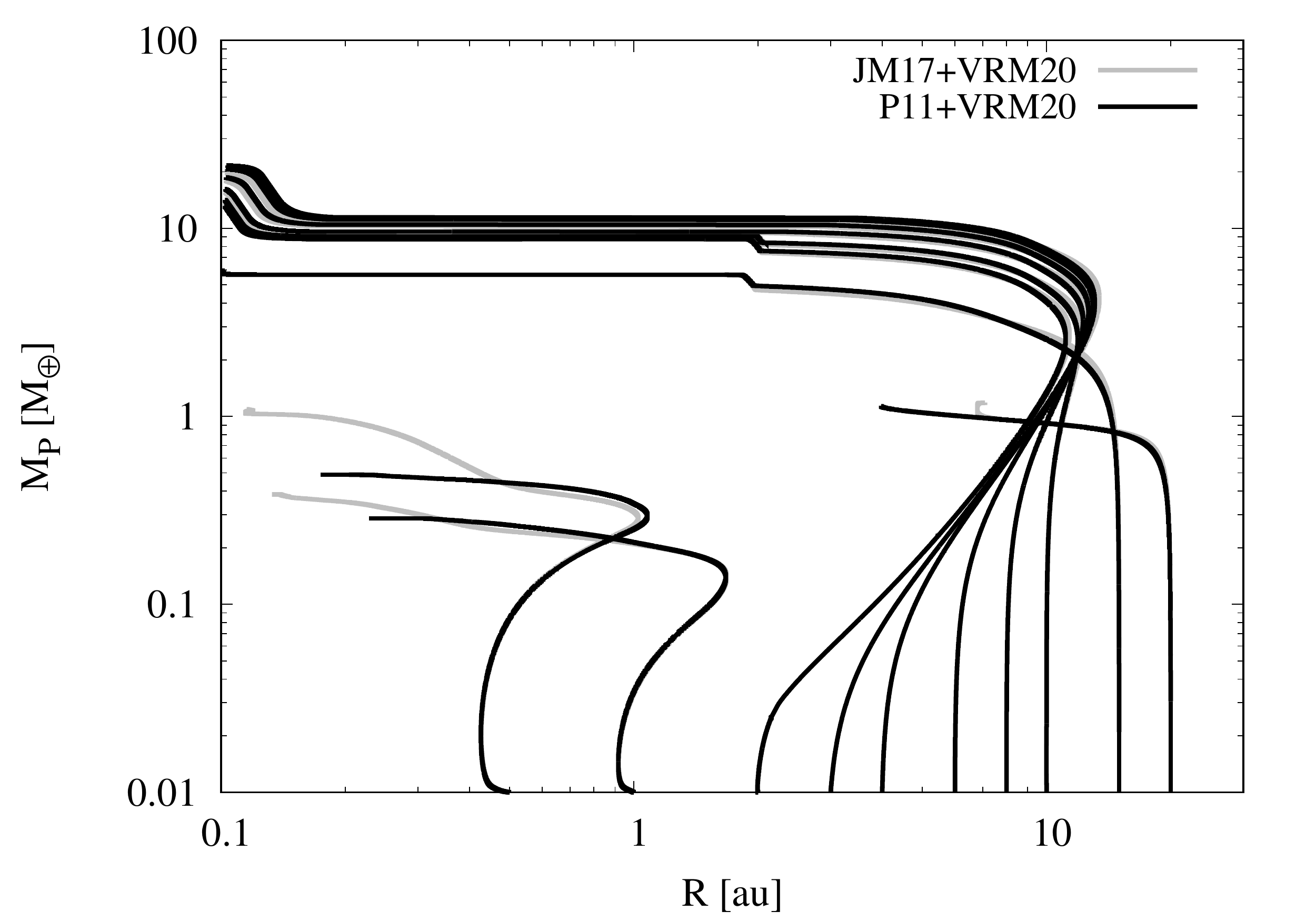} 
    \caption{Planet formation tracks using different type I migration recipes. Simulations correspond to our fiducial disc using $\alpha=10^{-4}$ and PIM from \citet{Lambrechts14}. The top panel correspond for the simulations where thermal torque is not considered. The grey lines correspond to the simulations where the migration recipes from JM17 are used. The black lines represent the simulations using the migration recipes from \citet[][P11]{Paardekooper2011}. The bottom panel shows the planets formation tracks when the thermal torque recipes from VRM20 are combined with the previous type I migration recipes.}
    \label{fig1_appex2}
\end{figure}

In the top panel of Fig.\ref{fig1_appex2}, we plot the planet formation tracks for the simulations where thermal torque is not considered. We can see that for the planets initially located beyond the ice line, planet formation tracks are basically the same. This is due to the fact that, as me showed before, formation timescales are very short for these planets. In these simulations corotation torques do not seem to play an important role. On the contrary, for the planets initially located inside the ice,  at 0.5 and 1~au, corotation torques generate a small outward migration slightly changing their final masses when the migration recipes from \citet{Paardekooper2011} are used (black lines). When the thermal torque is considered (the botton panel of Fig.\ref{fig1_appex2}), we find basically the same results as before. For those planets initially located beyond the ice line formation tracks are practically the same. However, small differences associated to the corotation torques are visible for the planets initially located at 0.5 and 1~au. Thus, these different tipe I migration recipes do not seem to generate different planet formation tracks for planet growing purely by pebble accretion. We note that \citet{Baumann2020} also found a similar results for planets growing purely by pebble accretion without considered the dust growth and evolution. 

\bsp	
\label{lastpage}
\end{document}